\documentclass[journal=jctcce,manuscript=article]{achemso}

\usepackage[version=3]{mhchem} % Formula subscripts using \ce{}
\usepackage[T1]{fontenc}       % Use modern font encodings
\usepackage{lmodern} % avoid compilation problems

\usepackage{color}
\usepackage{algpseudocode}
\usepackage{algorithm}
\usepackage{mathtools}

\usepackage{graphicx}
\usepackage{caption}
\usepackage{subcaption}
\usepackage{amsmath}
\usepackage{paralist}

\DeclareMathOperator*{\tr}{trace}

\newcommand\upperBound{\textrm{in}}
\newcommand\lowerBound{\textrm{out}}

\newcommand\homo{h}
\newcommand\lumo{l}

\newcommand\homoF{\homo}
\newcommand\lumoF{\lumo}

\author{Anastasia Kruchinina}
\affiliation{Division of Scientific
	Computing, Department of Information Technology, Uppsala University, Sweden}
\email{anastasia.kruchinina@it.uu.se}
\author{Elias Rudberg}
\email{elias.rudberg@it.uu.se}
\affiliation{Division of Scientific
	Computing, Department of Information Technology, Uppsala University, Sweden}
\author{Emanuel H. Rubensson}
\email{emanuel.rubensson@it.uu.se}
\affiliation{Division of Scientific
	Computing, Department of Information Technology, Uppsala University, Sweden}

\title[On-the-fly computation of frontal orbitals]{On-the-fly computation of frontal orbitals in
	\\density matrix expansions}

\date{\today}

\begin{document}
\begin{abstract}
	Linear scaling density matrix methods typically do not provide individual eigenvectors and eigenvalues of the Fock/Kohn-Sham matrix, so additional work has to be performed if they are needed. Spectral transformation techniques facilitate computation of frontal (homo and lumo) molecular orbitals. In the purify-shift-and-square method the convergence of iterative eigenvalue solvers is improved by combining recursive density matrix expansion with the folded spectrum method [J. Chem. Phys. 128, 176101 (2008)]. However, the location of the shift in the folded spectrum method and the iteration of the recursive expansion selected for eigenpair computation may have a significant influence on the iterative eigenvalue solver performance and eigenvector accuracy. In this work, we make use of recent homo and lumo eigenvalue estimates [SIAM J. Sci. Comput. 36, B147 (2014)] for selecting shift and iteration such that homo and lumo orbitals can be computed in a small fraction of the total recursive expansion time and with sufficient accuracy. We illustrate our method by performing self-consistent field calculations for large scale systems.
\end{abstract}
\maketitle

%%%%%%%%%%%%%%%%%%%%%%%%%%%%
%  INTRODUCTION
%%%%%%%%%%%%%%%%%%%%%%%%%%%%

\section{Introduction}

Computing interior eigenvalues of large matrices is one of
the most difficult problems in the area of numerical linear algebra and it appears in many applications. This work is inspired by and focuses on the area of electronic structure calculations, in particular Hartree--Fock~\cite{Roothaan} and Kohn--Sham density functional theory~\cite{hohen,KohnSham65}.  Let the eigenvalues of the Fock/Kohn-Sham matrix $F$ be arranged in ascending order
\begin{equation}
	\lambda_1 \leq \lambda_2 \leq  \cdots \leq \lambda_\homoF <
	\lambda_\lumoF \leq \cdots  \leq \lambda_{N-1} \leq \lambda_{N},
\end{equation}
where $\lambda_1, \ldots, \lambda_\homoF$ correspond to occupied electron orbitals, $\lambda_\lumoF, \ldots, \lambda_N$ correspond to unoccupied electron orbitals, and $N$ is the number of basis functions.
The highest occupied molecular orbital (homo) and the lowest unoccupied molecular orbital (lumo), corresponding to the eigenvalues $\lambda_\homoF$  and $\lambda_\lumoF$, respectively, are also known as frontal orbitals in the literature.
We assume here that there is a non-zero gap
$\lambda_\lumoF - \lambda_\homoF > 0$
between eigenvalues corresponding to occupied and unoccupied orbitals.

The analysis of molecular orbitals around the homo-lumo gap helps to describe the structure and properties of chemical compounds~\cite{arjunan2013crystal, echegaray2013pursuit, SureshKumar2015, Kurt2016, Toy2016, Yang2015}. The computation of the homo and lumo orbitals of a given Fock/Kohn-Sham matrix for large systems is of particular interest~\cite{Andermatt2016, inadomi2002definition, watanabe2009fragment}.

The one-electron density matrix $D$ corresponding to a given Fock or
Kohn--Sham matrix $F$ is the matrix for orthogonal projection onto the
subspace spanned by eigenvectors of $F$ that correspond to occupied
orbitals:
\begin{align}
	Fy_i        & = \lambda_i y_i, \label{eq:eig_problem} \\
	D \coloneqq & \sum_{i=1}^{n_\textrm{occ}} y_iy_i^T,
	\label{eq:D_def}
\end{align}
where $n_\textrm{occ}$ is the number of occupied orbitals.

A straightforward approach to obtain the density matrix is a diagonalization of the matrix $F$. It gives the full spectrum of the matrix $F$, but leads to a cubical scaling of the computational cost with system
size, restricting calculations to rather small systems.

Numerous approaches were developed for computing the density matrix with a linear complexity with respect to the number of basis functions~\cite{Bowler_2012}.
With such approaches linear scaling can be achieved for systems with non-zero gap if a technique for imposing matrix sparsity is used~\cite{benzi_decay}.
In recursive density matrix expansion methods the computation of the density matrix is viewed as the problem of evaluating the matrix function
\begin{equation} \label{eq:stepfun}
	D = \theta(\mu I -F),
\end{equation}
where $I$ is the identity matrix, $\theta$ is the Heaviside step function and $\mu$ is located
between the homo and lumo eigenvalues, which makes~\eqref{eq:stepfun}
equivalent to the definition in
\eqref{eq:D_def}~\cite{GoedeckerColombo1994}. Since there are no
eigenvalues in the homo-lumo gap the Heaviside step function can be
efficiently approximated by a function smooth in the homo-lumo gap.
A regularized approximation of the step function can be obtained
using the recursive polynomial expansion
\begin{align}
	D \approx X_n = f_n(f_{n-1}(\ldots f_0(F) \ldots)),
	\label{eq:rec_exp_general}
\end{align}
where $f_i$, $i=0, \ldots, n$ are low-order polynomials.  The
polynomial $f_0$ defined as
\begin{align}
	f_0(x) = \frac{\lambda_\textrm{max} - x}
	{\lambda_\textrm{max} - \lambda_\textrm{min}}
	\label{eq:initial_mapping}
\end{align}
maps the spectrum of the matrix $F$ into the interval $[0, \, 1]$ in
reverse order. Here $\lambda_\textrm{min}$ and $\lambda_\textrm{max}$
are the extremal eigenvalues of $F$ or bounds thereof, i.e.
\begin{equation}
	\lambda_\textrm{min} \leq \lambda_1 \textrm{ and }
	\lambda_\textrm{max} \geq \lambda_N.
\end{equation}
The polynomials $f_i$ are chosen such that they recursively push
occupied eigenvalues to 1 and unoccupied to 0.  Here we focus on the
second-order spectral projection polynomials (SP2) $f_i(x) = x^2$ or
$2x-x^2$, $i=1, \ldots, n$. In the original algorithm proposed by
Niklasson the polynomials used in each iteration are chosen based on
the trace of the matrix~\cite{Nikl2002}. Here, we base our choice of polynomials on estimates of the homo and lumo eigenvalues~\cite{interior_eigenvalues_2014}. A common
approach to decide when to stop recursive expansion iterations is to check when some quantity, measuring how far the matrix is from idempotency, goes below a predetermined threshold value. New parameterless stopping criteria for recursive polynomial expansions were recently proposed in Ref.~\citenum{stop_crit_2016}. Such criteria entail automatic detection of when numerical errors coming from removal of matrix
elements or rounding errors are becoming dominant, therefore preventing stopping if a significantly better result might be obtained and
avoiding failure to converge or performing iterations which do not notably improve the result.

Linear scaling methods avoid explicit computation of the eigenvectors, so to compute homo and lumo eigenpairs (pairs of eigenvalues and the corresponding eigenvectors) additional work has to be performed.
Several techniques have been proposed and used to facilitate the computation of interior  eigenpairs of real symmetric matrices near a given value. Examples include spectrum transformation methods like shift-and-square ~\cite{morgan1991computing, wang1994electronic}, shift-and-invert~\cite{ericsson1980spectral, matyus2009variational} and shift-and-project~\cite{Gomes2017,xiang2007linear} techniques and various subspace methods targeting interior eigenvectors satisfying a given property~\cite{petrenko2017new, tackett2002targeting}. Eigenpairs in a given interval can be obtained with methods using some filtering technique for approximation of the invariant subspace spanned by the eigenvectors of interest~\cite{bekas2008computation,di2016efficient,Levitt2015,li2015thick,Zhou2014770,zhou2006self}.

The shift-and-square technique is based on the transformation $(X-\sigma I)^2$ where $X$ is the matrix of interest and the shift $\sigma$ is a value close to the eigenvalues of interest. This method, also called the folded spectrum method in the literature, has been applied in various applications, including electronic structure calculations~\cite{Wang1994,wang1994electronic,WU1999ThickRestart}, analysis of complex network models~\cite{farkas2001spectra} and for matrices given in a compressed storage scheme~\cite{Mach2013}. In Ref.~\citenum{wang1994electronic} the shift-and-square method was applied directly to the Fock/Kohn-Sham matrix and the shift $\sigma$ was placed within the homo-lumo gap.

In previous works the shift $\sigma$ is chosen rather arbitrarily~\cite{Wang1994,wang1994electronic}, giving no guarantee as to how many eigenvectors will be computed from the occupied and unoccupied parts of the spectrum. Moreover, as noted in Refs.~\citenum{Mach2013} and~\citenum{wang1994electronic}, the
smallest eigenvalues of the transformed problem are much more clustered than the corresponding eigenvalues in the original problem. In addition, spectrum folding may produce multiple eigenvalues in the transformed  problem.
If just homo or lumo eigenpairs are to be obtained, the shift should be located within the homo-lumo gap. However, if the shift is chosen very close to the desired eigenvalue, the eigenvalues from the corresponding invariant subspace will be mapped to poorly separated eigenvalues, resulting in a large number of iterations of the chosen iterative eigenvalue solver, and therefore requiring the execution of many matrix-vector products.

As mentioned earlier the density matrix $D$ is an orthogonal projection onto the occupied subspace which in the shift-and-project technique is utilized for extracting occupied or unoccupied parts of the spectrum~\cite{Gomes2017, xiang2007linear}.
Occupied eigenvalues of $F$ near the homo-lumo gap correspond to the largest eigenvalues of $D(F-\lambda_{\textrm{min}} I)$. Unoccupied eigenvalues of $F$ near the homo-lumo gap correspond to the smallest eigenvalues of $(I-D)(F-\lambda_{\textrm{max}}I)$. Homo and lumo eigenpairs can be found by computing the largest eigenpair of $D(F-\lambda_{\textrm{min}} I)$ and the smallest eigenpair of $(I-D)(F-\lambda_{\textrm{max}}I)$, respectively. 
We will come back briefly to this method in section~\ref{sec:shift_and_project}.

\begin{figure}[ht!]
	\centering
	\includegraphics[width=\textwidth]{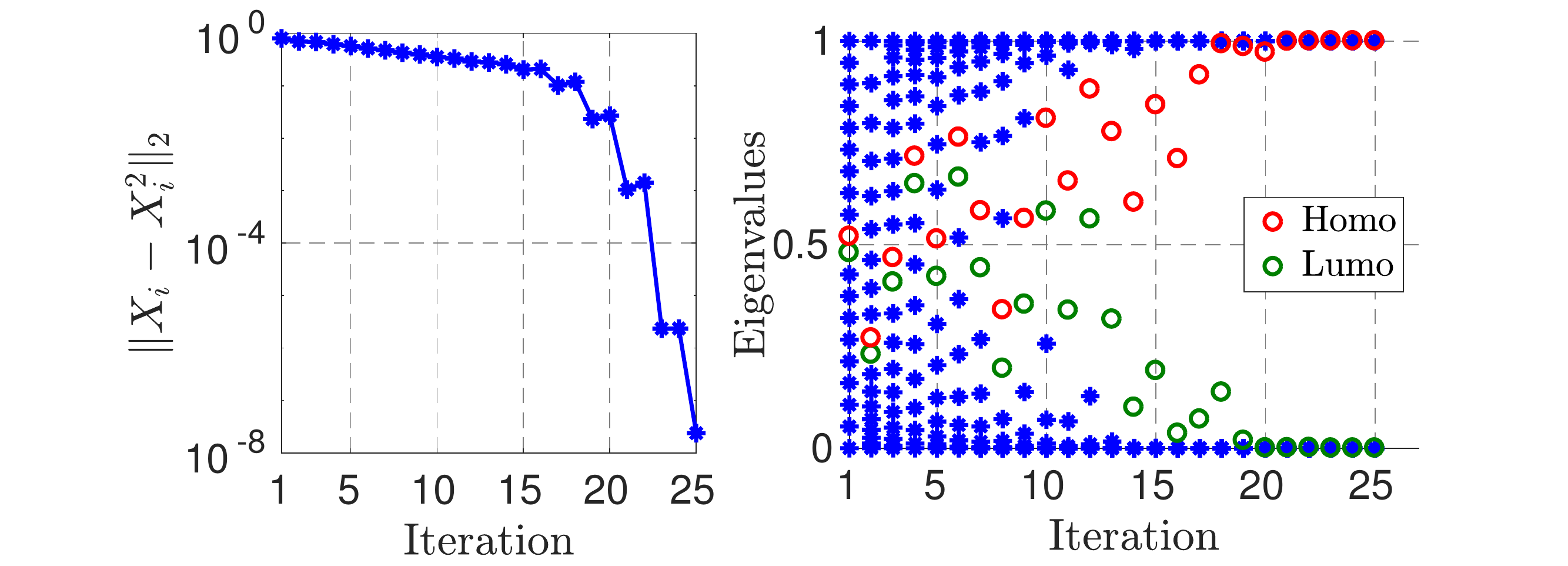}
	\caption{Idempotency error $\|X_i-X_i^2\|$ and eigenvalues of
		$X_i$ in every iteration of the SP2 recursive
		expansion. The matrix $X_0$ is a diagonal matrix with equidistant eigenvalues in the intervals $[0, \,0.48]$ and $[0.52, \,1]$.
    Homo and lumo eigenvalues are represented here
		as red and green empty circles, respectively. }
	\label{fig:rec_exp_separation}
\end{figure}

\begin{figure}[ht!]
	\centering
	\includegraphics[width=0.7\textwidth]{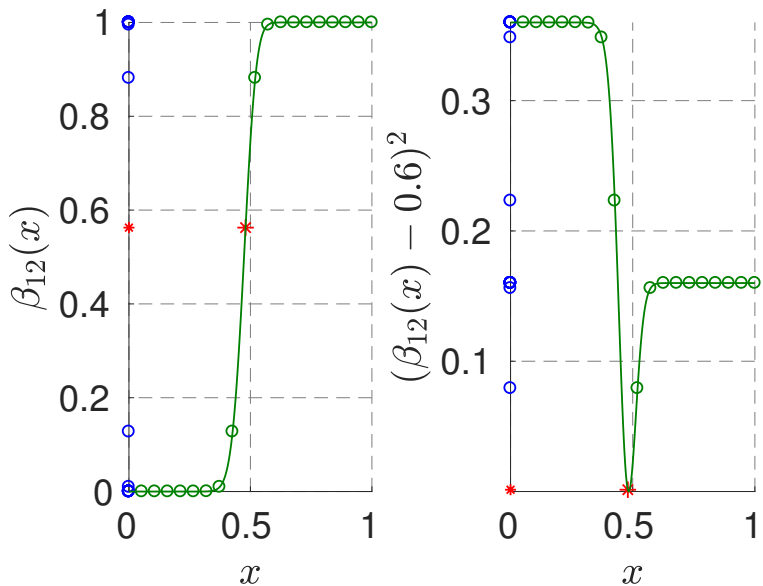}
	\caption{Illustration of the approximation polynomial $\beta_{12}(x)$ and the polynomial filter $(\beta_{12}(x)-\sigma)^2$ obtained in the purify-shift-and-square method in iteration $12$ for a particular choice of $\sigma = 0.6$ in the recursive expansion presented in Figure~\ref{fig:rec_exp_separation}. The lumo eigenvalue is represented using a red star marker.}
	\label{fig:filtering}
\end{figure}

Shift-and-square and shift-and-project techniques can be combined with the SP2
recursive expansion into the methods referred to as purify-shift-and-square and
purify-shift-and-project, respectively~\cite{interior_eigenvalues_2008}. In this
work we will further improve and simplify the purify-shift-and-square method
of Ref.~\citenum{interior_eigenvalues_2008}. The matrix $X_i$ in each iteration of the
recursive expansion can be represented as a matrix function of the matrix $F$,
see~\eqref{eq:rec_exp_general}, and therefore in exact arithmetics $F$ and
$X_i$, $i=0, 1, \ldots$ have a common set of eigenvectors. An important property
of the recursive polynomial expansion is that separation of the eigenvalues near
the homo-lumo gap from the rest of the
spectrum is increasing during the initial phase of the recursive expansion, see
Figure~\ref{fig:rec_exp_separation}~\cite{interior_eigenvalues_2008}.
The polynomial obtained in the SP2 recursive expansion is used here as a filter separating eigenvalues near the homo-lumo gap and consequently the number of eigensolver iterations required to compute corresponding eigenvectors is reduced.
An advantage of the purify-shift-and-square method is that no additional matrix-matrix multiplications are required. The matrix square in the transformation $(X_i-\sigma I)^2 = X_i^2 - \sigma X_i + \sigma^2I$ is anyway needed for the computation of $X_{i+1}$ in the next recursive expansion iteration.
In a chosen iteration $i$ of the
recursive expansion we obtain a polynomial
\begin{align}
	\beta_i(x) := f_i(f_{i-1}(\ldots f_1(x) \ldots)), \quad x\in[0, \, 1].
	\label{eq:approx_poly}
\end{align}
Composition of the polynomial $\beta_i(x)$ with the square polynomial used in the shift-and-square method gives the filter $(\beta_i(x)-\sigma I)^2$
illustrated in Figure~\ref{fig:filtering}, where $\sigma$ is chosen to be inside the homo-lumo gap. The figure shows the sharpness of the obtained filter.

In this work we address questions concerning selection of shift in the homo-lumo gap and an SP2 recursive expansion iteration for computation of the eigenvector of interest using the purify-shift-and-square method. The shift and iteration should ideally be chosen such that the smallest possible number of iterations in an iterative eigenvalue solver is needed while preserving a sufficient eigenvector accuracy.
Our method will make use of the homo and lumo eigenvalue estimates proposed in Ref.~\citenum{interior_eigenvalues_2014}.

\section{Homo and lumo eigenvalue estimates}
\label{sec:eigv_eistimates}

Accurate estimates for homo and lumo eigenvalues can be obtained using information extracted from the recursive expansion. The recent approach described in Ref.~\citenum{interior_eigenvalues_2014} requires only the
evaluation of Frobenius norms and traces of the matrices $X_i-X_i^2$ during the course of a recursive expansion. This approach capitalizes on the fact that the homo and lumo eigenvalues dominate the Frobenius norm and trace of $X_i-X_i^2$ at the end of the recursive expansion. This can be understood from the right panel of Figure~\ref{fig:rec_exp_separation}.

\begin{figure}[ht!]
	\centering
	\captionsetup[subfigure]{justification=centering}
	\begin{subfigure}[b]{0.4\textwidth}
		\includegraphics[width=\textwidth]{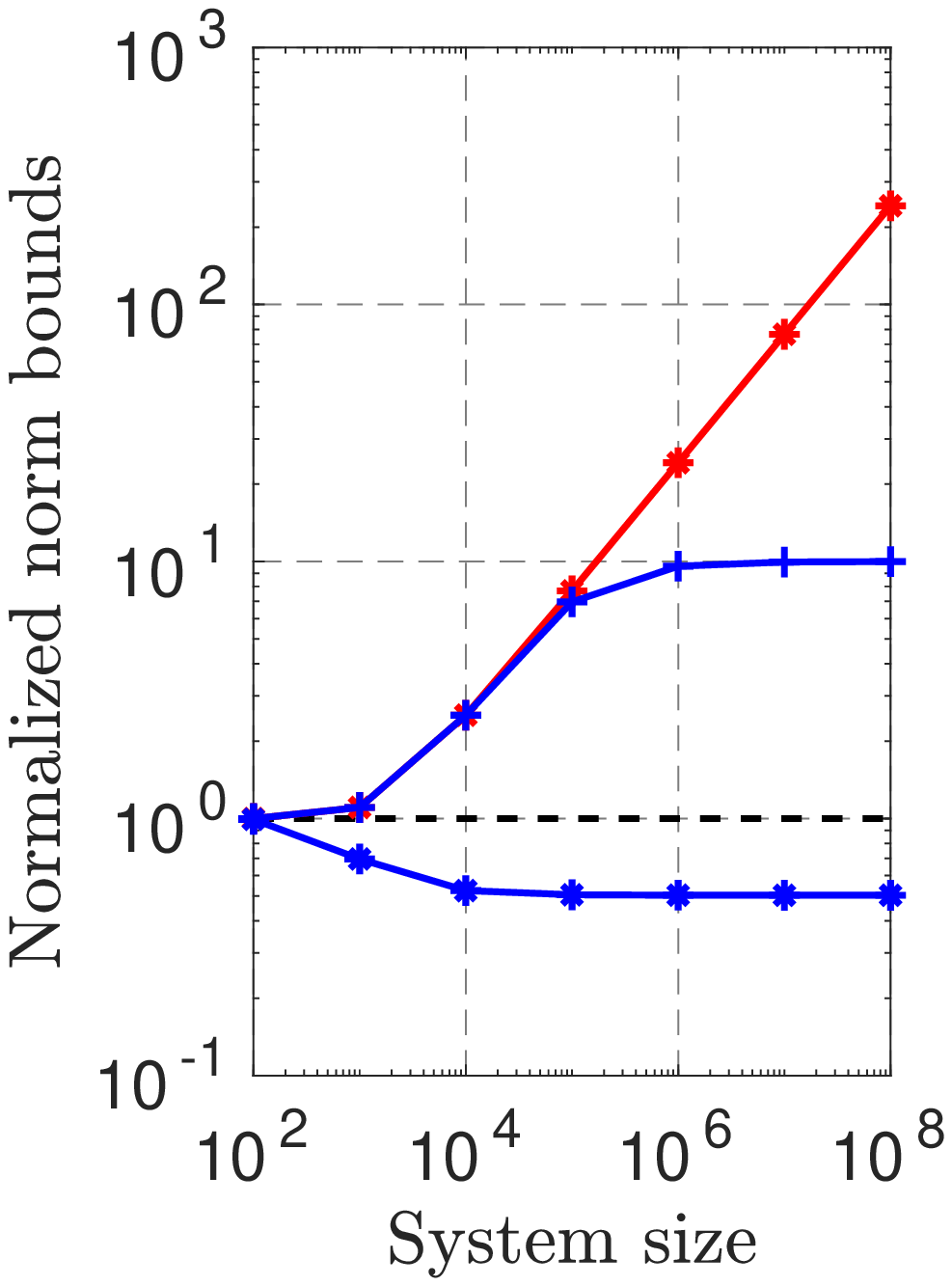}
	\end{subfigure}
	\begin{subfigure}[b]{0.4\textwidth}
		\includegraphics[width=\textwidth]{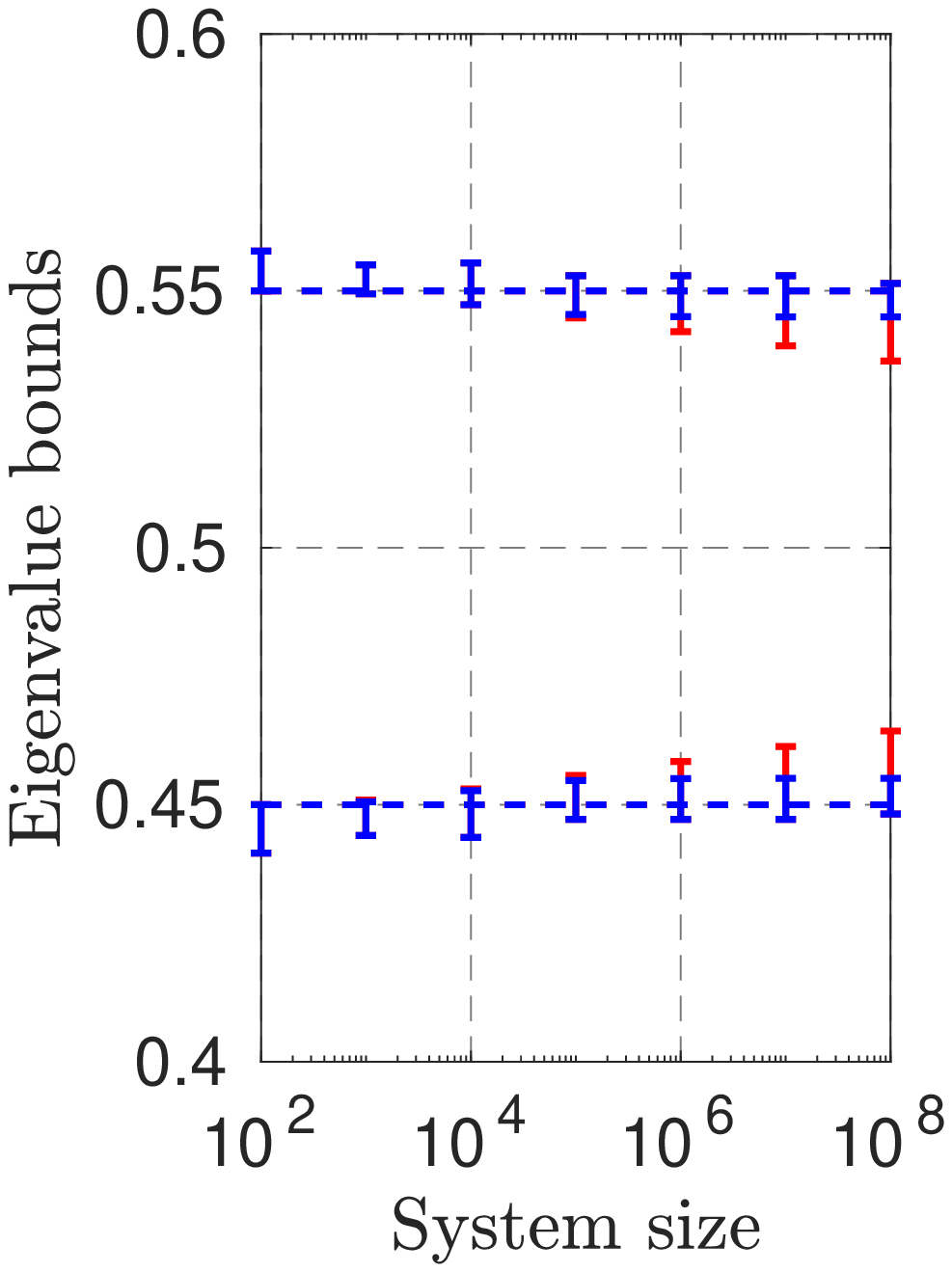}
	\end{subfigure}
	\caption{Illustration of the dependency of the spectral norm bounds and
	eigenvalue bounds on system size. The SP2 algorithm is applied to a  diagonal
	matrix of increasing size with equidistant eigenvalues in $[0, \, 0.45]$ and
	$[0.55, \, 1]$.  Left panel: (from top to bottom) upper bounds obtained using
	the Frobenius norm, upper bounds obtained using the mixed norm, lower bounds for the spectral norm.
	All values have been normalized so that the spectral norm in each case
	corresponds to a value of 1. The bounds are plotted for a matrix $X_i - X_i^2$
	obtained  2 iterations before the end of the recursive expansion. The
	submatrix size for the mixed norm is 100. Right panel: Intervals containing
	homo and lumo eigenvalues obtained using mixed and Frobenius norms for the
	inner bound are shown in blue and red, respectively. The outer bounds are the
	same for both cases.}
	\label{fig:norm_bounds}
\end{figure}

For the eigenvalue $\eta_i$ of $X_i$ that is closest to 0.5 in iteration $i$ the inequalities
\begin{align}
	\frac{v_i^2}{w_i}\leq \eta_i - \eta_i^2 \leq v_i, \quad i =
	1,\ldots, n
	\label{eq:bounds_ineq}
\end{align}
hold, where $v_i = \|X_i-X_i^2\|_F$ is the Frobenius norm of $X_i-X_i^2$ and $w_i = \tr(X_i-X_i^2)$ is the trace of $X_i-X_i^2$.  The approximation of the spectral norm by the Frobenius norm becomes more accurate when the matrix
has a lot of zero eigenvalues. Therefore, we expect the rightmost
bound in~\eqref{eq:bounds_ineq} to become more tight in the last
iterations of the recursive expansion. However, depending on the eigenvalue distribution, the upper bound given by the Frobenius norm may deteriorate with increasing system size, see the left panel of Figure~\ref{fig:norm_bounds} for an example. There is a simple remedy.
The Frobenius norm in the rightmost inequality in~\eqref{eq:bounds_ineq} can be replaced with the so
called mixed norm $\|\cdot \|_m$ first defined in Ref.~\citenum{mixedNormTrunc},
which satisfies
\begin{align}
	\|X_i-X_i^2\|_2 \leq \|X_i-X_i^2\|_m \leq \|X_i-X_i^2\|_F.
\end{align}
The mixed norm can be obtained by dividing the matrix into square submatrices of equal size, padding with zeros if needed, constructing a matrix with
the Frobenius norms of the obtained submatrices and computing the spectral norm of the constructed matrix. Thus the mixed norm is significantly cheaper to compute than the spectral norm for larger submatrix sizes. The mixed norm computed with a fixed submatrix size has the same asymptotic behavior as the spectral norm, and therefore the tightness of the rightmost bound in~\eqref{eq:bounds_ineq} with $m_i:=\|X_i-X_i^2\|_m$  instead of $v_i$ will not deteriorate with increasing system size, see the left panel in Figure~\ref{fig:norm_bounds} for an illustration.

Let $\lambda_\homoF^\textrm{out}$ and
$\lambda_\homoF^\textrm{in}$ be bounds of the homo eigenvalue
such that
\begin{equation}
	\lambda_\homoF^\textrm{out} \leq
	\lambda_\homoF \leq
	\lambda_\homoF^\textrm{in},
\end{equation}
and let $\lambda_\lumoF^\textrm{in}$ and
$\lambda_\lumoF^\textrm{out}$ be bounds of the lumo eigenvalue
such that
\begin{equation}
	\lambda_\lumoF^\textrm{in} \leq
	\lambda_\lumoF \leq \lambda_\lumoF^\textrm{out}.
\end{equation}

The complete algorithm for computing homo and lumo eigenvalue bounds using the mixed norm as described above is given in
Algorithm~\ref{alg:homo_lumo_bounds}, which is used after completion of the recursive expansion. Note that the algorithm is equivalent to Algorithm~3 presented in Ref.~\citenum{interior_eigenvalues_2014}, but here it is written for the SP2 recursive expansion and uses the mixed norm for the computation of inner bounds. In Algorithm~\ref{alg:homo_lumo_bounds}, $n$ is the total number of recursive expansion iterations and $p_i$ is the sequence of taken polynomials, where $p_i = 1$
if $f_i(x) = x^2$ and $p_i = 0$ if $f_i(x) = 2x-x^2$.

\begin{algorithm}\caption{Homo and lumo estimates obtained using information extracted from SP2
		\label{alg:homo_lumo_bounds}}
	\begin{algorithmic}[1]
		\State \textbf{input:} $n$, $\lambda_\textrm{max}$, $\lambda_\textrm{min}$, $v_i$, $m_i$, $w_i$,  $p_i$, $i=1,2,\dots,n$
		\State $\gamma = (3-\sqrt{5})/2$
		\State $i=n$
		\State $x_1 = 1$, $x_2 = 1$, $x_3 = 0$, $x_4 = 0$
		\While{ $v_i < \gamma - \gamma^2$ } \label{line:v_i_to_m_i}
		\State $z_1 = \frac{1}{2} \left(1-\sqrt{1-4v_i^2/w_i}\right)$
		\State $z_2 = \frac{1}{2} \left(1-\sqrt{1-4m_i}\right)$
		\State $z_3 = \frac{1}{2} \left(1+\sqrt{1-4m_i}\right)$
		\State $z_4 = \frac{1}{2} \left(1+\sqrt{1-4v_i^2/w_i}\right)$
		\For{$j = i, i-1, \ldots, 1$}
		\If{$p_j = 1$}
		\State $z_k = \sqrt{z_k}$, \quad $k=1, 2, 3, 4$
		\Else
		\State $z_k = 1-\sqrt{1-z_k}$, \quad $k=1, 2, 3, 4$
		\EndIf
		\EndFor
		\State $x_k = \min{(x_k, z_k)}$, \quad $k=1, 2$ \label{line:outer_bound_1}
		\State $x_k = \max{(x_k, z_k)}$, \quad $k=3, 4$ \label{line:outer_bound_2}
		\State $i = i-1$
		\EndWhile
		\State $\lambda_\homoF^\textrm{out} = \lambda_\textrm{max} - (\lambda_\textrm{max}-\lambda_\textrm{min})x_4$
		\State $\lambda_\homoF^\textrm{in} = \lambda_\textrm{max} - (\lambda_\textrm{max}-\lambda_\textrm{min})x_3$
		\State $\lambda_\lumoF^\textrm{in} = \lambda_\textrm{max} - (\lambda_\textrm{max}-\lambda_\textrm{min})x_2$
		\State $\lambda_\lumoF^\textrm{out} = \lambda_\textrm{max} - (\lambda_\textrm{max}-\lambda_\textrm{min})x_1$
		\State \textbf{output:} $\lambda_\homoF^\textrm{out}$, $\lambda_\homoF^\textrm{in}$, $\lambda_\lumoF^\textrm{in}$, $\lambda_\lumoF^\textrm{out}$
	\end{algorithmic}
\end{algorithm}

See the right panel in Figure~\ref{fig:norm_bounds} for a comparison of the eigenvalue bounds obtained using $m_i$ instead of $v_i$ in the rightmost inequality in~\eqref{eq:bounds_ineq}. We can see that the inner bounds computed using the mixed norm is tighter than the bounds computed using the Frobenius norm for large system sizes.
In each iteration in Algorithm~\ref{alg:homo_lumo_bounds} the most loose outer bounds are taken,
see lines~\ref{line:outer_bound_1} and~\ref{line:outer_bound_2}. However, the  Frobenius norm significantly overestimates the spectral norm for non-trivial large matrices, so for larger matrix sizes fewer iterations are taken into account due to the condition on line~\ref{line:v_i_to_m_i}.
Therefore, in the right panel in Figure~\ref{fig:norm_bounds} the outer bounds become more tight for larger system sizes. For even larger systems the condition on line~\ref{line:v_i_to_m_i} may never be satisfied. To be able to handle also such cases one may use the mixed norm $m_i$ instead of the Frobenius norm $v_i$ on line~\ref{line:v_i_to_m_i}, i.e. one may use the condition $m_i < \gamma - \gamma^2$.

The intervals obtained in Algorithm~\ref{alg:homo_lumo_bounds}  are propagated between self-consistent field (SCF) cycles using Weyl's theorem for eigenvalue movement as described in Refs.~\citenum{interior_eigenvalues_2014} and~\citenum{m-accPuri}. Propagated bounds are used for selecting polynomials in the next SCF cycle.

Let the inner and outer bounds for the homo eigenvalue of  $X_i$ be denoted with $\homo_\upperBound^{i}$ and $\homo_\lowerBound^{i}$ respectively, and the inner and outer bounds for the lumo eigenvalue with $\lumo_\upperBound^{i}$ and
$\lumo_\lowerBound^{i}$ respectively.  In
Algorithm~\ref{alg:homo_lumo_bounds_every_iter} we compute bounds for
each iteration and use them to determine the sequence of
polynomials which will be used in the recursive expansion.  Here $\lambda_\homoF^\textrm{out}$, $\lambda_\homoF^\textrm{in}$,
$\lambda_\lumoF^\textrm{in}$, $\lambda_\lumoF^\textrm{out}$ are bounds for the homo and lumo eigenvalues of the matrix $F$ propagated from the previous SCF cycle. With $\varepsilon_M$ we denote the machine epsilon.

\begin{algorithm}\caption{Computation of the polynomial sequence in SP2~\label{alg:homo_lumo_bounds_every_iter}}
	\begin{algorithmic}[1]
		\State \textbf{input:} $\lambda_\textrm{min}$, $\lambda_\textrm{max}$,
		$\lambda_\homoF^\textrm{out}$,
		$\lambda_\homoF^\textrm{in}$,
		$\lambda_\lumoF^\textrm{in}$,
		$\lambda_\lumoF^\textrm{out}$

		\State $\widehat{\homo}_\upperBound^{0} = 1-
		\frac{\lambda_\textrm{max} - \lambda_\homoF^\textrm{in}}
		{\lambda_\textrm{max} - \lambda_\textrm{min}}$, \quad
		$\widehat{\homo}_\lowerBound^{0} = 1-
		\frac{\lambda_\textrm{max} - \lambda_\homoF^\textrm{out}}
		{\lambda_\textrm{max} - \lambda_\textrm{min}}$
		\State $\lumo_\upperBound^{0} =
		\frac{\lambda_\textrm{max} - \lambda_\lumoF^\textrm{in}}
		{\lambda_\textrm{max} - \lambda_\textrm{min}}$, \quad
		$\lumo_\lowerBound^{0} =
		\frac{\lambda_\textrm{max} - \lambda_\lumoF^\textrm{out}}
		{\lambda_\textrm{max} - \lambda_\textrm{min}}$
		\State $i = 0$
		\While {($\widehat{\homo}_\upperBound^{i} > \varepsilon_M $ or $\lumo_\upperBound^{i} > \varepsilon_M$)}
		\State $i = i + 1$
		\If{$\lumo_\upperBound^{i-1} \geq \widehat{\homo}_\upperBound^{i-1}$}
		\State $\lumo_\lowerBound^{i} = \left(\lumo_\lowerBound^{i-1}\right)^2,
		\quad
		\lumo_\upperBound^{i} =\left(\lumo_\upperBound^{i-1}\right)^2$

		\State $\widehat{\homo}_\lowerBound^{i} = 2\widehat{\homo}_\lowerBound^{i-1}-\left(\widehat{\homo}_\lowerBound^{i-1}\right)^2,
		\quad
		\widehat{\homo}_\upperBound^{i} = 2\widehat{\homo}_\upperBound^{i-1}-\left(\widehat{\homo}_\upperBound^{i-1}\right)^2$
		\State $p_{i} = 1$
		\Else
		\State $\lumo_\lowerBound^{i} = 2\lumo_\lowerBound^{i-1}-\left(\lumo_\lowerBound^{i-1}\right)^2,
		\quad
		\lumo_\upperBound^{i} = 2\lumo_\upperBound^{i-1}-\left(\lumo_\upperBound^{i-1}\right)^2$
		\State $\widehat{\homo}_\lowerBound^{i} = \left(\widehat{\homo}_\lowerBound^{i-1}\right)^2,
		\quad
		\widehat{\homo}_\upperBound^{i} =  \left(\widehat{\homo}_\upperBound^{i-1}\right)^2$
		\State $p_{i} = 0$
		\EndIf
		\State $\homo_\upperBound^{i} = 1-\widehat{\homo}_\upperBound^{i}$, $\homo_\lowerBound^{i} = 1-\widehat{\homo}_\lowerBound^{i}$
		\EndWhile
		\State $n_{\textrm{max}} = i$
		\State \textbf{output:} $n_{\textrm{max}}$, $p_i$,
		$\homo_\lowerBound^{i}$,
		$\homo_\upperBound^{i}$,
		$\lumo_\lowerBound^{i}$,
		$\lumo_\upperBound^{i}$,
		$i=0,1,\dots,n_{\textrm{max}}$
	\end{algorithmic}
\end{algorithm}

\section{Computation of homo and lumo eigenvectors}
\label{sec:folded_method}

In this section we will consider a purify-shift-and-square approach for computing homo and lumo eigenpairs of the matrix $F$. We will use the homo and lumo eigenvalue bounds obtained using the method discussed in section~\ref{sec:eigv_eistimates} to
select the shifts and the recursive expansion iterations for computing homo and lumo eigenvectors.

Consider a symmetric matrix $X_i$ in a given iteration $i$ of the recursive expansion and note that its spectrum is located in the interval
$[0, \, 1]$.  Assume that the eigenvalue of $X_i$ closest to a given value $\sigma_i\in[0,\ 1]$ is unique and that there is no other eigenvalue at the same distance from $\sigma_i$. The eigenvector of $X_i$ corresponding to this eigenvalue is equal to the eigenvector corresponding to the smallest eigenvalue of
\begin{align}
	g(X_i, \sigma_i)=(X_i-\sigma_i I)^2.
	\label{eq:mapping_g}
\end{align}
After this transformation, the eigenvalues of $X_i$ near the shift $\sigma_i$ have the same eigenvectors as the smallest eigenvalues of $g(X_i, \sigma_i)$. Computation of extremal eigenpairs is favorable for iterative eigensolvers such as the Lanczos method.  If homo and lumo estimates are available, we can choose $\sigma_i$ such that the homo or lumo eigenpair is computed. Then, using the Rayleigh quotient we compute the corresponding homo or lumo eigenvalue of the matrix $F$.

Let the lumo and homo eigenvalues of the matrix $X_i$ be uniquely close to the shifts $\sigma^i_\lumo$ and $\sigma^i_\homo$, respectively. Let $i_\lumo$ and $i_\homo$ be the chosen recursive expansion iterations for computing lumo and homo eigenvectors. We summarize the purify-shift-and-square approach in Algorithm~\ref{alg:folded_method}. In line~\ref{line:truncation} the removal of small matrix elements, or truncation, is written as an addition of an explicit perturbation matrix $E_i$ to $X_i$.
Particular choices of shifts and iterations will be suggested in the following sections.

% Algorithm:
\begin{algorithm}\caption{Purify-shift-and-square approach for computing homo and lumo eigenpairs of the matrix $F$\label{alg:folded_method}}
	\begin{algorithmic}[1]
		\State \textbf{input:} $F$, $p_i$,
		$i_\lumo$, $i_\homo$, $\sigma^i_\lumo$, $\sigma^i_\homo$, $\lambda_\textrm{max}$, $\lambda_\textrm{min}$
		\State $X_0 =  \frac{\lambda_\textrm{max}I - F}
		      {\lambda_\textrm{max} - \lambda_\textrm{min}}$
    \State $i=1$
		\While{stopping criterion is not fulfilled}
		\If{$p_i = 1$}
		%\State $X_i = ((1-\alpha_i)I+\alpha_i\widetilde{X}_{i-1})^2$
		\State $X_i = \widetilde{X}_{i-1}^2$
		\Else
		%\State $X_i = 2\alpha_i\widetilde{X}_{i-1}-(\alpha_i\widetilde{X}_{i-1})^2$
		\State $X_i = 2\widetilde{X}_{i-1}-\widetilde{X}_{i-1}^2$
		\EndIf
		\State $\widetilde{X}_i = X_i + E_i$ \label{line:truncation}
		\If{$i = i_\homo$}
		\State $y_\homoF$ := eigenvector for smallest eigenvalue of   $(\widetilde{X}_i-\sigma^i_\homo I)^2$
		\State $\lambda_\homoF =  (y_\homoF^T F y_\homoF) / \|y_\homoF\|$
		\EndIf
		\If{$i = i_\lumo$}
		\State $y_\lumoF$ := eigenvector for smallest eigenvalue of  $(\widetilde{X}_i-\sigma^i_\lumo I)^2$
		\State $\lambda_\lumoF = (y_\lumoF^T F y_\lumoF )/ \|y_\lumoF\|$
		\EndIf
    \State $i=i+1$
		\EndWhile
		\State \textbf{output:} $(\lambda_\lumoF,
		y_\lumoF)$, $(\lambda_\homoF, y_\homoF)$
	\end{algorithmic}
\end{algorithm}

\subsection{Selecting shifts}

Above we have discussed the purify-shift-and-square method for computing homo and lumo eigenpairs of the Fock/Kohn-Sham matrix, where the shift-and-square method is applied to the matrix $X_i$ in a given iteration $i$ of the recursive expansion. In this subsection we will propose particular choices of shift $\sigma_i$ in the mapping~\eqref{eq:mapping_g}. We denote with
$\sigma^{i}_{\homo}$ and $\sigma^{i}_{\lumo}$ the shifts chosen in iteration $i$ for computation of homo and lumo eigenvectors, respectively. We base our choice of shifts on the homo and lumo eigenvalue bounds obtained as described in section~\ref{sec:eigv_eistimates}.
Let $\homo^{i}$ and $\lumo^{i}$ be the exact homo and lumo eigenvalues, respectively, of the matrix $X_i$.
In each iteration of Algorithm~\ref{alg:homo_lumo_bounds} the inner bounds for homo and lumo remain the same or improve, so one may expect to have tight inner bounds for homo and lumo of the matrix $F$. On the contrary, to improve the reliability of the outer bounds for $F$, the looser outer bounds for homo and lumo are taken in each iteration of Algorithm~\ref{alg:homo_lumo_bounds}. Therefore, in practice, the outer bounds of $F$ are looser than the inner bounds. Thus we will assume that
\begin{align}
  \lumo_\lowerBound^{i} < \lumo^{i} \leq \lumo_\upperBound^{i} \text{ and } \homo_\upperBound^{i} \leq \homo^{i} < \homo_\lowerBound^{i},
\label{eq:assump_eig_bounds}
\end{align}
which is true in all but trivial cases.

\begin{figure}[ht!]
	\centering
	\includegraphics[width=0.6\textwidth]{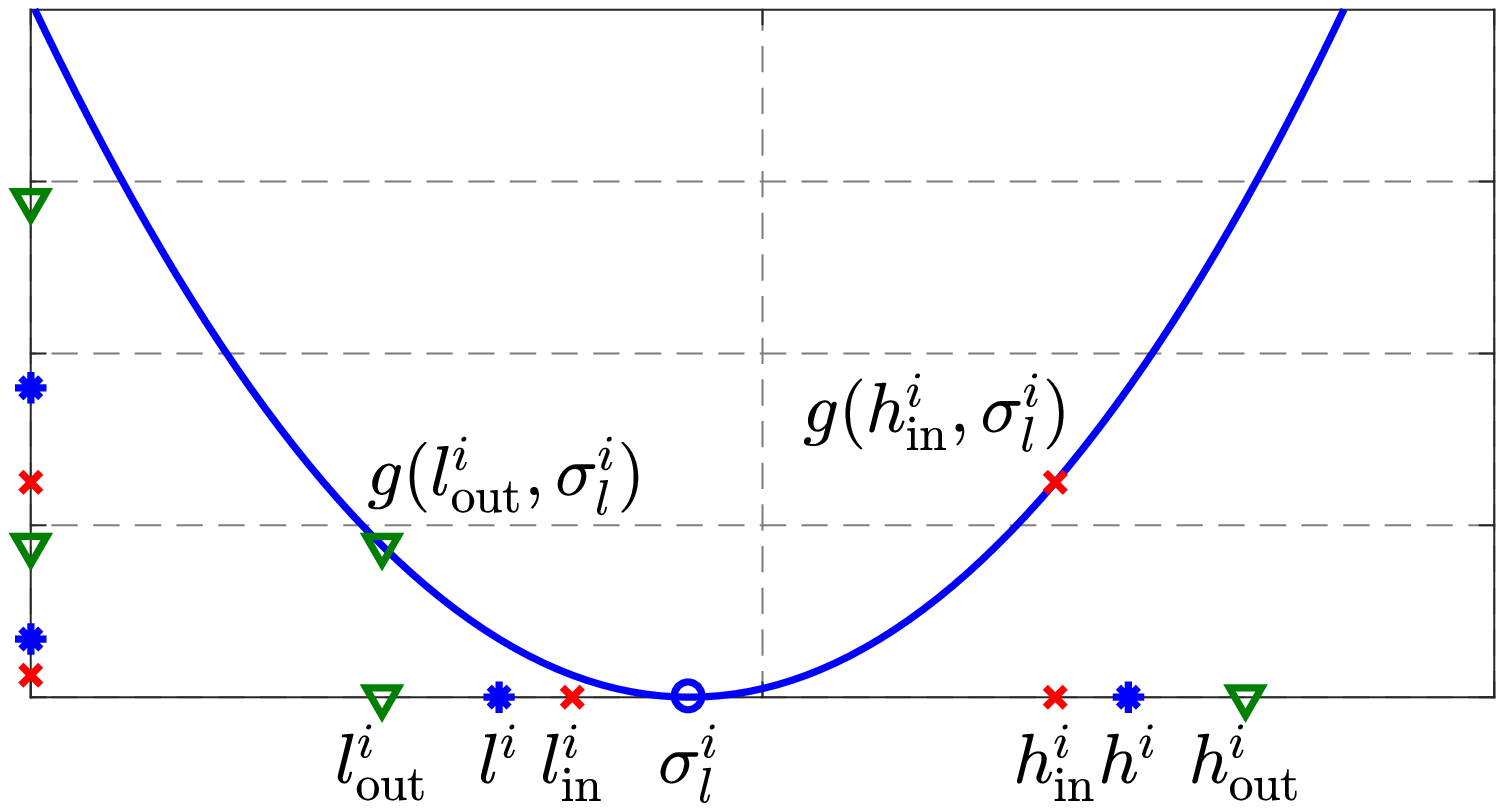}
	\caption{Mapping of homo and lumo eigenvalues and their estimates by the $g(x, \sigma^{i}_{\lumo})$ function. To ensure that the smallest eigenvalue of $g(X_i, \sigma^{i}_{\lumo})$ and the lumo eigenvalue of $X_i$ have the same eigenvector the two conditions~\eqref{eq:cond_lumo_1} and~\eqref{eq:cond_lumo_2} must be satisfied.}
	\label{fig:mapping_g_general}
\end{figure}

The shift $\sigma^{i}_{\lumo}$ should be chosen in such a way that the smallest eigenvalue of the matrix $g(X_i, \sigma^i_{\lumo})$ is simple and corresponds to the same eigenvector as the lumo eigenvalue of $X_i$.
To achieve this, two conditions should be satisfied:
\begin{align}
	\lumo_\upperBound^{i}                        & \leq \sigma^{i}_{\lumo} \label{eq:cond_lumo_1}                            \\
	g(\lumo_\lowerBound^{i}, \sigma^{i}_{\lumo}) & \leq g(\homo_\upperBound^{i},  \sigma^{i}_{\lumo}) \label{eq:cond_lumo_2}
\end{align}
See Figure~\ref{fig:mapping_g_general} for an illustration.
The condition~\eqref{eq:cond_lumo_1} ensures that no eigenvalue from the unoccupied part of the spectrum except lumo  will be transformed to the smallest eigenvalue of $g(X_i, \sigma^i_{\lumo})$. The condition~\eqref{eq:cond_lumo_2} ensures that no eigenvalue from the occupied part of the spectrum  will be transformed to the smallest eigenvalue of $g(X_i, \sigma^i_{\lumo})$.
Condition~\eqref{eq:cond_lumo_2} is equivalent to
\begin{align}
	\sigma^{i}_{\lumo} \leq \frac{\homo_\upperBound^{i}+\lumo_\lowerBound^{i}}{2}.
\end{align}
Therefore by selecting in the recursive expansion iteration $i$  shift $\sigma^{i}_{\lumo}$ such that
\begin{align}
	\lumo_\upperBound^{i} & \leq \sigma^{i}_{\lumo} \leq \frac{\homo_\upperBound^{i}+\lumo_\lowerBound^{i}}{2}
	\label{eq:shift_bounds_lumo}
\end{align}
we know a priori that by computing the smallest eigenpair of the matrix $g(X_i, \sigma^i_{\lumo})$ we will obtain the lumo eigenvector.

Similarly, we get the condition
\begin{align}
	\frac{\lumo_\upperBound^{i}+\homo_\lowerBound^{i}}{2}\leq\sigma^i_{\homo}
	\leq \homo_\upperBound^{i}
	\label{eq:shift_bounds_homo}
\end{align}
for the shift $\sigma^{i}_{\homo}$ ensuring that by computing the smallest eigenpair of the matrix $g(X_i, \sigma^i_{\homo})$ we will obtain the homo eigenvector.

We have narrowed down the possible choices of shifts.  Now we will propose particular choices of shifts keeping in mind bounds ~\eqref{eq:shift_bounds_lumo} and ~\eqref{eq:shift_bounds_homo}.

It is well known that iterative eigensolvers are sensitive to the relative eigenvalue separation~\cite{GolubMatrixComp,stathopoulos1998dynamic}. For a simple eigenvalue $\eta_1$ such that
$\eta_1 < \eta_2 \leq \cdots  \leq \eta_N$ we define the relative eigenvalue distance as
\begin{align}
	\delta_{\textrm{rel}}(\eta ) = \frac{\eta_2 - \eta_1}{\eta_N - \eta_1}
  \label{eq:rel_sep_formula}
\end{align}
If $\delta_{\textrm{rel}}(\eta )$ is high, then iterative eigensolvers typically require much fewer iterations than in case $\delta_{\textrm{rel}}(\eta )$ is small.
Thus we would like to choose  $\sigma^{i}_{\homo}$ and  $\sigma^{i}_{\lumo}$ giving large ratio $\delta_{\textrm{rel}}$ for the smallest eigenvalues of  $g(X_i, \sigma^{i}_{\homo})$ and $g(X_i, \sigma^{i}_{\lumo})$, respectively. Below we discuss the choice of $\sigma^{i}_{\lumo}$ and provide illustrative examples. A similar discussion can be done for the choice of $\sigma^{i}_{\homo}$.

\begin{figure}[ht!]
	\centering
	\captionsetup[subfigure]{justification=centering}
	\centering
	\begin{subfigure}[t]{0.3\textwidth}
		\centering
		\includegraphics[width=\textwidth]{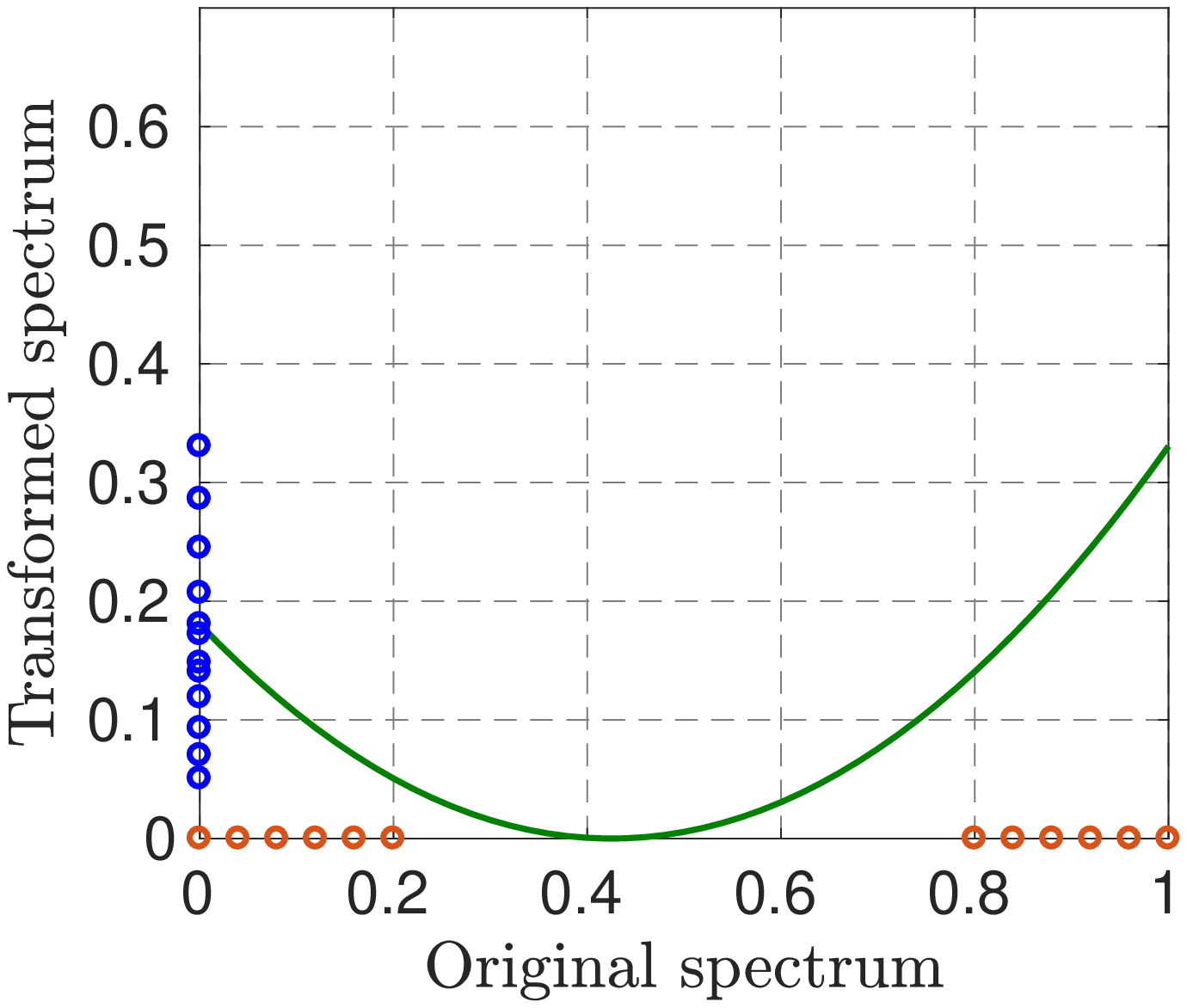}
		\caption{$\sigma = 0.425$}
	\end{subfigure}
	\centering
	\begin{subfigure}[t]{0.3\textwidth}
		\centering
		\includegraphics[width=\textwidth]{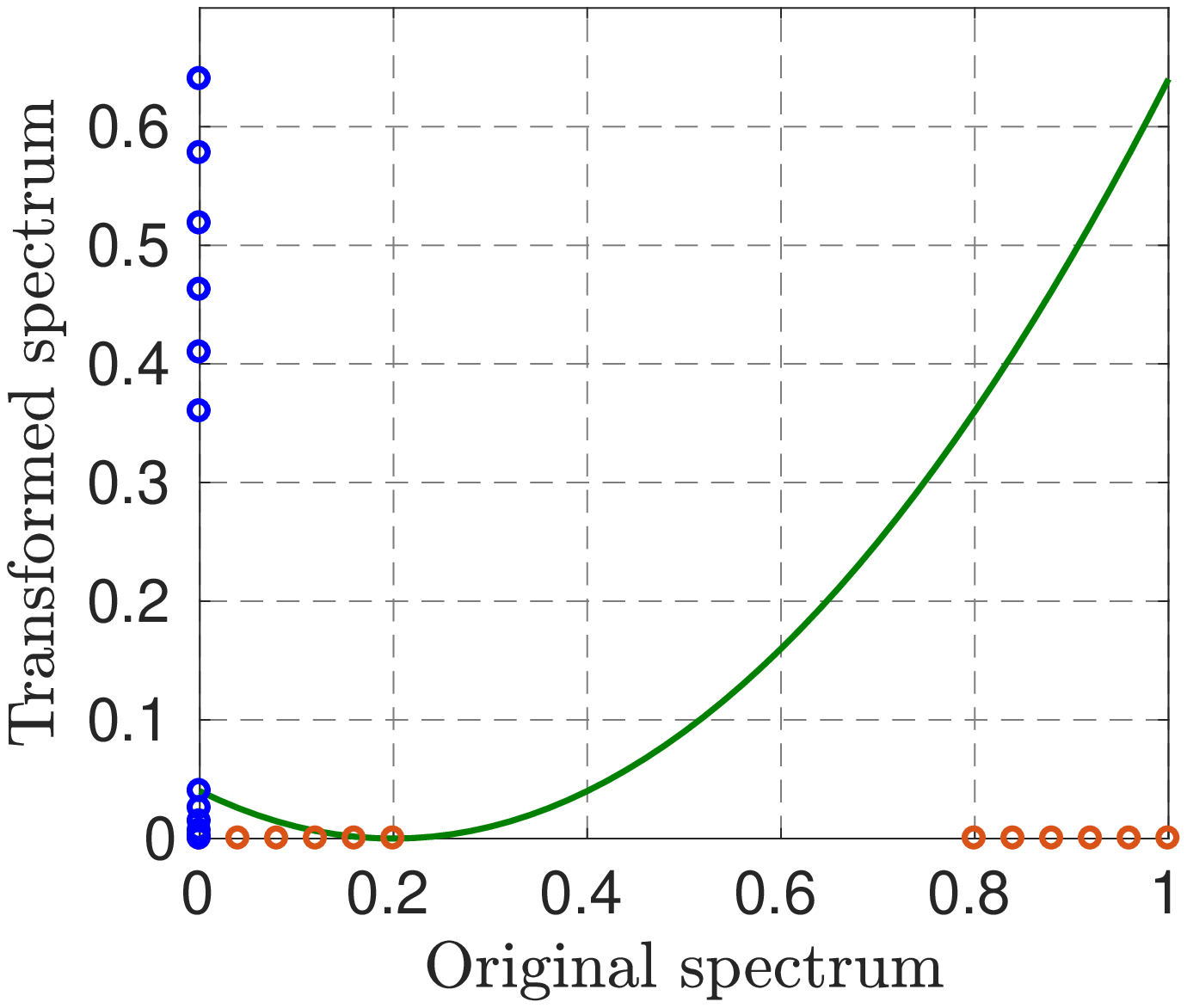}
		\caption{$\sigma = 0.2$}
	\end{subfigure}
  \begin{subfigure}[t]{0.3\textwidth}
    \centering
    \includegraphics[width=\textwidth]{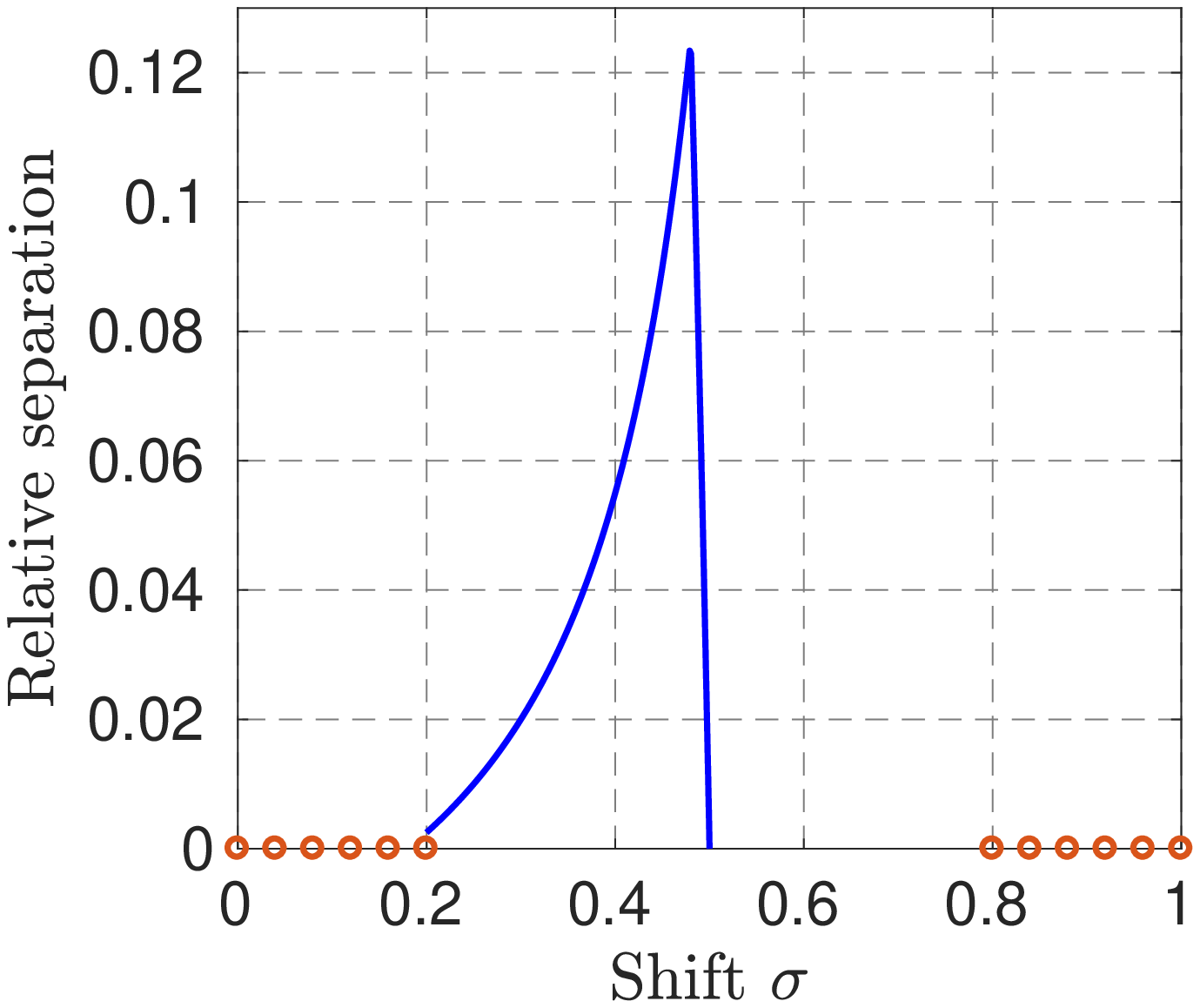}
    \caption{Relative separation}
  \end{subfigure}
	\caption{Illustration of spectrum transformation using the shift-and-square approach. The original matrix spectrum contains equidistant eigenvalues in the intervals $[0,\,0.2]$ and $[0.8,\,1]$ depicted as red circles on the x-axes. We assume that the lumo and homo eigenvalues are equal to 0.2 and 0.8, respectively.
  Panels (a) and (b) present the spectrum transformed using the polynomial $(x-\sigma)^2$ with two choices of shift $\sigma$. Note the compression of the smallest eigenvalues in the transformed spectrum.
  Panel (c) shows relative separation~\eqref{eq:rel_sep_formula} in the transformed spectrum for shifts satisfying~\eqref{eq:shift_bounds_lumo}.
   }
	\label{fig:mapping_shift_and_square}
\end{figure}

In Figure~\ref{fig:mapping_shift_and_square} we compare spectrum transformations using the polynomial $g(x, \sigma)$ with $\sigma = 0.425$ and $\sigma = 0.2$. The original spectrum is separated into occupied and unoccupied parts, with 0.5 located in the homo-lumo gap. The relative distance for the smallest eigenvalue of the transformed spectrum is significantly smaller if shift $\sigma = 0.2$ is used than when $\sigma = 0.425$ is used. The example suggests to choose shift equal to the right bound of the interval~\eqref{eq:shift_bounds_lumo}, i.e.
$\sigma = \frac{\homo_\upperBound^{i}+\lumo_\lowerBound^{i}}{2}$, when the left bound of~\eqref{eq:shift_bounds_lumo} is satisfied. Such a choice reduces the compression of the smallest eigenvalues in the transformed spectrum. Note that if in the example above shift 0.5 is chosen, the transformed spectrum will have multiple smallest eigenvalues, and the eigenvector of interest cannot be computed. However, such situation will not happen in practice if assumption~\eqref{eq:assump_eig_bounds} holds.

\begin{figure}[ht!]
	\centering
	\captionsetup[subfigure]{justification=centering}
	\centering
	\begin{subfigure}[t]{0.45\textwidth}
		\centering
		\includegraphics[width=\textwidth]{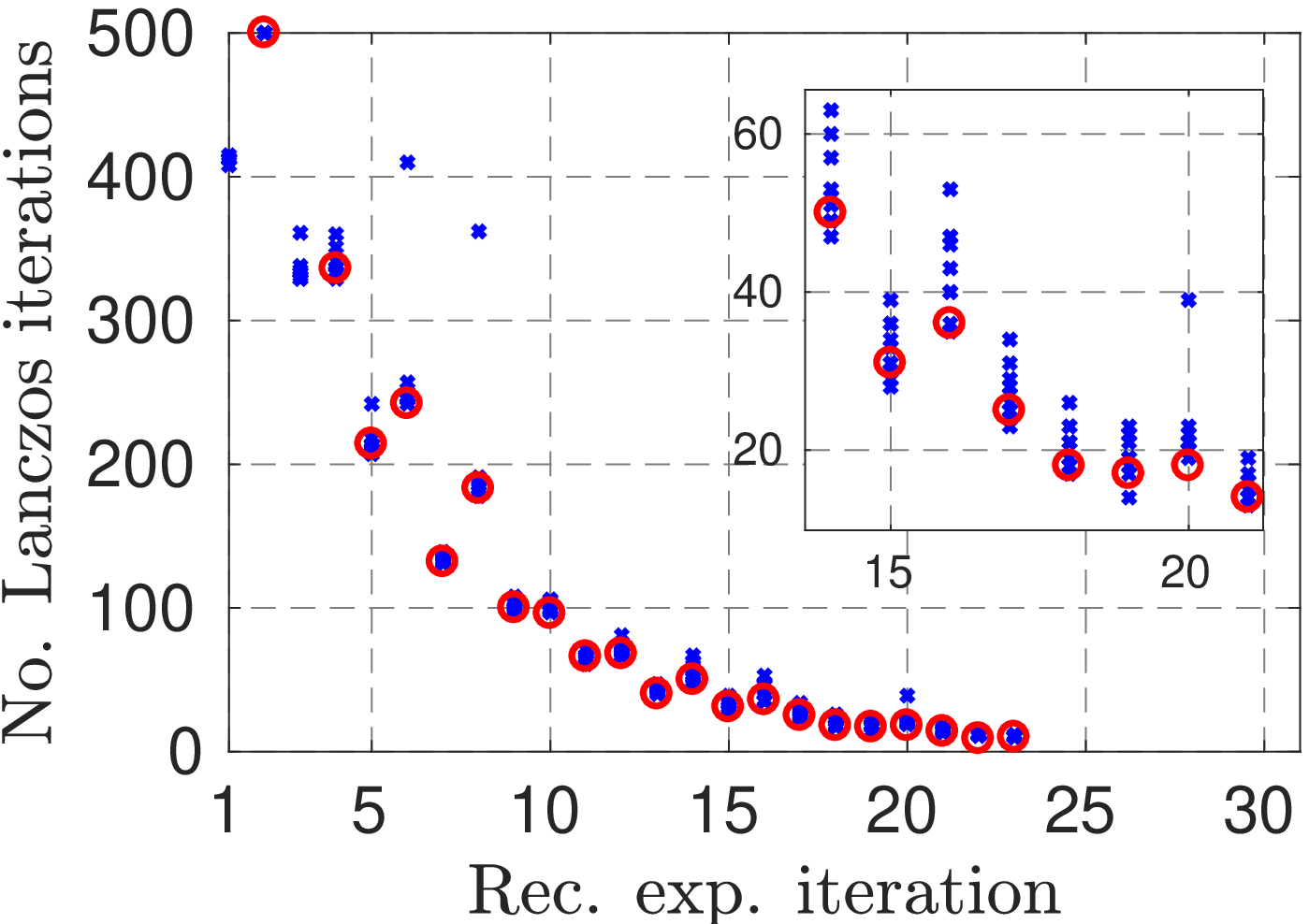}
		\caption{Homo eigenpair}
	\end{subfigure}
	\centering
	\begin{subfigure}[t]{0.45\textwidth}
		\centering
		\includegraphics[width=\textwidth]{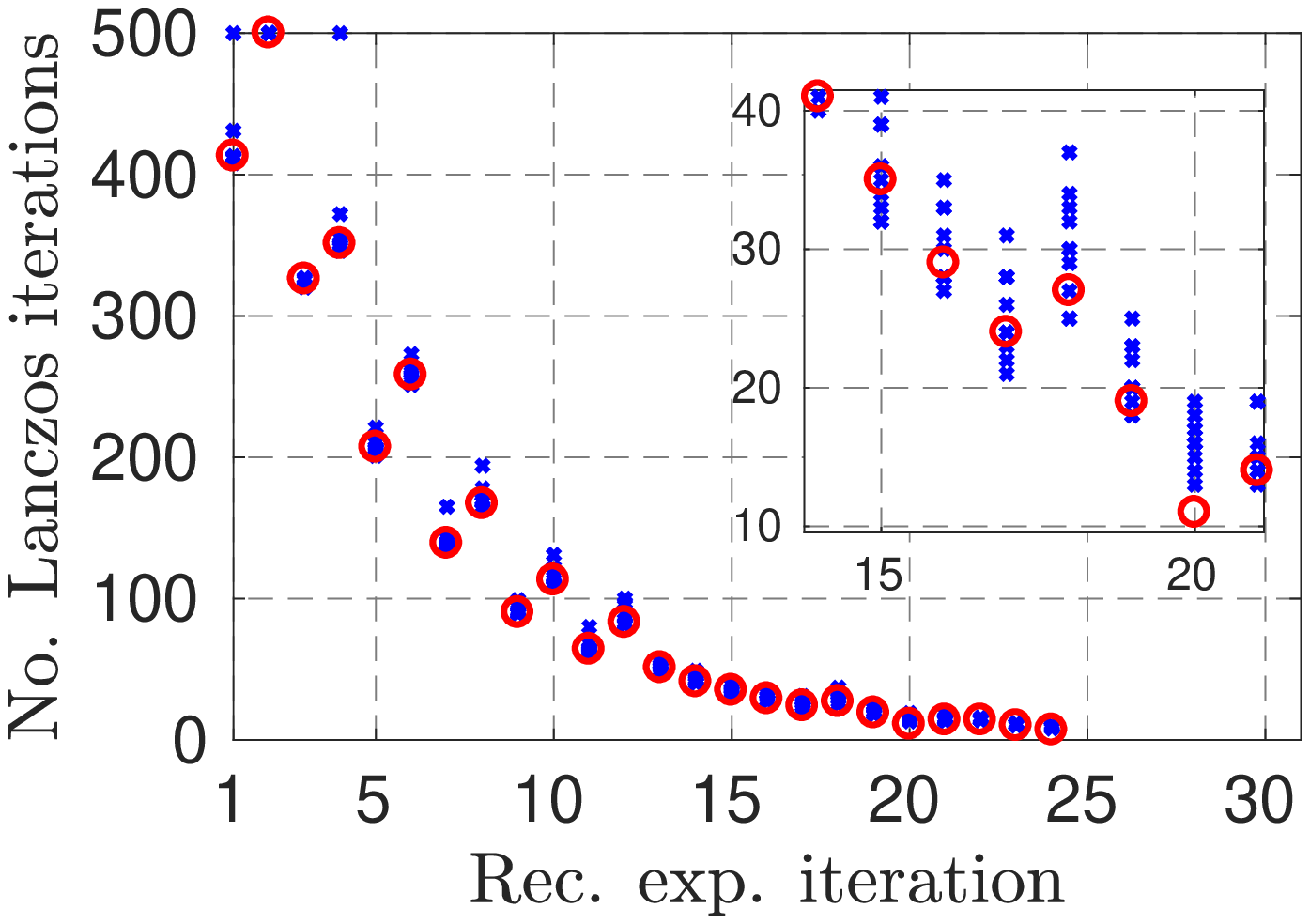}
		\caption{Lumo eigenpair}
	\end{subfigure}
	\caption{The SP2 recursive expansion is applied to a random symmetric dense matrix of size 300 with spectrum in $[0,\, 1]$, homo-lumo gap 0.01 located around 0.5, and 150 occupied orbitals. The number of Lanczos iterations required for computing homo and lumo eigenpairs for various shifts is presented. 20 shifts are chosen equidistantly inside the homo-lumo gap. If the maximum allowed number of Lanczos iterations 500 is reached, the result is not plotted. Numbers obtained with proposed shifts are marked with red circles.}
	\label{fig:lan_iter_various_sigma}
\end{figure}

Figure~\ref{fig:lan_iter_various_sigma} shows the number of Lanczos iterations required for computing homo and lumo eigenpairs in each iteration $i$ of the recursive expansion for various shifts.
We apply the SP2 expansion to a random symmetric dense matrix of size 300.
Occupied and unoccupied eigenvalues are equidistantly distributed in intervals $[0,\, 0.495]$ and $[0.505,\, 1]$, respectively. The eigenvectors of the matrix were taken from a QR factorization of a matrix with random elements from a normal distribution. The distance of the inner  bounds for homo and lumo eigenvalues of the matrix $X_0$ from the exact values is set to $10^{-3}$ and the distance of the outer bounds from the exact values is $10^{-2}$,~i.e.
$\lumo_\upperBound^{0} - \lumo^{0} = \homo^{0} - \homo_\upperBound^{0} =  10^{-3}$ and $\lumo^{0} - \lumo_\lowerBound^{0} = \homo_\lowerBound^{0} - \homo^{0} = 10^{-2}$.
We take 20 different shifts located inside the homo-lumo gap of the matrix $X_i$. Results obtained using chosen shifts $\sigma^{i}_{\lumo} = \frac{\homo_\upperBound^{i}+\lumo_\lowerBound^{i}}{2}$ and $\sigma^{i}_{\homo} = \frac{\lumo_{\upperBound}^{i}+\homo_{\lowerBound}^{i}}{2}$ are highlighted. If the maximum number of Lanczos iterations 500 is reached and no good approximation of the homo or lumo eigenvector is computed, we are not plotting the result.
The figure shows that the choice of shift may significantly influence the number of iterations in the beginning of the recursive expansion. The proposed shifts are almost optimal in terms of the number of Lanczos iterations in each recursive expansion iteration.

Based on the discussion above we choose shifts equal to
\begin{align}
	\sigma^{i}_{\lumo} = \frac{\homo_\upperBound^{i}+\lumo_\lowerBound^{i}}{2},
	% \quad \text{ if } \frac{\homo_\upperBound^{i}+\lumo_\lowerBound^{i}}{2}  \geq \lumo_\upperBound^{i}
	\label{eq:sigma_lumo}
\end{align}
where eligible iterations for computing the lumo eigenvector are iterations when $\sigma^{i}_{\lumo} \geq \lumo_\upperBound^{i}$ is satisfied,
and
\begin{align}
	\sigma^{i}_{\homo} = \frac{\lumo_{\upperBound}^{i}+\homo_{\lowerBound}^{i}}{2},
	% \quad \text{ if } \frac{\lumo_{\upperBound}^{i}+\homo_{\lowerBound}^{i}}{2}  \leq \homo_\upperBound^{i}.
	\label{eq:sigma_homo}
\end{align}
where eligible iterations for computing the homo eigenvector are iterations when $\sigma^{i}_{\homo} \leq \homo_\upperBound^{i}$ is satisfied.

\subsection{Selecting iterations}

In the previous subsection we discussed the choice of shift in the purify-shift-and-square method for a given recursive expansion iteration. For each iteration our choice of shift was aimed at improving performance of the iterative eigensolver. Here we assume that shifts $\sigma^{i}_{\lumo}$ and $\sigma^{i}_{\homo}$ are computed using
~\eqref{eq:sigma_lumo} and~\eqref{eq:sigma_homo}, respectively.
We will provide a way to select a recursive expansion iteration
such that in addition to reducing the number of eigensolver iterations, the eigenvector can be computed with sufficient accuracy.

% accuracy of eigenvector
The accuracy of an eigenvector corresponding to a simple eigenvalue depends on the spectrum separation around the eigenvalue. Let $\delta(\eta)$ be the absolute distance between a simple eigenvalue $\eta$ and the rest of the spectrum:
\begin{align}
	\delta(\eta ) = \min_{i, \eta \neq \eta_i} |\eta - \eta_i|.
\end{align}
The ratio $\frac{1}{\delta(\eta )}$ can be considered as a condition number of the problem of computing the eigenvector corresponding to the simple eigenvalue $\eta$~\cite{stewart1973}.
If an eigenvalue is far from 0 and it is well separated from the rest of the spectrum, i.e. $\delta(\eta)$ is large, the eigenvector is well conditioned and the norm of the difference between computed and exact eigenvectors will in general be small.

\begin{figure}
	\centering
	\captionsetup[subfigure]{justification=centering}
	\includegraphics[width=0.6\textwidth]{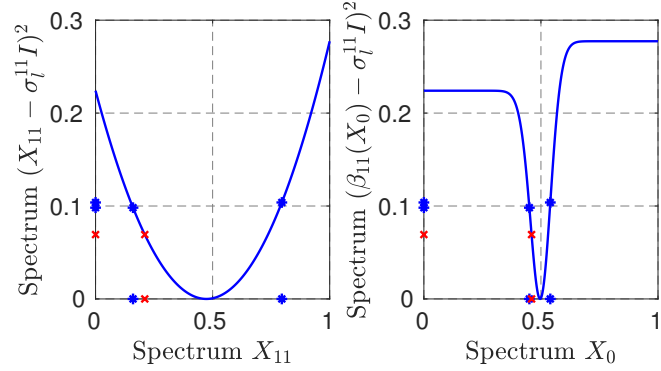}
	\caption{Illustration of the spectrum transformation near the lumo eigenvalue by the polynomials $g(x, \sigma^{11}_{\lumo})$ and $g(\beta_{11}(x),
\sigma^{11}_{\lumo})$. On the x-axis we denote with red cross the lumo
eigenvalue of the matrix $X_i$ in a given iteration $i$ (i.e. $l^{11}$ in the left
panel, $l^0$ in the right panel). With two blue stars we denote the two
closest eigenvalues to the lumo eigenvalue from occupied and unoccupied parts of the spectrum,
respectively. On the y-axis we present the obtained eigenvalues after mapping by the
corresponding polynomials.  For computation of the lumo eigenvector we select
the recursive expansion iteration where the polynomial $g(\beta_i(x),
\sigma^{i}_{\lumo})$ has the largest by absolute value slope at the lumo
eigenvalue of the matrix $X_0$.
  }
	\label{fig:compare_spectrum_filter}
\end{figure}

Let $l^0$ be the lumo eigenvalue of $X_0$ and let $l_2^0$ be the eigenvalue closest to $l^0$ such that $l_2^0 < l^0$.
By selecting the recursive expansion iteration where the absolute value of the slope of the polynomial $g(\beta_i(x), \sigma^{i}_{\lumo})$ at the lumo eigenvalue $l^0$ is  largest, we expect a large absolute separation between the eigenvalues $g(\beta_i(l^0), \sigma^{i}_{\lumo})$ and $g(\beta_i(l_2^0), \sigma^{i}_{\lumo})$. See the illustration in Figure~\ref{fig:compare_spectrum_filter} where we have plotted the polynomials $g(x, \sigma^{i}_{\lumo})$ and $g(\beta_i(x), \sigma^{i}_{\lumo})$. Recursive expansion is applied to a diagonal matrix of size 100 with 50 occupied orbitals and homo-lumo gap 0.08. 

The derivative of the approximation polynomial $\beta_{i}$ at a point $x\in[0,1]$ is the following:
\begin{align}
	\beta_0(x)  & = x, \quad \beta'_0(x)=1                               \\
	\beta_i'(x) & = f_i'(\beta_{i-1}(x)) \beta_{i-1}'(x), \quad i \geq 1
\end{align}
We have
\begin{align}
	g'(\beta_i(x), \sigma^{i}_{\lumo})= & 2\left(\beta_i(x)-\sigma^{i}_{\lumo}\right)\beta_i'(x).
                                      %
	                                    % & 2\left(\beta_i(x)-\sigma^{i}_{\lumo}\right)f_i'(\beta_{i-1}(x)) \beta_{i-1}'(x).
\end{align}
The eigenvalue bounds for the lumo eigenvalue may be used instead of the exact lumo eigenvalue of $X_0$. Under assumption~\eqref{eq:assump_eig_bounds} and according to the discussion in section~\ref{sec:eigv_eistimates}, the inner bounds are in practice more accurate than the outer bounds. Thus we will use the eigenvalue bound $\lumo_\upperBound^{0}$ instead of the exact lumo eigenvalue of $X_0$.

Summarizing the above discussion, to compute the lumo eigenvector we suggest to select a recursive expansion iteration $i_\lumo$ such that  $|g'(\beta_{i_\lumo}(\lumo_\upperBound^{0}), \sigma^{i_\lumo}_{\lumo})|$ is as large as possible. Note that $g'(\beta_{i_\lumo}(\lumo_\upperBound^{0}), \sigma^{i_\lumo}_{\lumo})$ has a negative value.
Correspondingly, to compute the homo eigenvector we will search for a recursive expansion iteration $i_\homo$ such that  $g'(\beta_{i_\homo}(\homo_\upperBound^{0}), \sigma^{i_\homo}_{\homo})$ is as large as possible.
The algorithm for selecting iterations $i_\homo$ and $i_\lumo$ of the recursive expansion for computing homo and lumo eigenpairs, respectively, is summarized in Algorithm~\ref{alg:determine_iter}.

\begin{algorithm}\caption{Determine iterations for computing homo and
		lumo eigenvectors in SP2~\label{alg:determine_iter}}
	\begin{algorithmic}[1]
		\State \textbf{input:} $n_{\textrm{max}}$,
    %$\homo_\lowerBound^{i}$,
    $\homo_\upperBound^{i}$,
		%$\lumo_\lowerBound^{i}$,
    $\lumo_\upperBound^{i}$,   $\sigma^{i}_{\homo}$, $\sigma^{i}_{\lumo}$, $i=0,1,\dots,n_{\textrm{max}}$
		\State %$x = \homo$, $y = \lumo$,
		$\beta'_{\lumo} = 1$, $\beta'_{\homo} = 1$, $G_\homo = -\infty$, $G_\lumo = -\infty$
		\State $i_\homo = -1$,  $i_\lumo = -1$
		\For{$i = 1, 2, \ldots, n_{\textrm{max}}$}

		\State $\beta_{\homo} = 1-\homo_\upperBound^{i-1}$
		\State $\beta_{\lumo} = \lumo_\upperBound^{i-1}$

		\If{$p_i =1$}
		\State $\beta'_{\homo} = 2\beta_{\homo} \beta'_{\homo}$
		\State $\beta'_{\lumo} = 2\beta_{\lumo} \beta'_{\lumo}$
		\Else
		\State $\beta'_{\homo} = 2(1 - \beta_{\homo})\beta'_{\homo}$
		\State $\beta'_{\lumo} = 2(1 - \beta_{\lumo})\beta'_{\lumo}$
		\EndIf

		\State $g'_{\homo} = 2\left(\beta_{\homo}-\sigma^{i}_{\homo}\right)\beta'_{\homo}$
		\State $g'_{\lumo} = 2\left(\beta_{\lumo}-\sigma^{i}_{\lumo}\right)\beta'_{\lumo}$

		\If{$\sigma^{i}_{\homo} \leq \homo_\upperBound^{i} $ \textbf{ and } $g'_{\homo} \geq G_{\homo}$}
		\State $G_{\homo} = g'_{\homo}$
		\State $i_\homo = i$
		\EndIf

		\If{$\lumo_\upperBound^{i} \leq \sigma^{i}_{\lumo}$ \textbf{ and } $|g'_{\lumo}| \geq G_{\lumo}$}
		\State $G_{\lumo} = g'_{\lumo}$
		\State $i_\lumo = i$
		\EndIf

		\EndFor
		\State \textbf{output:} $i_\lumo$, $i_\homo$
	\end{algorithmic}
\end{algorithm}

Let us summarize all the steps needed for computing homo and lumo eigenvectors during the recursive expansion:
\begin{compactitem}
	\item Use Algorithm~\ref{alg:homo_lumo_bounds_every_iter} to determine sequence of polynomials for the recursive expansion and homo and lumo estimates for the matrices $X_i$ in each iteration
	\item Use formulas~\eqref{eq:sigma_lumo} and \eqref{eq:sigma_homo} to compute shifts
	\item Use Algorithm~\ref{alg:determine_iter} to determine iterations of the recursive expansion for homo and lumo eigenvector computation
	\item Use Algorithm~\ref{alg:folded_method} to perform the recursive expansion and compute homo and lumo eigenvectors on-the-fly
	\item Use Algorithm~\ref{alg:homo_lumo_bounds} to compute new homo and lumo estimates for the matrix $F$, which will be propagated to the next SCF cycle
\end{compactitem}

%%%%%%%%%%%%%%%%%%%%%%%%%%%%%%%%%%%%%%%%%%%%%%%%%%%%%%%%%%%%%%%%%%%%%%%%%%%%%%%%%%%%%%%%
%%%%%%%%%%%%%%%%%%%%%%%%%%%%%%%%%%%%%%%%%%%%%%%%%%%%%%%%%%%%%%%%%%%%%%%%%%%%%%%%%%%%%%%%

\section{Application to self-consistent field calculations}

% tests in Ergo
In this section we apply the purify-shift-and-square method for computation of the homo and lumo orbitals in SCF calculations with the quantum
chemistry program Ergo \cite{ergo_web,Ergo2011}.  We used direct
inversion in the iterative subspace (DIIS) for SCF convergence
acceleration \cite{pulay_1980,pulay82} and stopped the calculations as
soon as the largest absolute element of $F'D'S-SD'F'$ was smaller than
a given threshold value $\tau_{\textrm{SCF}}$, where $S$ is the basis set overlap matrix and $F'$ and $D'$ are the Fock and density matrices in non-orthogonal basis. Here we perform Hartree-Fock calculations, but all the algorithms proposed in this paper can also be used in Kohn-Sham density functional theory calculations. The computation of the density matrix from the Fock matrix is done in orthogonal
basis. For transformation from non-orthogonal basis to orthogonal basis we
employ the inverse Cholesky factor of the overlap matrix. The hierarchical matrix library described in Ref.~\citenum{hierarchical_matrix_lib} was used for sparse matrix
operations. A block size of 32 was used at the lowest level in the
sparse hierarchical representation. The inner bounds for the homo and
lumo eigenvalues (see section~\ref{sec:eigv_eistimates}) are used for  determination of polynomial sequence in the
recursive expansion in Algorithm~\ref{alg:homo_lumo_bounds_every_iter},  error control, and determination of recursive expansion iterations
for computing homo and lumo eigenvectors in Algorithm~\ref{alg:determine_iter}. The recent stopping criterion developed in Ref.~\citenum{stop_crit_2016} was used for termination of the recursive expansion iterations. The mixed norm with block size 32
was used both in the stopping criterion and for removal of small
matrix elements. Truncation with control of the error in the occupied subspace
measured by the largest canonical angle between the exact and
approximate subspaces was performed as described in Refs.~\citenum{mixedNormTrunc} and~\citenum{m-accPuri} with a predefined tolerance $\tau_{\textrm{puri}}$.

Numerical tests were run on the Triolith cluster at the National
Supercomputer Centre (NSC) in Link{\"o}ping, Sweden using the Intel 16.0.2 C++ compiler
and Intel MKL version 11.3.2 for
matrix operations at the lowest level in the sparse hierarchical
representation. Each node on Triolith has
two 8-core Intel Xeon E5-2660 ``Sandy Bridge''
processors running at 2.2 GHz. The calculations were
performed on a node with 128 GB of memory.

All results presented in the following subsections were obtained in the last SCF cycle. All timing results represent elapsed wall time. The Lanczos algorithm was considered converged when the relative residual
$\frac{\|Ax-\lambda x\|}{|\lambda|}$ for the eigenpair approximation $(\lambda, x)$ was less than $10^{-12}$. We used random starting guess for Lanczos iterations, unless otherwise stated.

\subsection{Alkane chains and extended pyrimidine derivative molecule}
\label{sec:alkanes_and_mol}

In our first test we created a sequence of alkane chains C$_n$H$_{2n+2}$ with increasing $n$. The homo-lumo gap and spectrum width of the Fock matrix do not change significantly with increasing length of the chains, but eigenvalues near the homo-lumo gap tend to be more clustered and become nearly degenerate for larger system sizes. The distance between lumo and the rest of the spectrum decreases notably faster than the distance between homo and the rest of the spectrum. Thus for large systems we expect that computation of lumo eigenpairs will require more Lanczos iterations than the computation of homo eigenpairs.

Hartree-Fock calculations were performed on the alkane chains using the Gaussian basis set \mbox{6-311G*}. The initial guess density matrix was obtained from calculations with a smaller basis set \mbox{3-21G}. SCF convergence and truncation tolerances were set to $\tau_{\textrm{SCF}} = 10^{-3}$ and $\tau_{\textrm{puri}} = 10^{-3}$, respectively.

\begin{figure}[h!]
	\centering
	\captionsetup[subfigure]{justification=centering}
	\begin{subfigure}[b]{0.3\textwidth}
		\includegraphics[width=\textwidth]{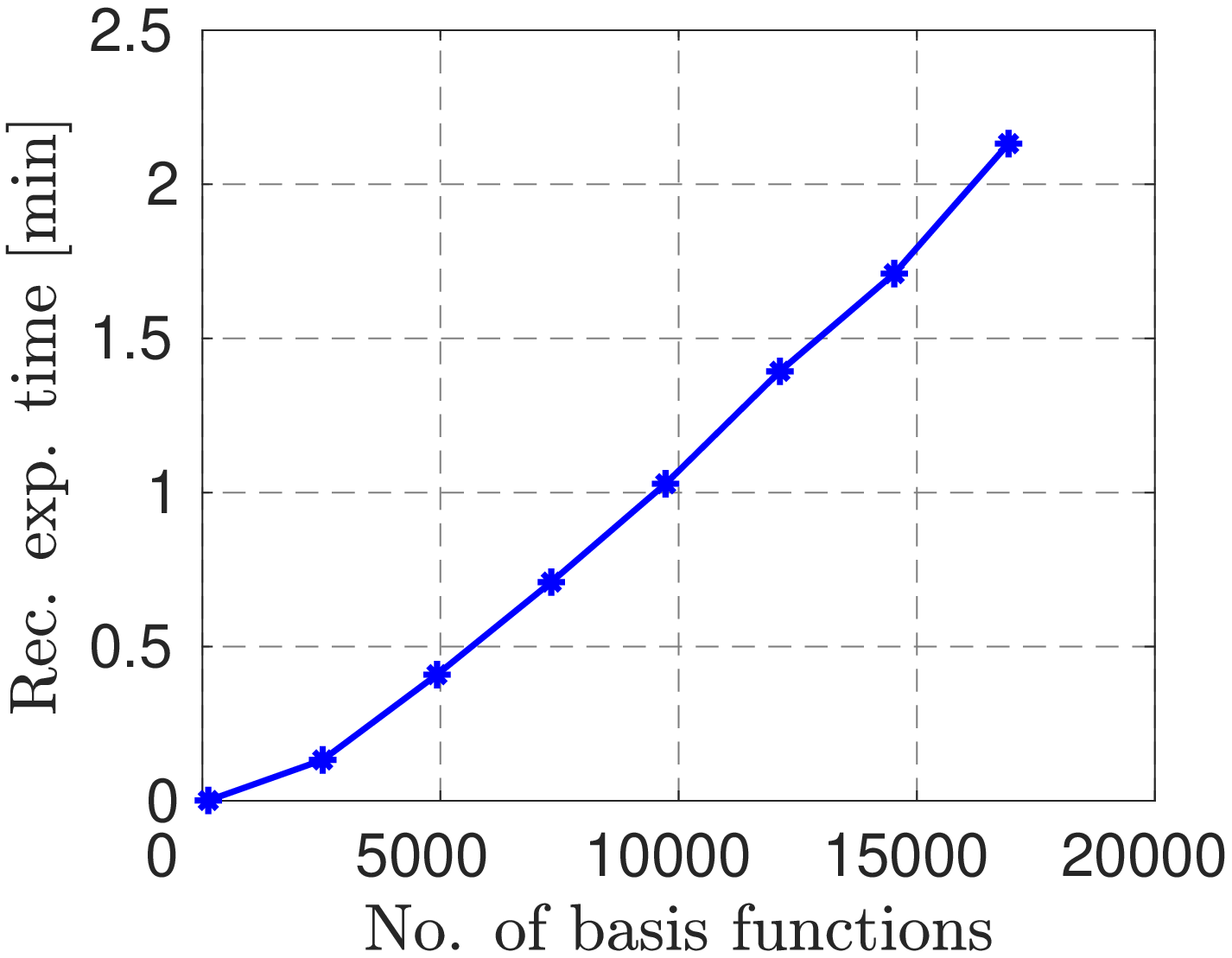}
	\end{subfigure}\quad
	\begin{subfigure}[b]{0.3\textwidth}
		\includegraphics[width=\textwidth]{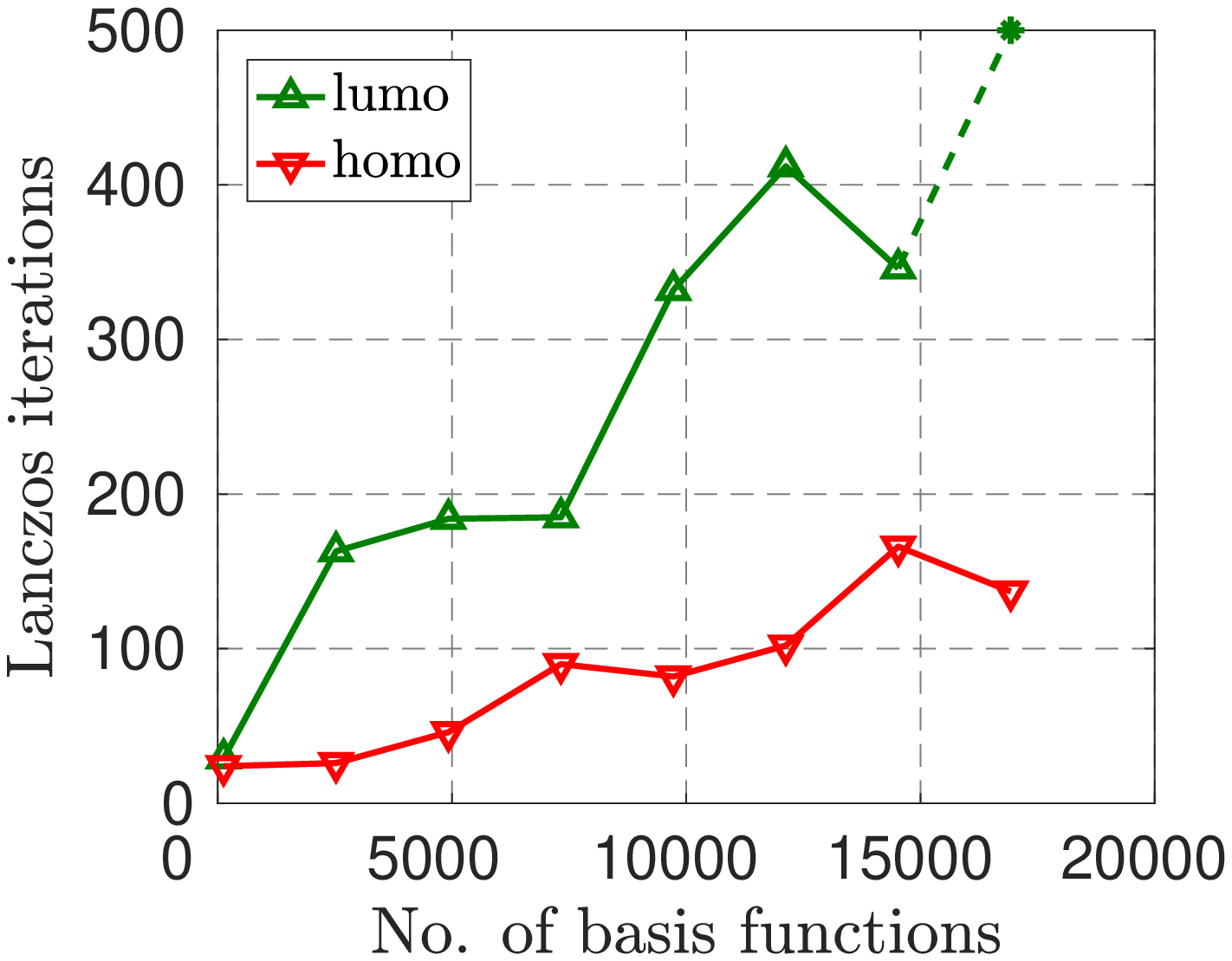}
	\end{subfigure}
	\begin{subfigure}[b]{0.3\textwidth}
		\includegraphics[width=\textwidth]{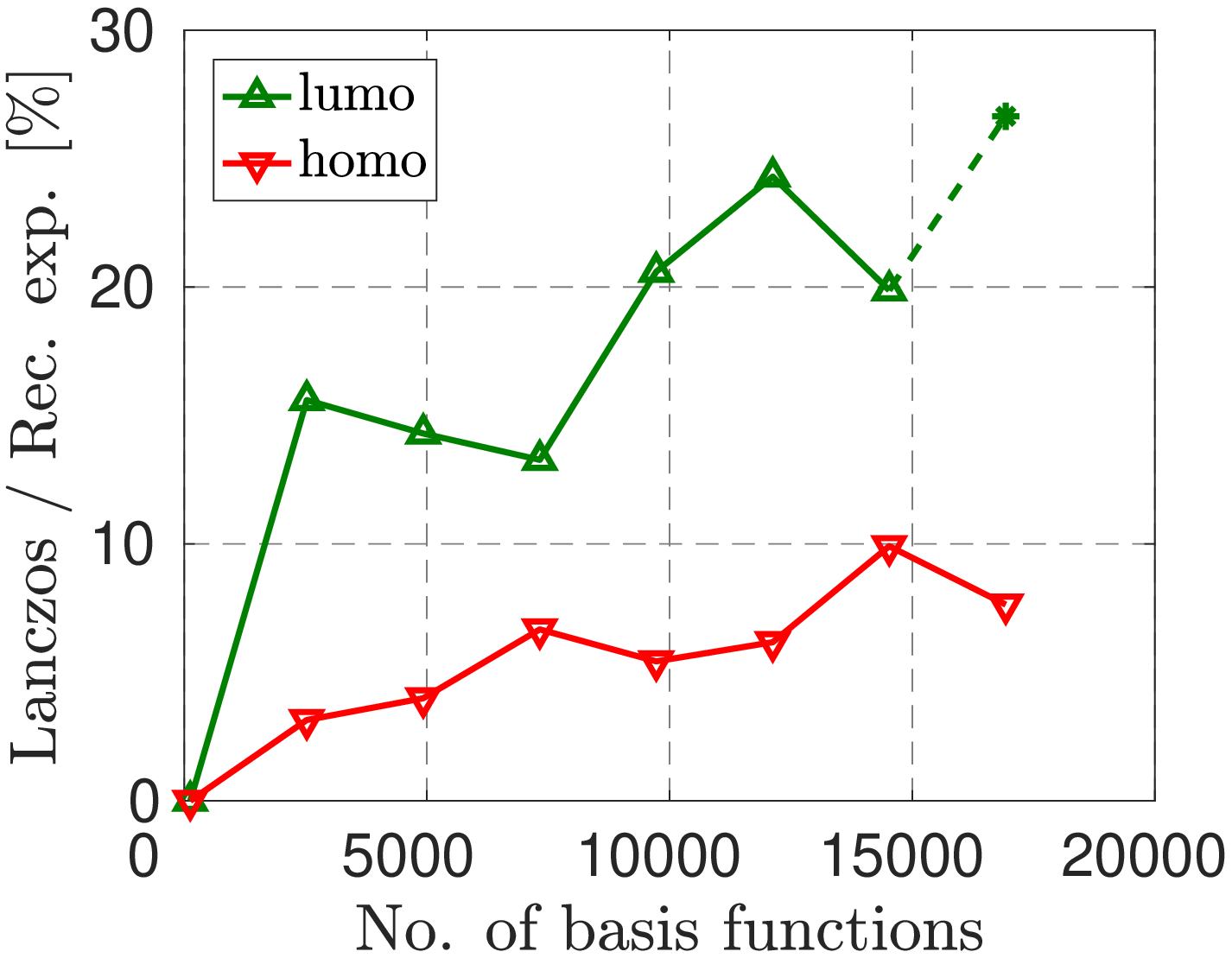}
	\end{subfigure}
	\caption{Recursive expansion in the last SCF cycle of \mbox{HF/6-311G*} calculations
	performed for the alkane chains. The execution time in the left panel includes computation of homo and lumo eigenvectors. In the central and right panels the computation of the lumo eigenpair where the maximum number of allowed Lanczos iterations is reached without convergence is indicated using a star marker.}
	\label{fig:alkane_all}
\end{figure}

In Figure~\ref{fig:alkane_all} we present total recursive expansion time, number of Lanczos iterations and fraction of time required for computing homo and lumo eigenvectors compared to the total recursive expansion time for alkane chains of increasing length. The recursive expansion execution time includes computation of eigenvectors.
The maximum number of Lanczos iterations was set to 500. For the largest presented system size the Lanczos algorithm did not converge within 500 iterations for the lumo eigenpair. The time spent on Lanczos iterations is in this case indicated with a star marker.

\begin{figure}[h!]
	\centering \captionsetup[subfigure]{justification=centering}
		\includegraphics[width=0.5\textwidth]{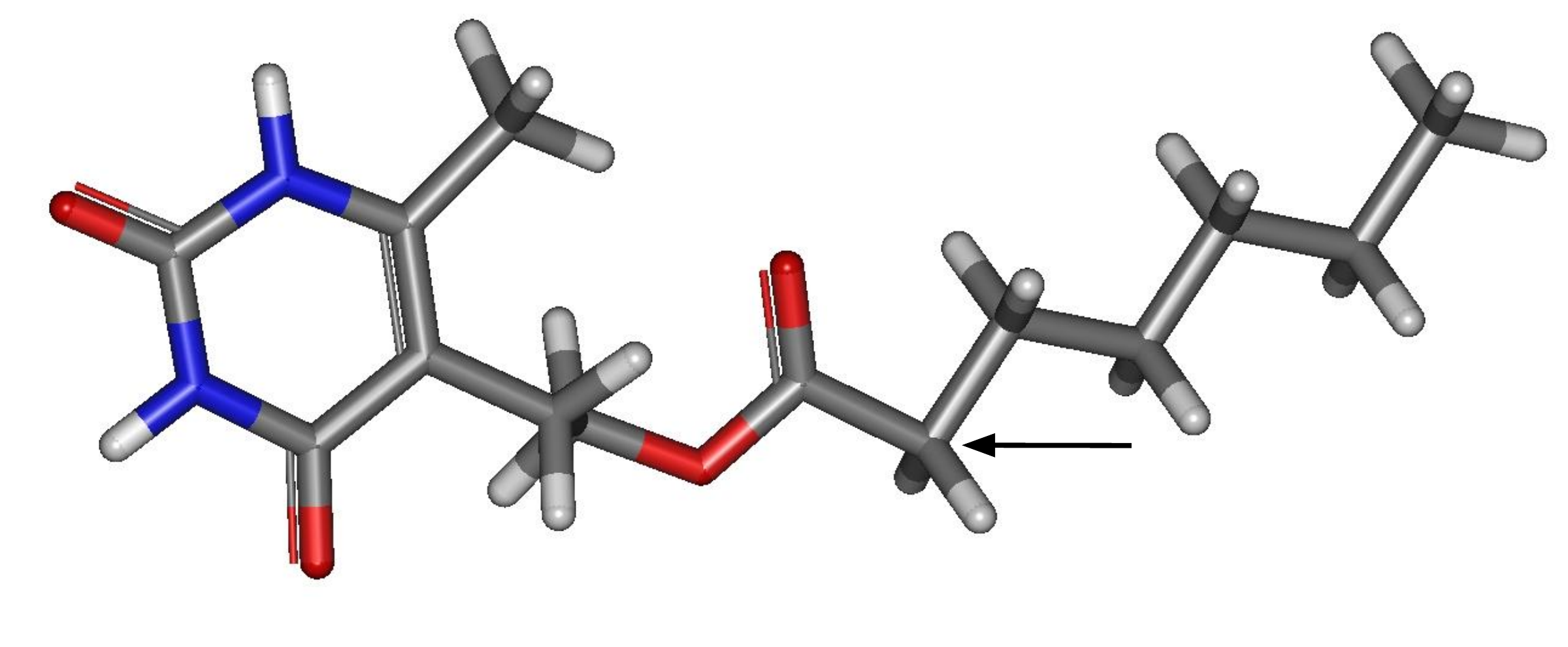}
	\caption{Extended C$_9$H$_{12}$N$_2$O$_4$ molecule by adding pentane %C$_5$H$_{12}$
  to the carbon atom indicated by the arrow.}
  \label{fig:extended_molecule_by_pentane}
\end{figure}

\begin{figure}[h!]
	\centering
	\captionsetup[subfigure]{justification=centering}
	\begin{subfigure}[b]{0.45\textwidth}
		\centering
		\includegraphics[width=\textwidth]{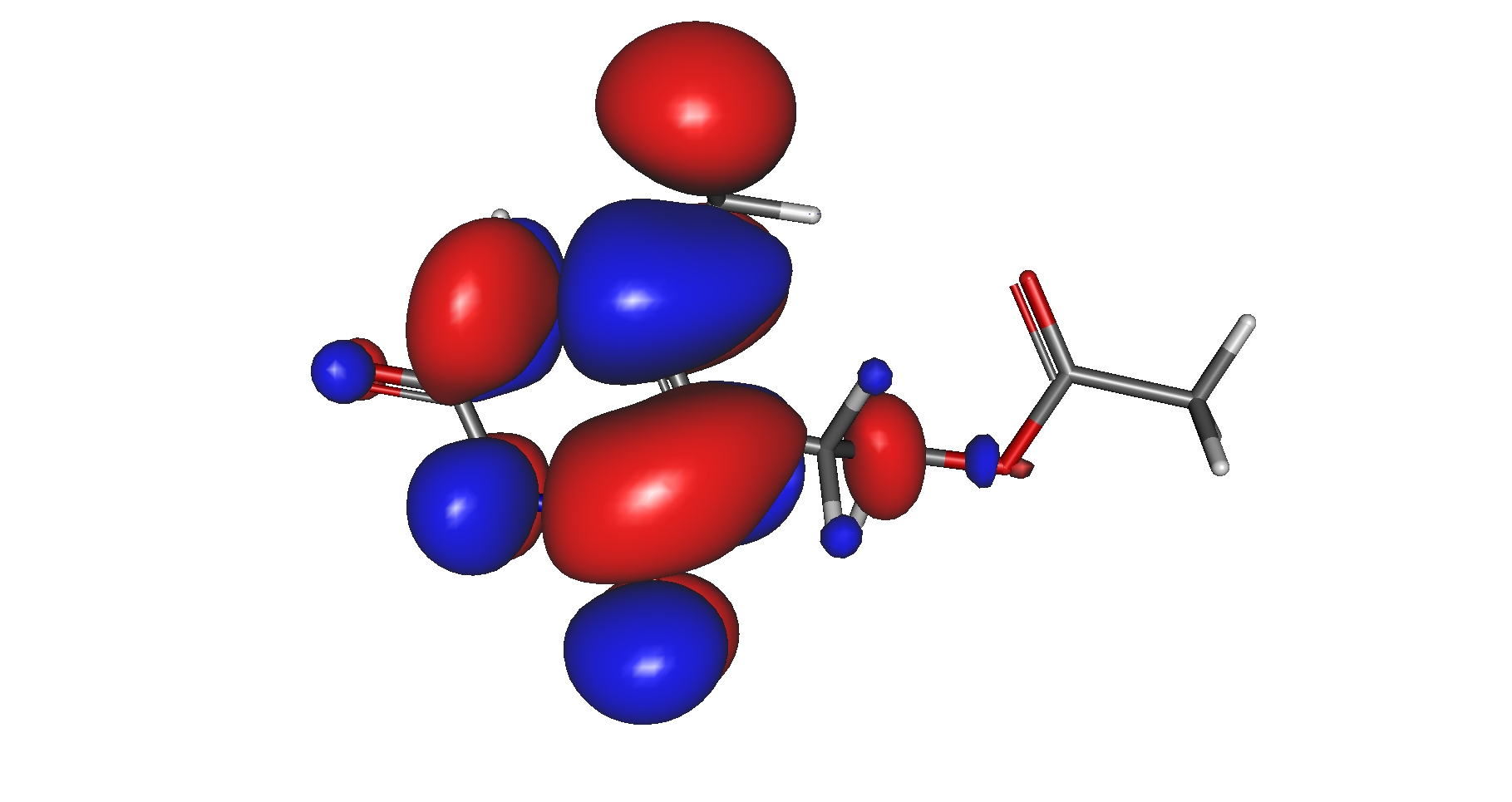}
		\caption{Lumo orbital}
	\end{subfigure}
	\begin{subfigure}[b]{0.4\textwidth}
		\centering
		\includegraphics[width=\textwidth]{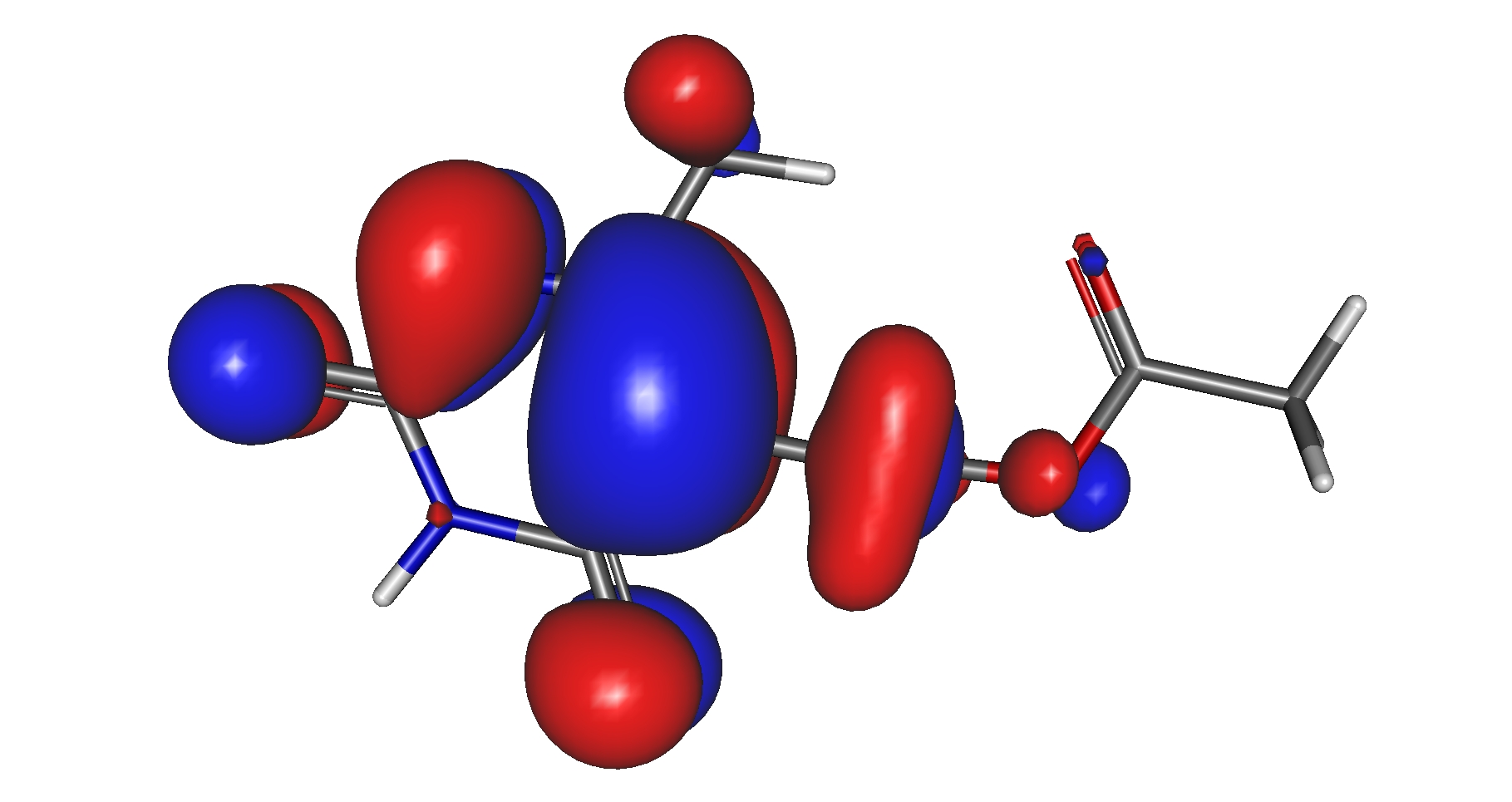}
		\caption{Homo orbital}
	\end{subfigure}
	\caption{Homo and lumo molecular orbitals for C$_9$H$_{12}$N$_2$O$_4$ obtained
	with \mbox{HF/6-311G*} calculations. Computed homo energy is $-0.347604$ a.u.,
	lumo energy is $0.100358$ a.u.}
	\label{fig:homo_and_lumo_orbitals_dione}
\end{figure}

\begin{figure}[h!]
	\centering
	\captionsetup[subfigure]{justification=centering}
  \begin{subfigure}[b]{0.45\textwidth}
    \includegraphics[width=\textwidth]{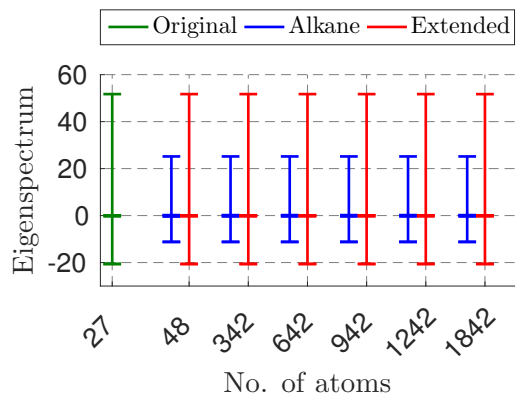}
    \caption{Obtained spectrum}
    \label{fig:molecule_full_spectrum}
  \end{subfigure}

	\begin{subfigure}[b]{0.45\textwidth}
		\includegraphics[width=\textwidth]{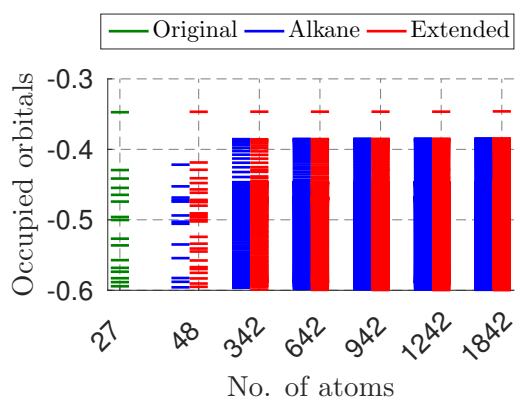}
		\caption{Occupied part of the spectrum}
	\end{subfigure}
	\begin{subfigure}[b]{0.45\textwidth}
		\includegraphics[width=\textwidth]{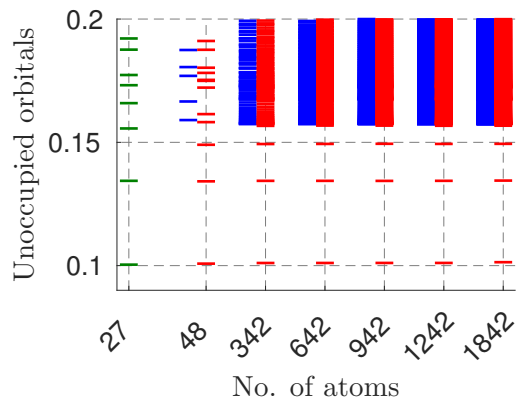}
		\caption{Unoccupied part of the spectrum}
	\end{subfigure}
	\caption{Illustration of the eigenspectrum for increasing size alkane chains,
	pyrimidin derivative and extended systems. Upper panel: Spectrum width and
	location of the homo-lumo gap. Bottom panels: Close-up on the eigenspectrum near the
	homo-lumo gap.}
	\label{fig:molecule_parts_spectrum}
\end{figure}

In our second test we combine alkane chains with a
5-(2-Acetoxyethyl)-6-methylpyrimidin-2,4-dione molecule
(\mbox{C$_9$H$_{12}$N$_2$O$_4$}). The crystal structure of \mbox{C$_9$H$_{12}$N$_2$O$_4$} was
synthesized by Kraljevi{\'{c}} et al.~\cite{Kraljevic} and can be obtained from
Cambridge Crystallographic Data Center (CCDC 749761). We optimized the molecular
structure of C$_9$H$_{12}$N$_2$O$_4$ in the ground state using the Gaussian 09W program~\cite{g09} by performing density functional theory calculations with the B3LYP functional and the \mbox{6-311++G**} standard Gaussian basis set.
The molecule C$_9$H$_{12}$N$_2$O$_4$ was extended by adding alkanes to the outermost carbon atom, as illustrated in Figure~\ref{fig:extended_molecule_by_pentane}.
We will refer to the new obtained systems as extended with alkane chains pyrimidin derivative, or
simply extended systems. Figure~\ref{fig:homo_and_lumo_orbitals_dione}
illustrates homo and lumo orbitals of C$_9$H$_{12}$N$_2$O$_4$ computed using the Ergo program by performing Hartree-Fock calculations with the \mbox{6-311G*} basis set, and illustrated using the Gabedit GUI~\cite{gabedit}.
%pictures from Gabedit
A spectrum comparison between C$_9$H$_{12}$N$_2$O$_4$, alkane chains and
extended molecules is given in Figure~\ref{fig:molecule_parts_spectrum}.

\begin{figure}[h!]
	\centering
	\captionsetup[subfigure]{justification=centering}
	\begin{subfigure}[b]{0.3\textwidth}
		\includegraphics[width=\textwidth]{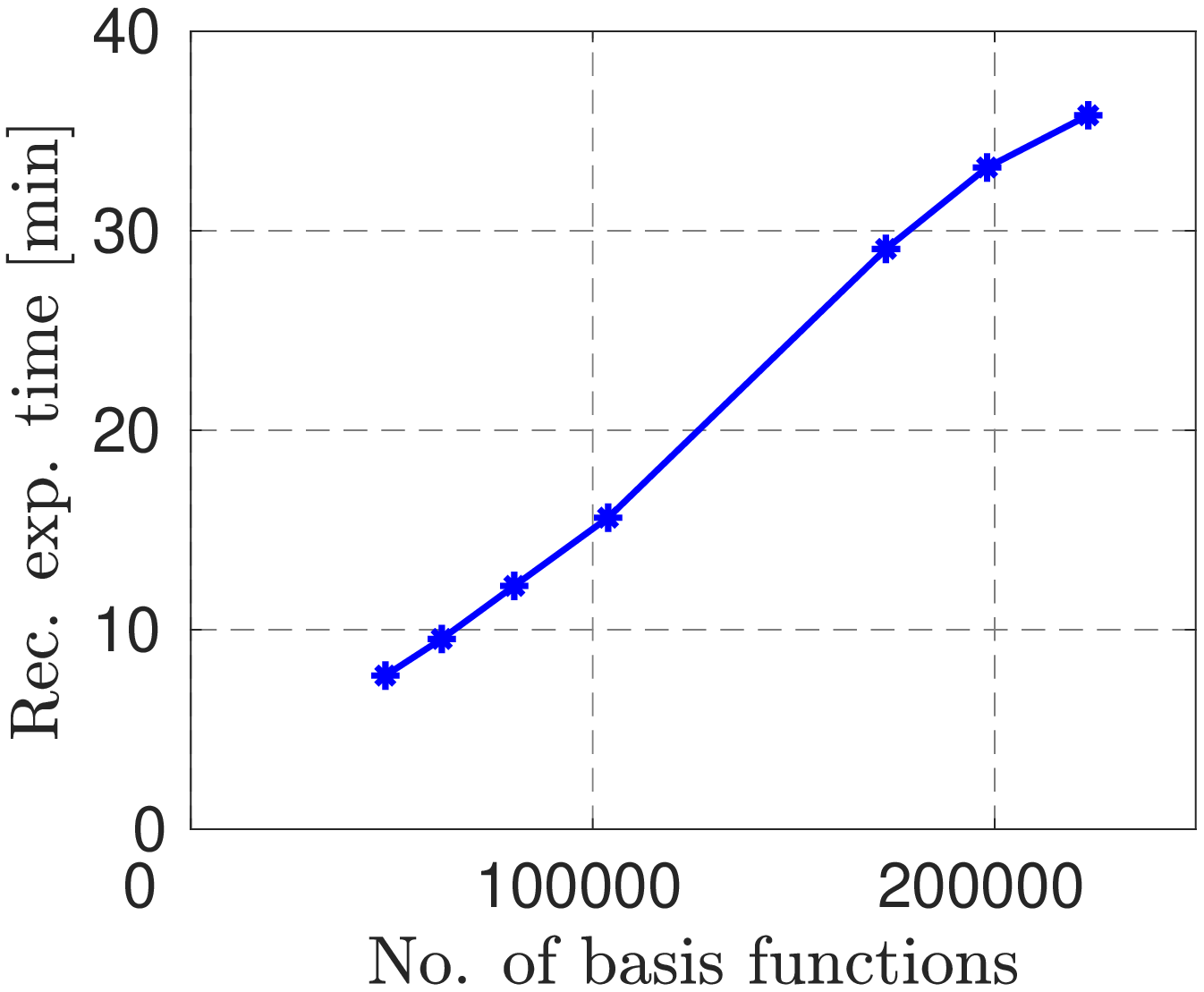}
		%   \caption{Total execution time of the recursive expansion}
	\end{subfigure}
	\begin{subfigure}[b]{0.3\textwidth}
		\includegraphics[width=\textwidth]{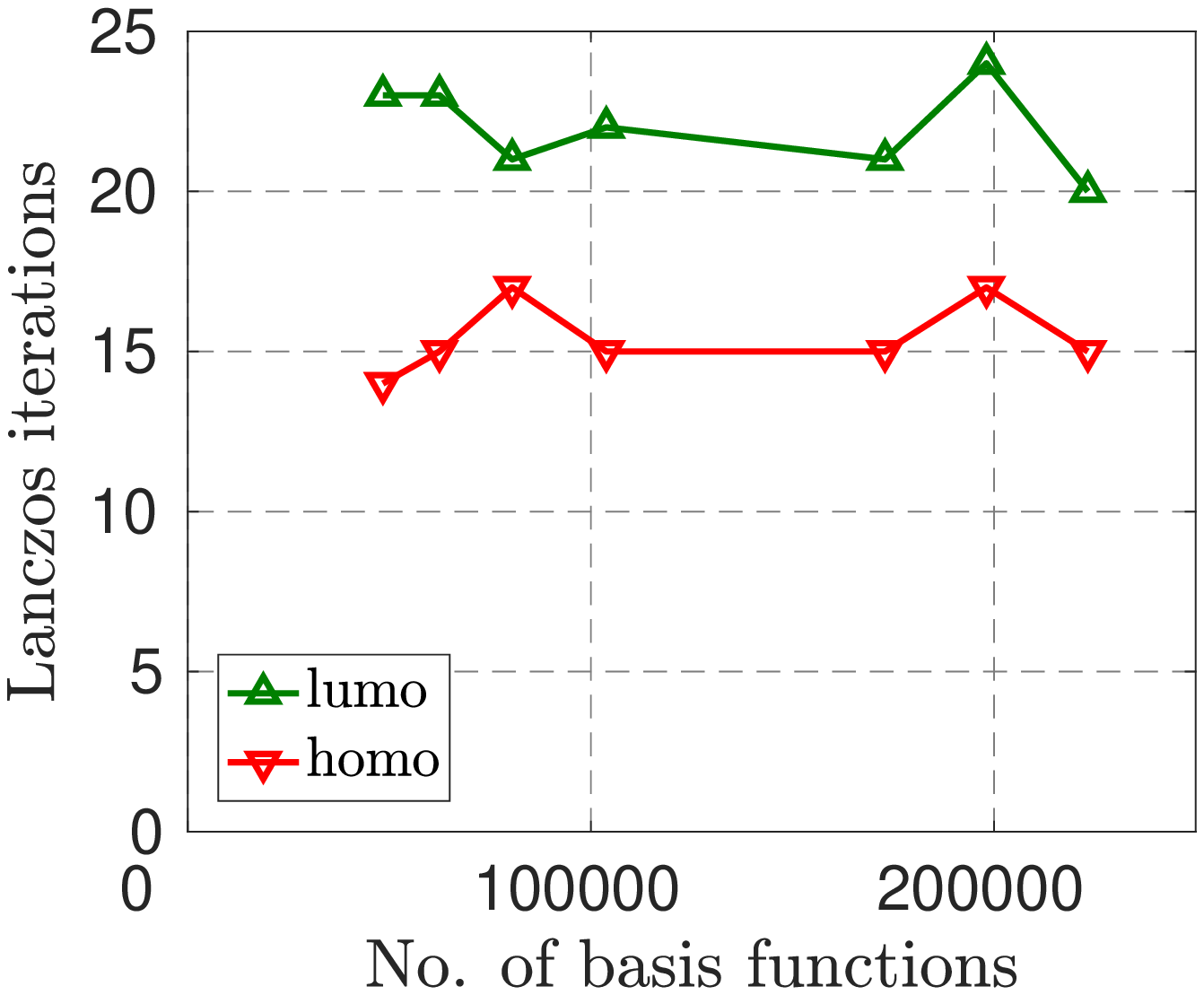}
		%  \caption{Number of Lanczos iterations for homo and lumo eigenpairs}
	\end{subfigure}
	\begin{subfigure}[b]{0.3\textwidth}
		\includegraphics[width=\textwidth]{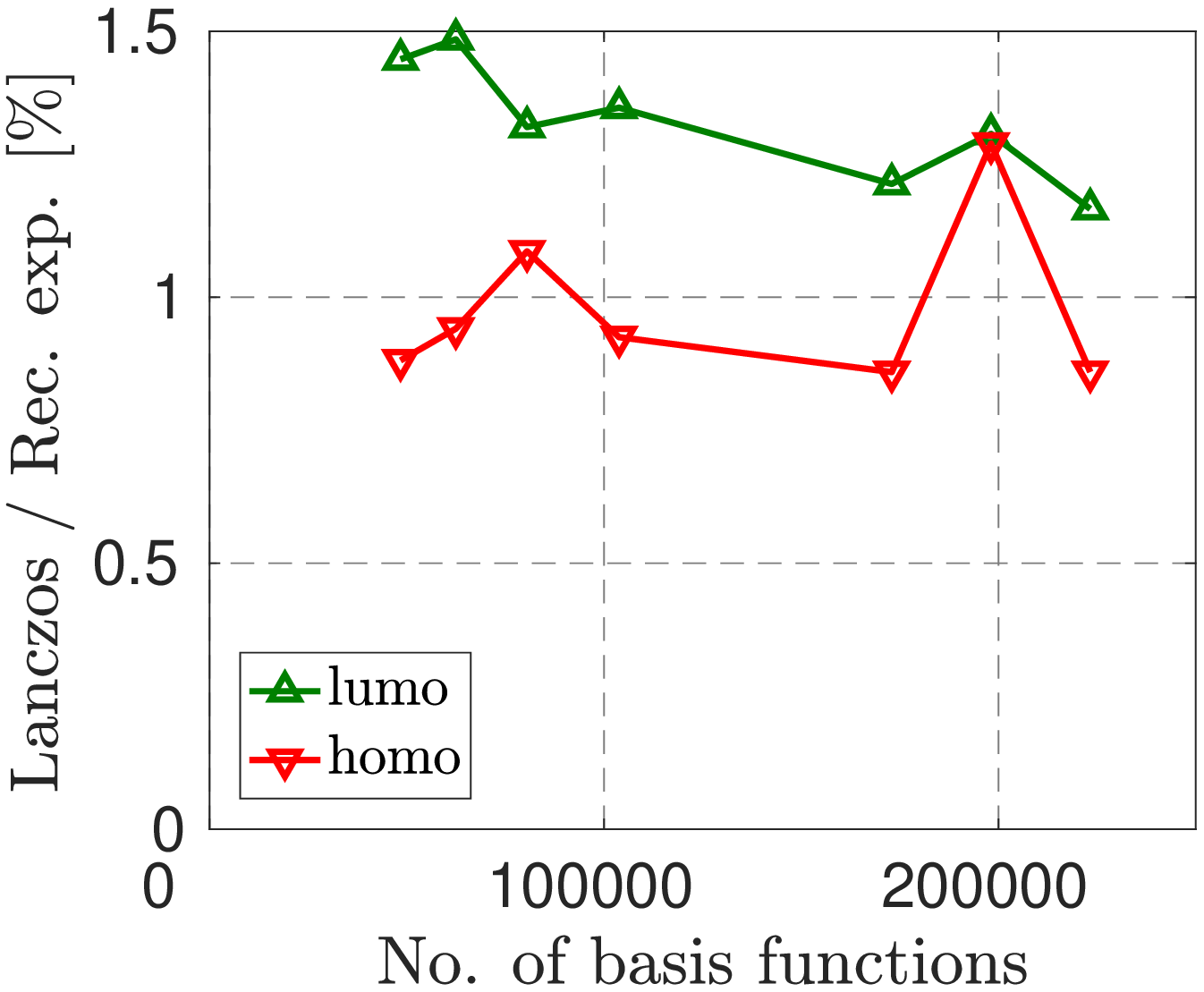}
		%  \caption{Ratio of times for Lanczos iterations and the recursive expansion}
	\end{subfigure}
  \caption{Recursive expansion in the last SCF cycle of
	\mbox{HF/6-311G*} calculations performed for the C$_9$H$_{12}$N$_2$O$_4$
	molecule extended with alkane chains. The execution time in the left panel
	includes computation of homo and lumo eigenvectors. }
	\label{fig:molext_nnzrow_puritime}
\end{figure}

We perform Hartree-Fock calculations with the same basis set and parameters as described for the alkane chains. Results are given in Figure~\ref{fig:molext_nnzrow_puritime}. In the extended systems the spectrum width and the part of the spectrum close to the homo-lumo gap of the Fock matrix is determined by the C$_9$H$_{12}$N$_2$O$_4$ molecule, such that the relative distance between homo, lumo and the rest of the spectrum does not change significantly for increasing system sizes. Indeed, the central panel in Figure~\ref{fig:molext_nnzrow_puritime} shows that almost the same  number of Lanczos iterations is required for computation of homo and lumo eigenpairs for various system sizes.
%
% Molecule + alkane chains

In this subsection we have considered two systems. The alkane chains represent ``hard'' systems since the eigenvalues of the Fock matrix become nearly degenerate already for relatively small chain lengths, resulting in a very high condition number for the problem of computing eigenvectors.
In such cases, however, it may not make sense to compute individual eigenvectors.
The extended molecules can be seen as ``easy'' systems since the eigenvalue distribution near the homo-lumo gap does not depend on the system size. The fraction of time needed for computation of homo and lumo eigenpairs relative to the total time of the recursive expansion is around 2-2.5\%.

\subsection{Water clusters and protein-water molecular systems}

\subsubsection{Water clusters}

In this subsection we describe results of performing Hartree-Fock calculations on water clusters and protein-water molecular systems. These calculations are examples of homo and lumo eigenpair computation for 3D systems of increasing size.

Water cluster geometries were generated from a large molecular dynamics simulation of bulk water at standard temperature and pressure by including all water molecules within spheres of varying radii. The xyz coordinate files can be obtained from \url{http://www.ergoscf.org}.

Hartree-Fock calculations were performed on the water clusters using the Gaussian basis set 3-21G. Initial guess density matrices were obtained from  calculations with a smaller basis set STO-2G. SCF convergence and truncation tolerances were set to $\tau_{\textrm{SCF}} = 10^{-3}$ and $\tau_{\textrm{puri}} = 10^{-3}$, respectively.

\begin{figure}[h!]
	\centering
	\captionsetup[subfigure]{justification=centering}
	\begin{subfigure}[b]{0.45\textwidth}
		\includegraphics[width=\textwidth]{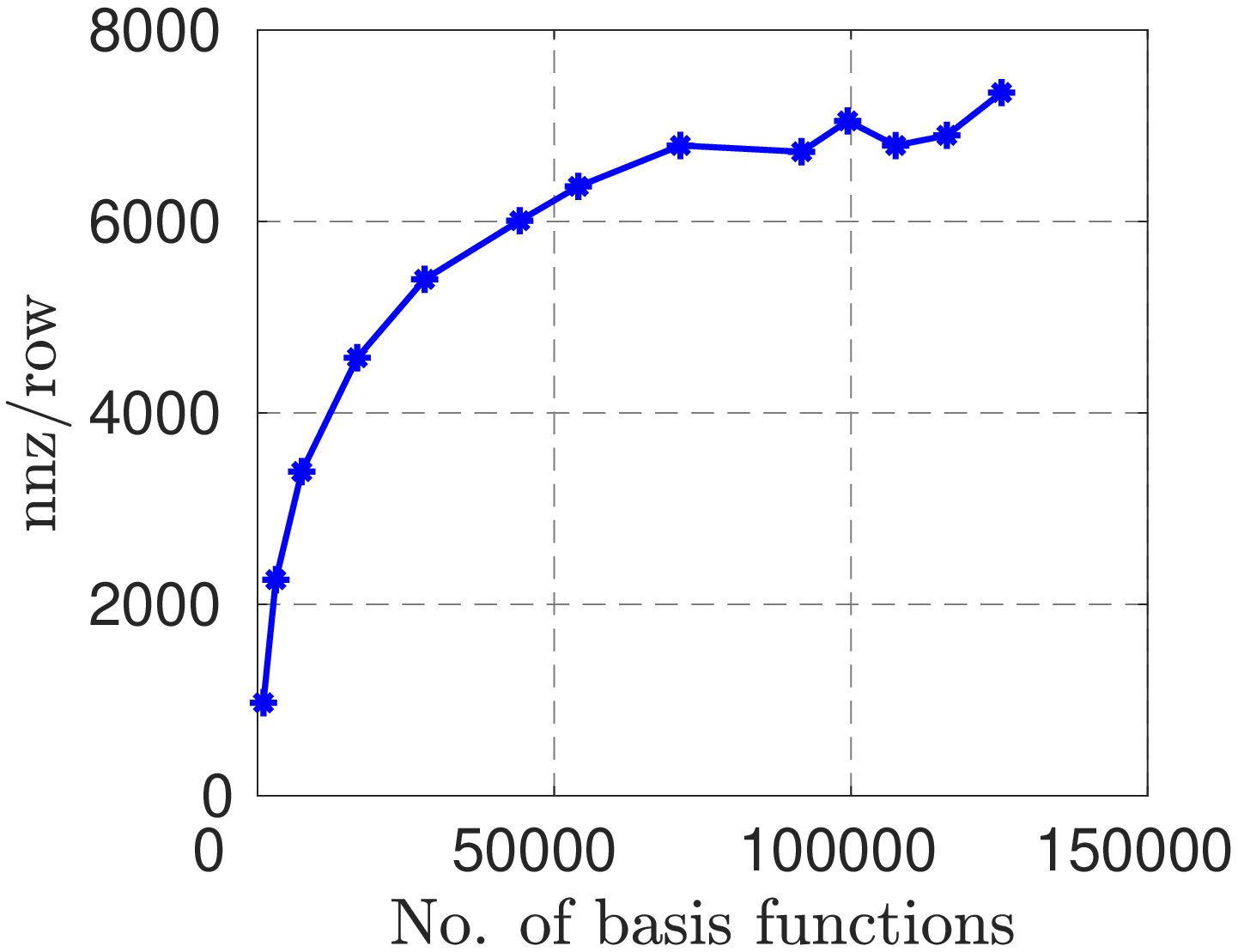}
		%\caption{}
	\end{subfigure}
	\begin{subfigure}[b]{0.45\textwidth}
		\includegraphics[width=\textwidth]{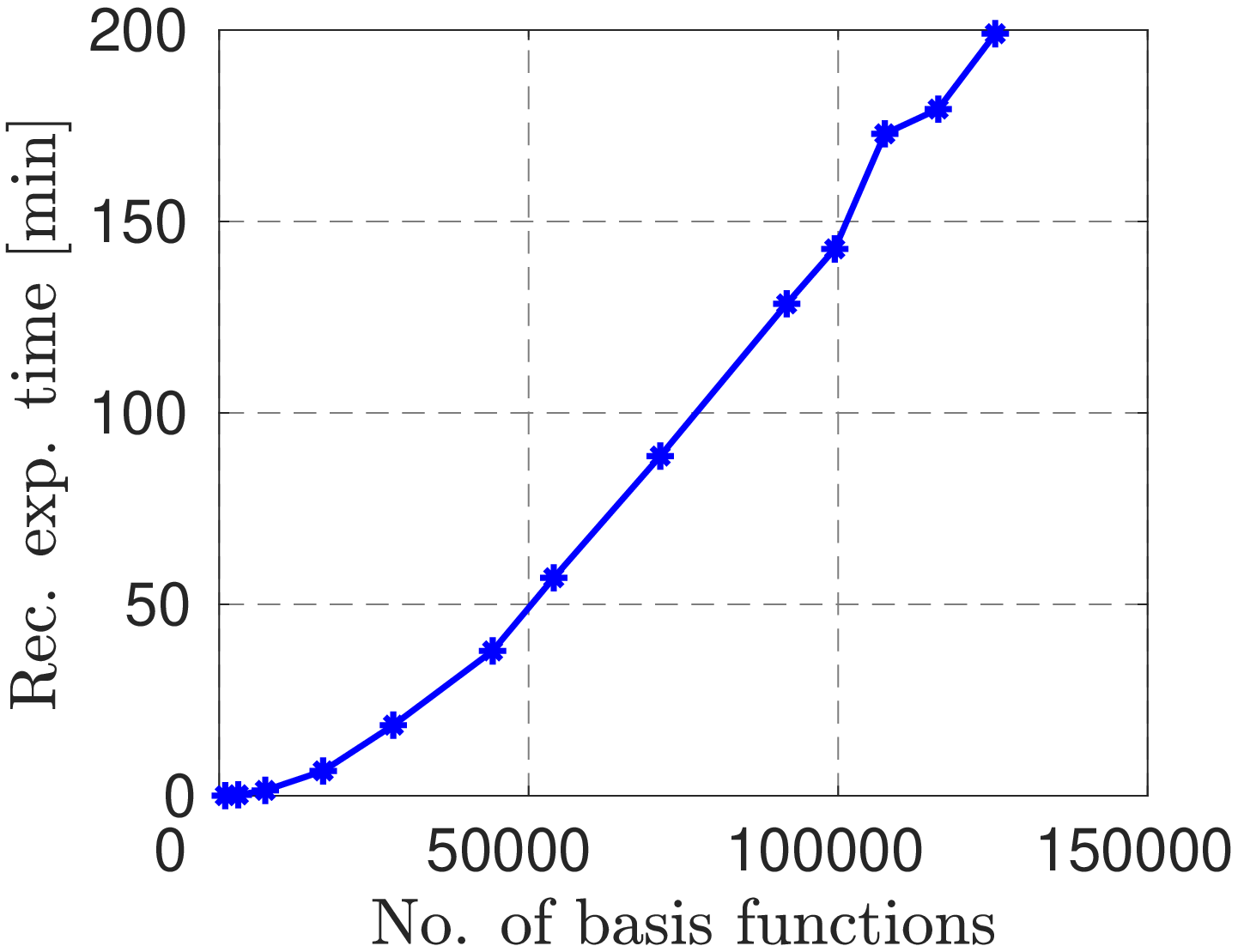}
		%\caption{}
	\end{subfigure}
	\caption{Recursive expansion in the last SCF cycle of \mbox{HF/3-21G} calculations performed  for water clusters. (Left panel) Number of non-zero elements per row in the density matrix. (Right panel) Recursive expansion execution time including computation of eigenvectors.}
	\label{fig:water_nnzrow_puritime}
\end{figure}

\begin{figure}[h!]
	\centering
	\captionsetup[subfigure]{justification=centering}
	\begin{subfigure}[b]{0.45\textwidth}
		\includegraphics[width=\textwidth]{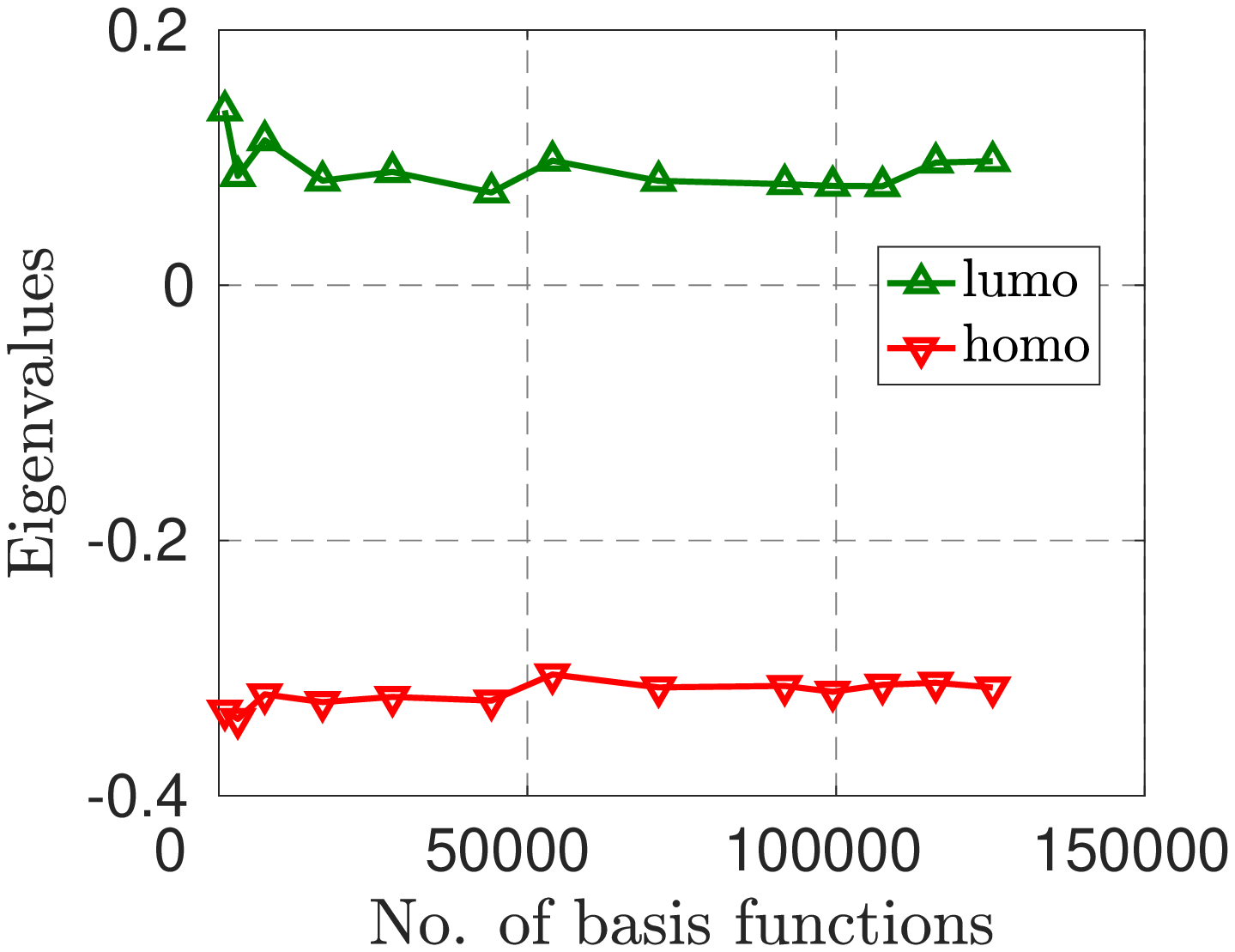}
		%\caption{}
	\end{subfigure}
	\begin{subfigure}[b]{0.45\textwidth}
		\includegraphics[width=\textwidth]{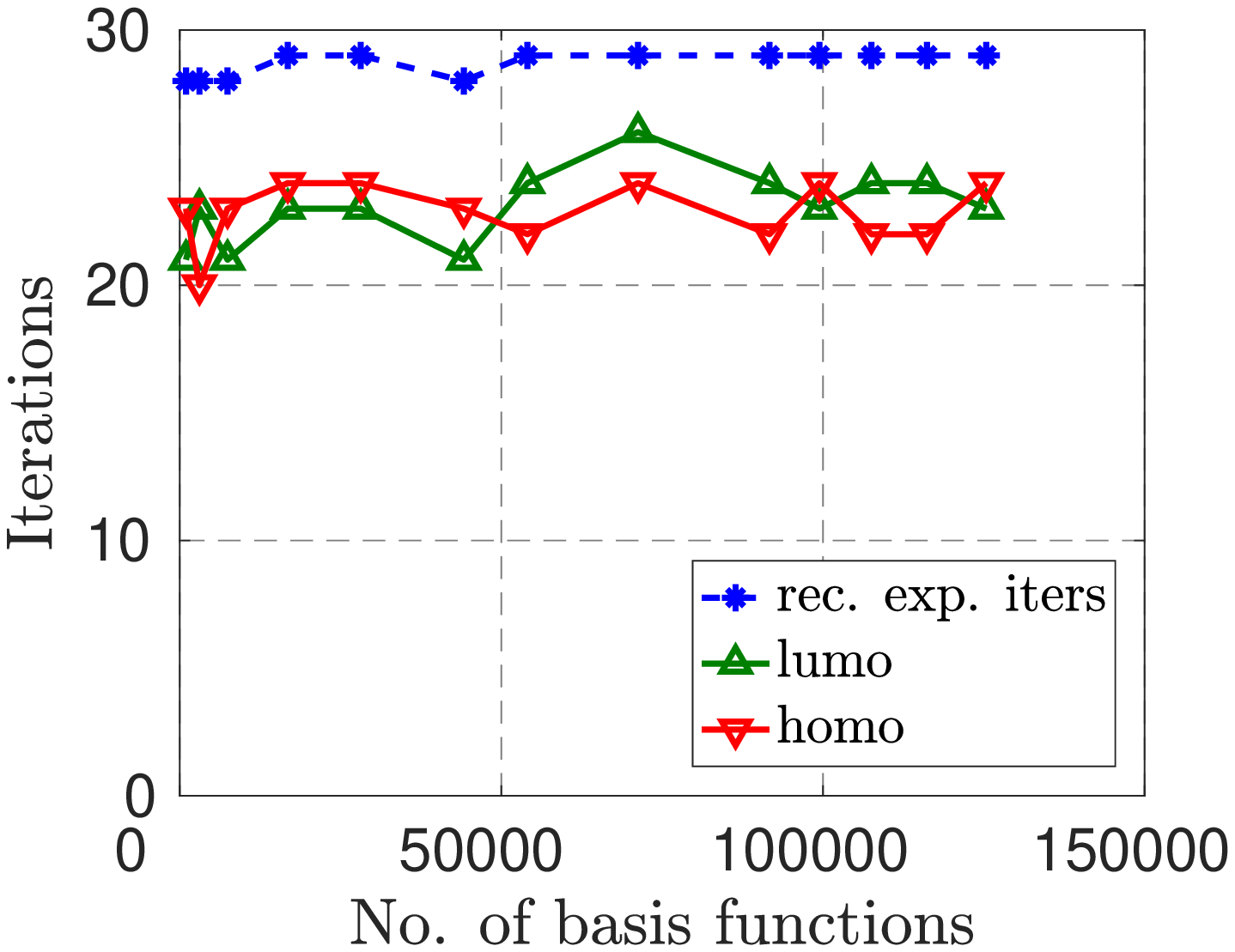}
		%\caption{}
	\end{subfigure}
	\caption{Recursive expansion in the last SCF cycle of \mbox{HF/3-21G} calculations performed  for water clusters. The homo-lumo gap (left panel) and the number of recursive expansion iterations (blue dashed line on the right panel) do not change significantly with systems size. In addition, the right panel presents recursive expansion iterations chosen for computing homo and lumo eigenpairs.
	}
	\label{fig:water_homo_lumo_gap}
\end{figure}

The number of non-zeros per row in the density matrix and total time of the recursive expansion are presented in Figure~\ref{fig:water_nnzrow_puritime}.
The homo-lumo gap does not change significantly with increasing system size, and therefore the total number of recursive expansion iterations does not change significantly either, see the right panel in Figure~\ref{fig:water_homo_lumo_gap}.

\begin{figure}[h!]
	\centering
	\captionsetup[subfigure]{justification=centering}
	\begin{subfigure}[b]{0.45\textwidth}
		\includegraphics[width=\textwidth]{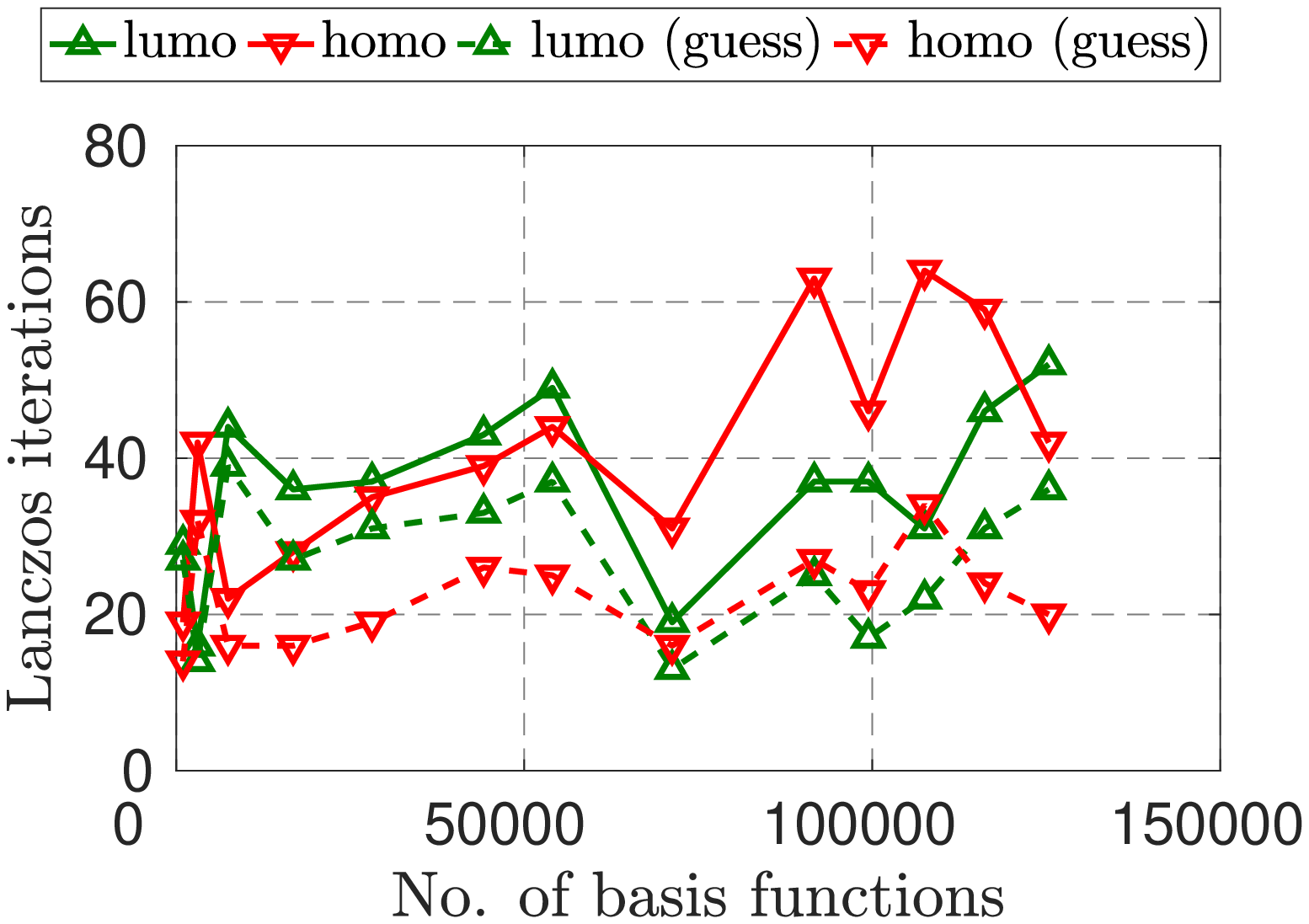}
	\end{subfigure}
	\begin{subfigure}[b]{0.45\textwidth}
		\includegraphics[width=\textwidth]{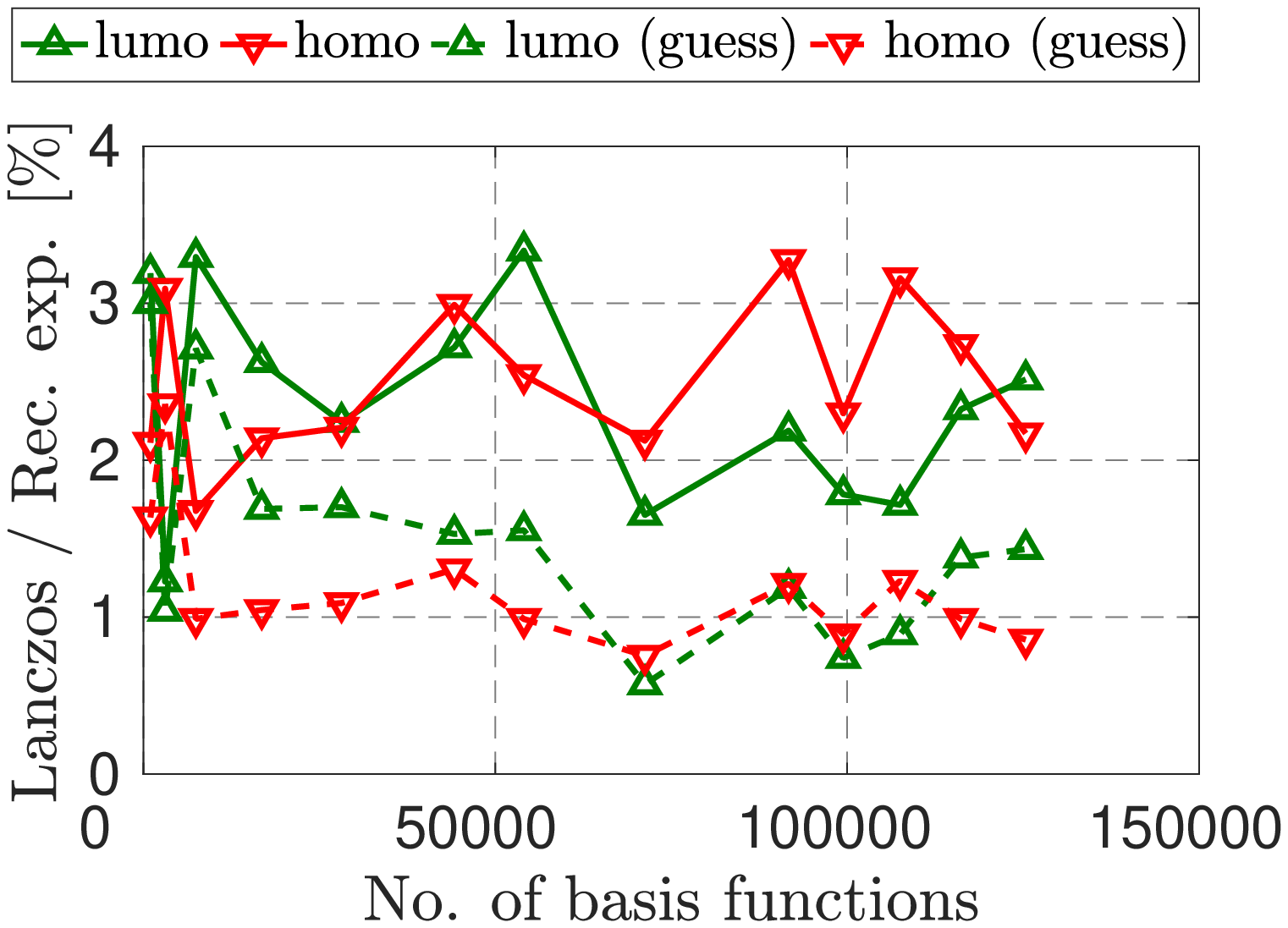}
		%\caption{}
	\end{subfigure}
	\caption{Recursive expansion in the last SCF cycle of \mbox{HF/3-21G} calculations performed  for water clusters. Left panel: Number of Lanczos iterations obtained with random initial guesses (solid lines) and eigenvectors from the previous SCF cycle as initial guesses (dashed lines). Right panel: Corresponding fraction of time spent on performing Lanczos iteration relative to the total execution time of the recursive expansion.
	}
	\label{fig:water_lanczos_iters}
\end{figure}

The number of Lanczos iterations and fraction of time required for the Lanczos algorithm relative to the total recursive expansion time is given in Figure~\ref{fig:water_lanczos_iters}. In the previous calculations we used random starting guess for Lanczos iterations.
If the SCF calculation is near convergence, then, if available, the homo and lumo eigenvectors from the previous SCF cycle can be used as starting guesses for the Lanczos algorithm in the current cycle.
In Figure~\ref{fig:water_lanczos_iters} we compare results obtained with random initial guesses and results with eigenvectors from the previous SCF cycle as guesses. The number of Lanczos iterations were reduced almost by a factor of two when non-random initial guess was used. However, one should be careful with using eigenvectors from  previous SCF cycles as initial guesses. If the calculations are still far from  self-consistency, the eigenvalue locations of two Fock matrices in consecutive cycles might differ significantly,
resulting in worse performance of the Lanczos algorithm compared to calculations with random initial guess or even misconvergence.

\subsubsection{Protein-water molecular systems}

We created a set of molecular systems by adding water molecules around a protein fragment 1RVS with  geometry taken  from the  Protein  Data  Bank  (PDB). We used the structure labeled ``model 1'' in the PDB file. We have used Gromacs to add water around the protein and performed a molecular dynamics simulation on the obtained system. Then all the atoms of the protein fragment and the surrounding water  located  within a certain distance from the protein were extracted. By increasing distance we include more atoms and thus obtain molecular systems of increasing size. The protein-water system was embedded in a set of classical point charges
as described in Ref.~\citenum{RudbergProteins}.

Hartree-Fock calculations were performed on the protein fragment surrounded by water molecules within various distances using the Gaussian basis set 6-31G**. Initial guesses for density matrices were obtained from calculations with the smaller basis set 3-21G. SCF convergence and truncation tolerances were set to $\tau_{\textrm{SCF}} = 3\cdot10^{-4}$ and $\tau_{\textrm{puri}} = 10^{-3}$, respectively.

\begin{figure}[h!]
	\centering
	\captionsetup[subfigure]{justification=centering}
	\begin{subfigure}[b]{0.45\textwidth}
		\includegraphics[width=\textwidth]{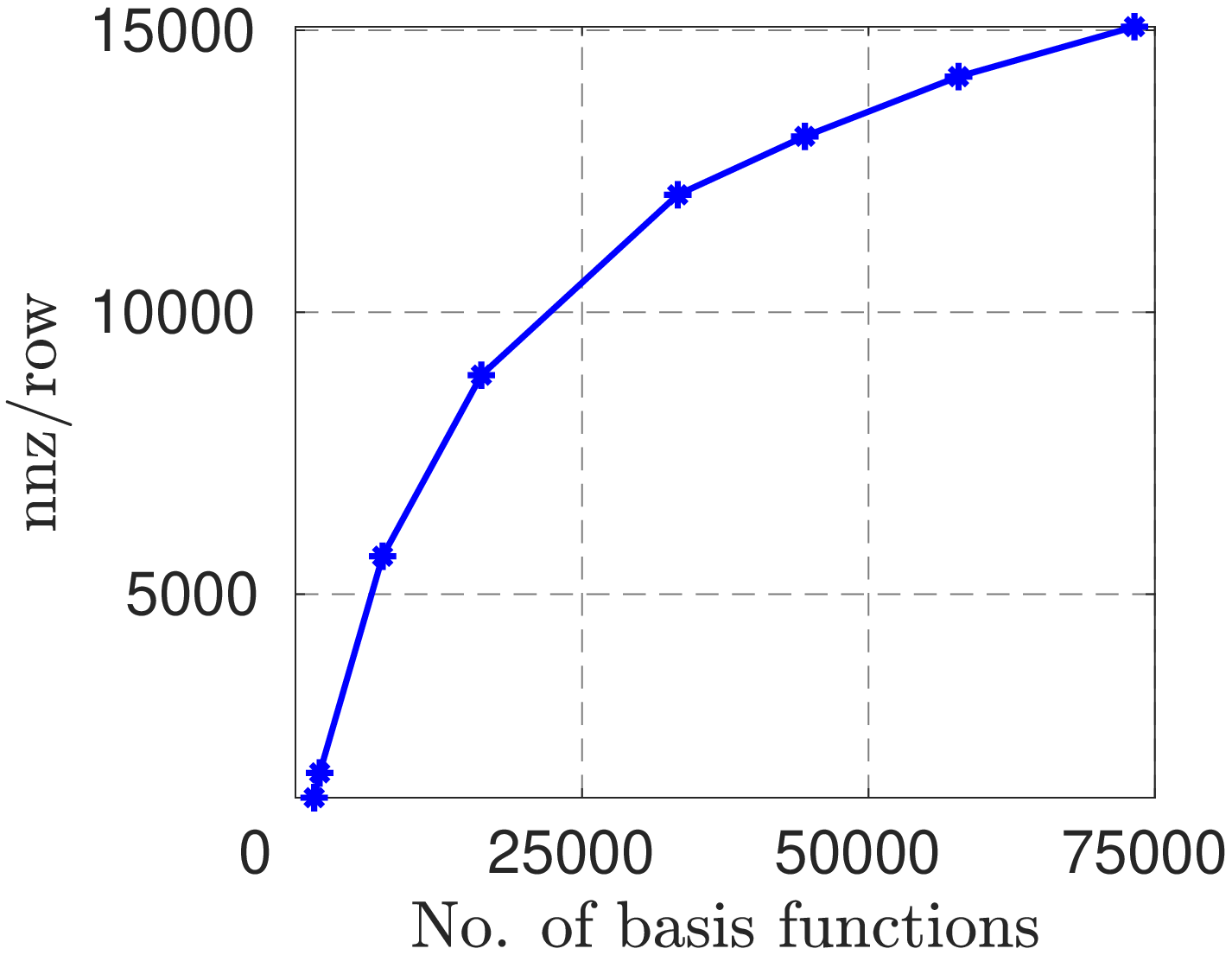}
		%\caption{}
	\end{subfigure}
	\begin{subfigure}[b]{0.45\textwidth}
		\includegraphics[width=\textwidth]{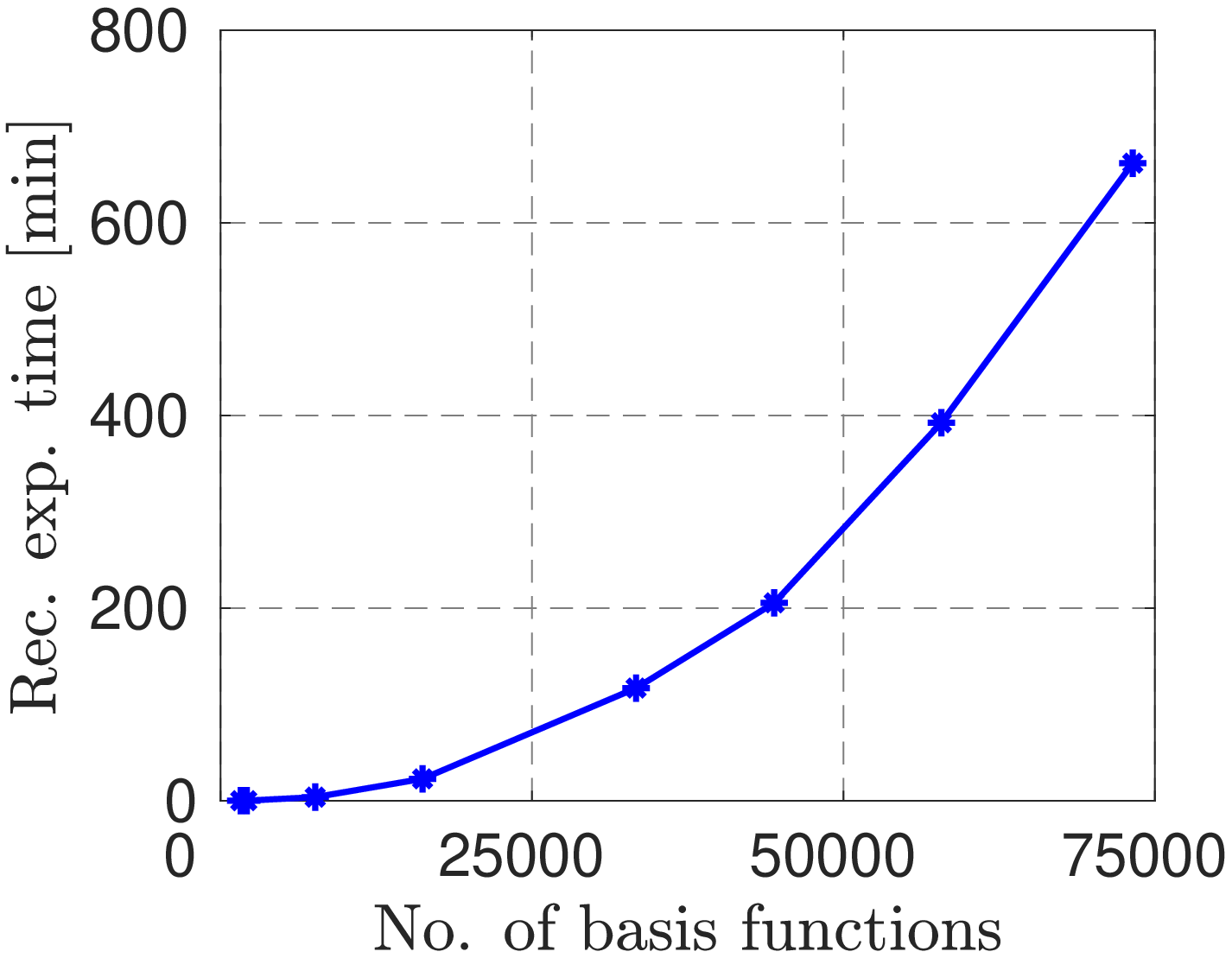}
		%\caption{Unoccupied part of the spectrum}
	\end{subfigure}
	\caption{Recursive expansion in the last SCF cycle of \mbox{HF/6-31G**} calculations performed  for protein-water systems. Left panel: Number of non-zero elements per row in the density matrix. Right panel: Recursive expansion execution time including computation of eigenvectors.}
	\label{fig:protein_nnzrow_puritime}
\end{figure}

\begin{figure}[h!]
	\centering
	\captionsetup[subfigure]{justification=centering}
	\begin{subfigure}[b]{0.45\textwidth}
		\includegraphics[width=\textwidth]{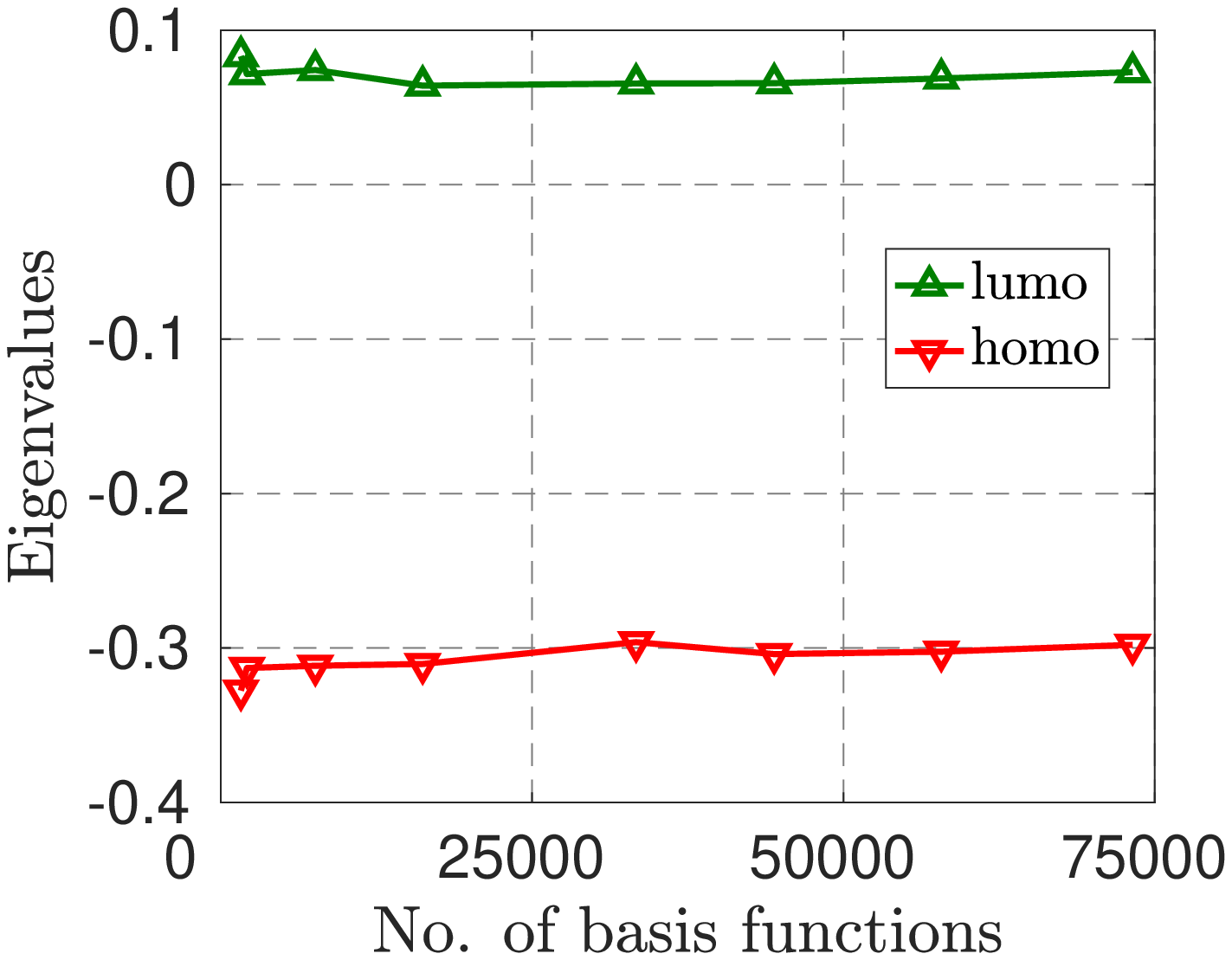}
		%\caption{Number of Lanczos iterations}
	\end{subfigure}
	\begin{subfigure}[b]{0.45\textwidth}
		\includegraphics[width=\textwidth]{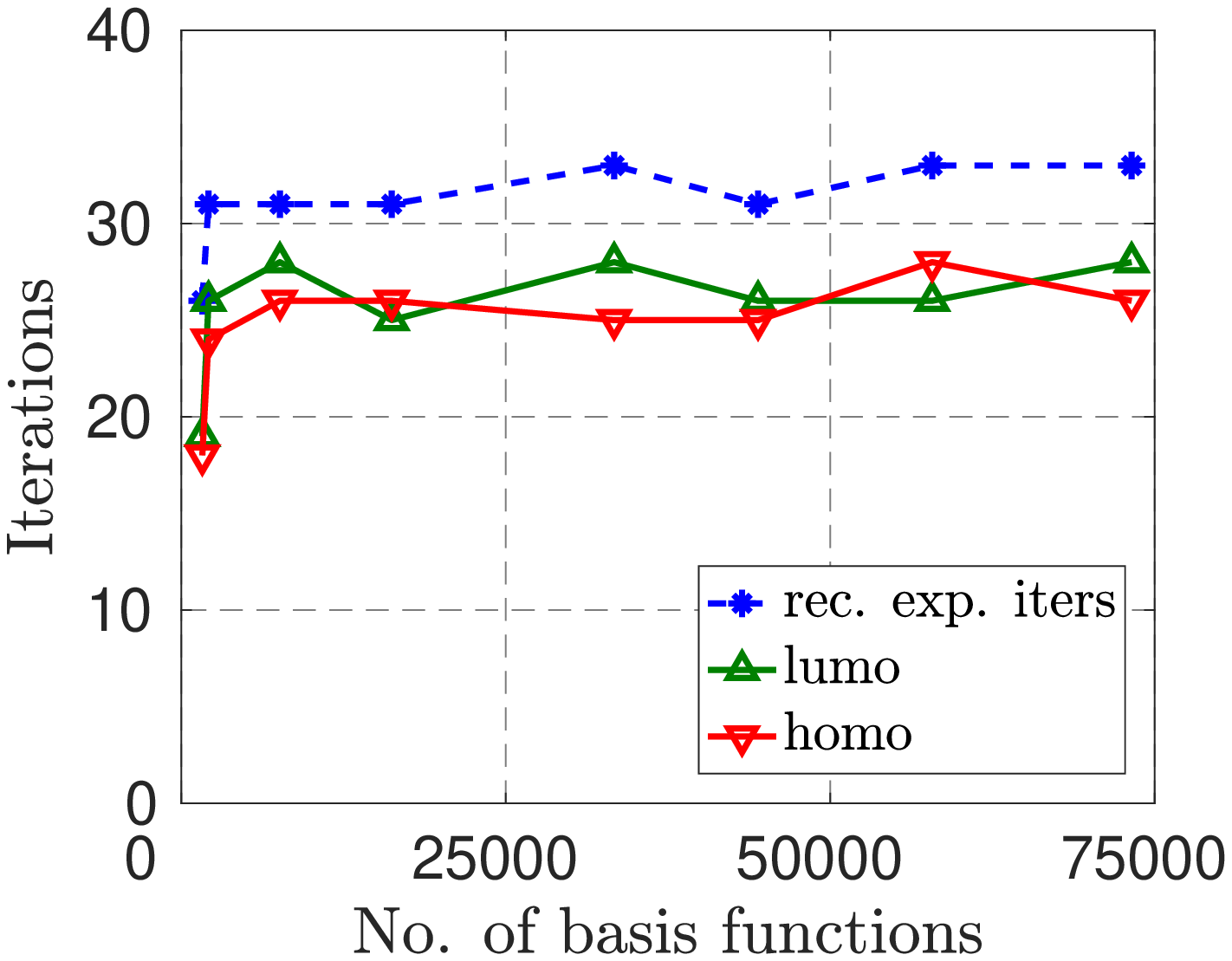}
		%\caption{Chosen recursive expansion iteration}
	\end{subfigure}
	\caption{Recursive expansion in the last SCF cycle of \mbox{HF/6-31G**} calculations performed  for protein-water systems. The homo-lumo gap (left panel) and  the number of recursive expansion iterations (blue dashed line in the right panel) do not change significantly with systems size. In addition, the right panel presents recursive expansion iterations chosen for computing homo and lumo eigenpairs.}
	\label{fig:protein_homo_lumo_gap}
\end{figure}

\begin{figure}[h!]
	\centering
	\captionsetup[subfigure]{justification=centering}
	\begin{subfigure}[b]{0.45\textwidth}
		\includegraphics[width=\textwidth]{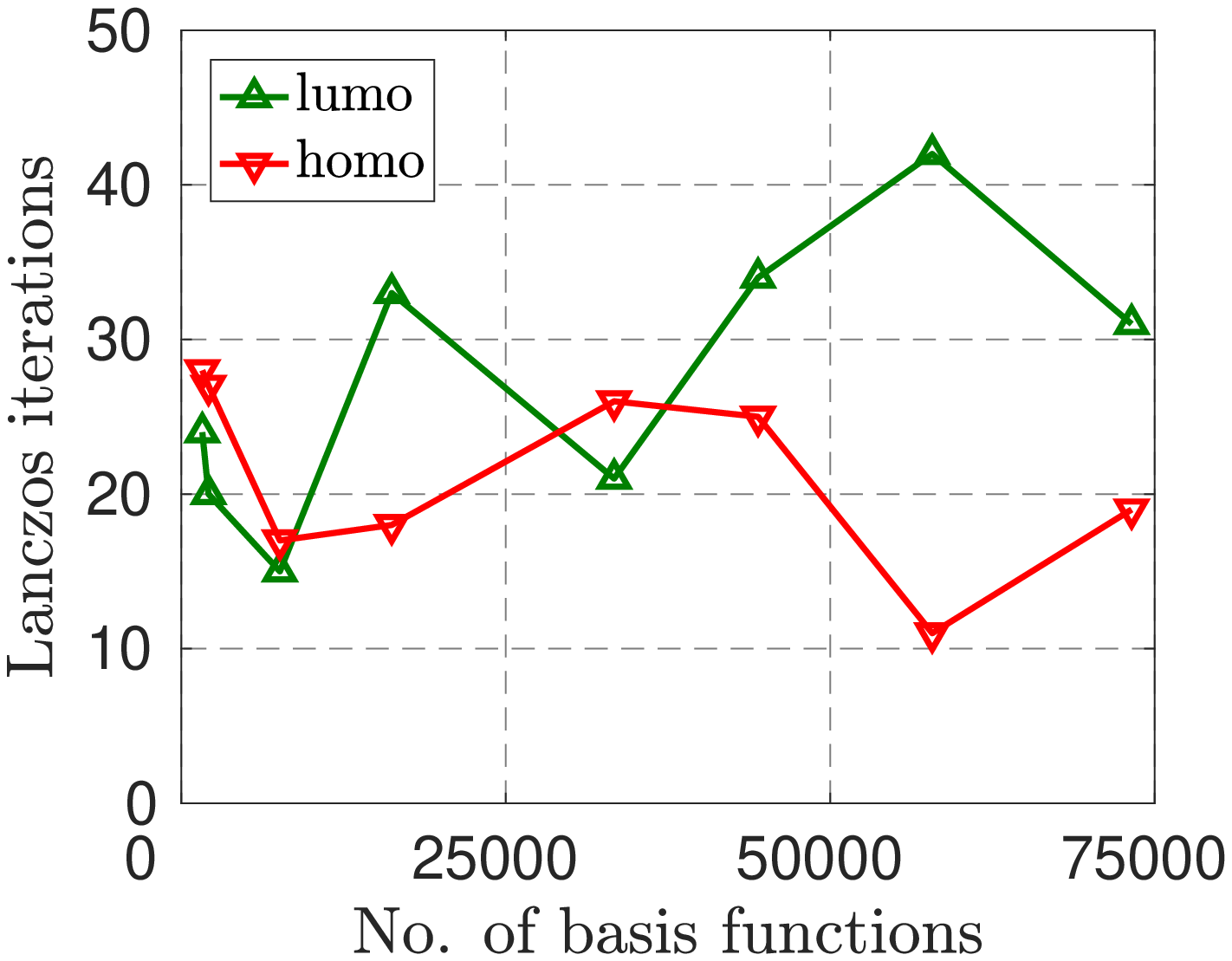}
		%\caption{Number of Lanczos iterations}
	\end{subfigure}
	\begin{subfigure}[b]{0.45\textwidth}
		\includegraphics[width=\textwidth]{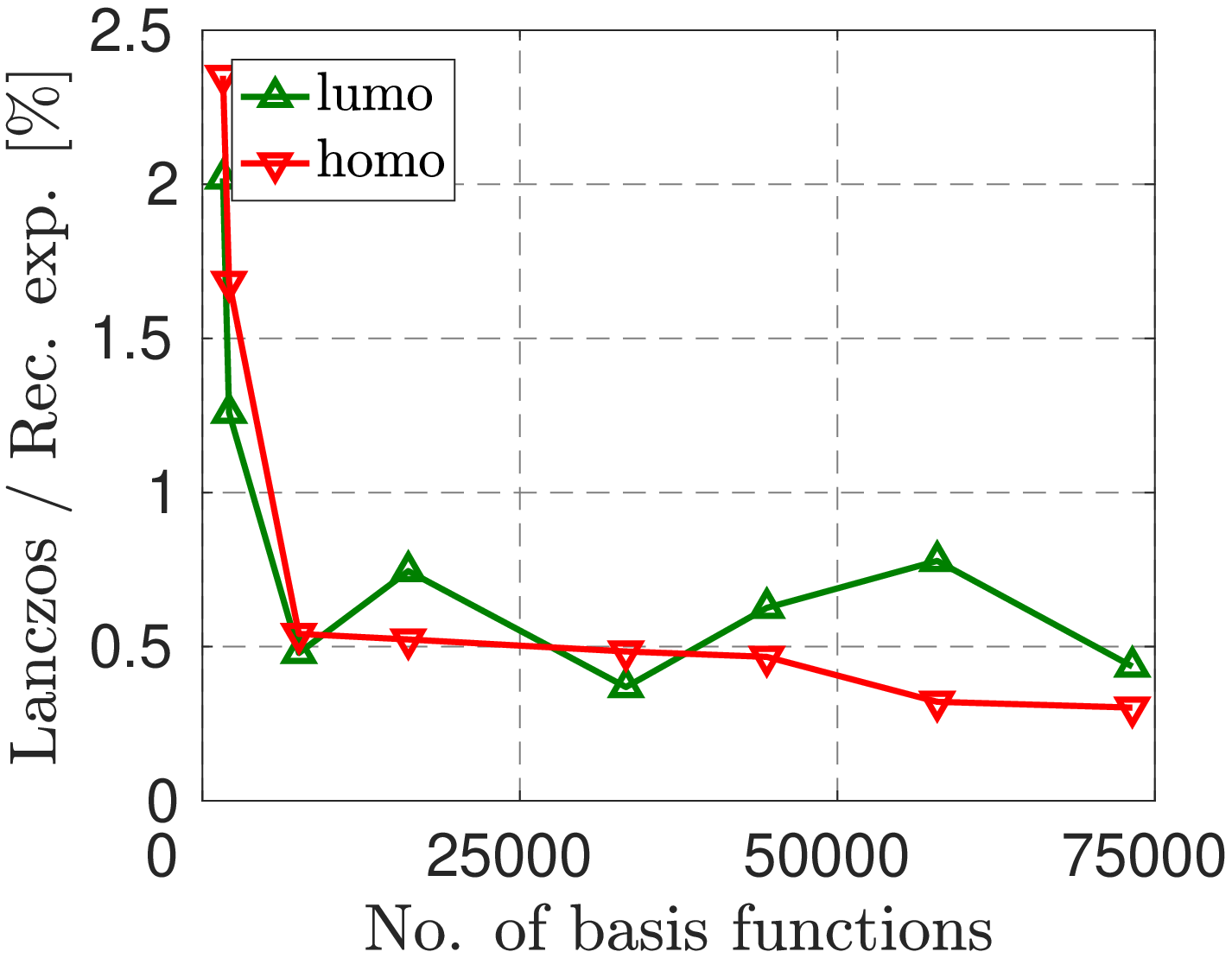}
		%\caption{Ratio of times for Lanczos iterations and the recursive expansion}
	\end{subfigure}
	\caption{Recursive expansion in the last SCF cycle of \mbox{HF/6-31G**} calculations performed  for protein-water systems. Left panel: Number of Lanczos iterations required for computing homo and lumo eigenpairs. Right panel: Corresponding fraction of time spent on performing Lanczos iteration relative to the total execution time of the recursive expansion. }
	\label{fig:protein_lan_iter_time}
\end{figure}

The number of non-zeros per row in the density matrix and total time of the recursive expansion are presented in Figure~\ref{fig:protein_nnzrow_puritime}.
Intermediate matrices in the recursive expansion are significantly dense and the number of nonzero elements per row is increasing with system size, which means that the linear scaling regime is not yet reached for the considered system sizes.
As in the case with water clusters, the number of recursive expansion iterations does not change significantly for varying system size, see the right panel in  Figure~\ref{fig:protein_homo_lumo_gap}.
The number of Lanczos iterations and the fraction of time required for the Lanczos algorithm to the total recursive expansion time is given in Figure~\ref{fig:protein_lan_iter_time}. Computation of homo and lumo eigenvectors takes less than 1 percent of the total recursive expansion time.

\subsection{Comparison with related algorithms}

\subsubsection{Shift-and-square method}

Originally, the shift-and-square method was applied to the matrix $F$~\cite{wang1994electronic}. We perform Hartree-Fock calculations on the alkane chains containing 17 and 317 atoms and compare the number of Lanczos iterations required for the shift-and-square and the purify-shift-and-square methods. We apply the shift-and-square method to the scaled Fock matrix $X_0=p_0(F)$. Since this transformation does not change the relative distribution of eigenvalues, it should not influence significantly the number of required Lanczos iterations.
The shift in the shift-and-square method should be located in the homo-lumo gap, but it is usually chosen arbitrarily. Therefore we test 16 different shifts distributed equidistantly inside the homo-lumo gap.
The number of Lanczos iterations for each shift is presented in Tables~\ref{table:sigma_numiter} and~\ref{table:sigma_numiter2} for alkane chains containing 17 and 317 atoms, respectively. The maximum allowed number of Lanczos iterations was 5000. Shifts resulting in the smallest number of Lanczos iterations required for computing homo and lumo eigenpairs are emphasized with bold font. For comparison, the purify-shift-and-square method requires 24 and 30 Lanczos iterations for computing homo and lumo eigenpairs, respectively, in the case of 17 atoms, and 26 and 163 Lanczos iterations, respectively, in the case of 317 atoms.

We also applied the shift-and-square method for computing homo and lumo eigenvectors of the C$_9$H$_{12}$N$_2$O$_4$ molecule extended with
alkane chains (6042 atoms in total). We again pick 16 different shifts distributed equidistantly inside the homo-lumo gap, but the Lanczos algorithm did not converge in 5000 iterations for any of them. In comparison, using the purify-shift-and-square method we computed the homo and lumo eigenpairs in 14 and 20 Lanczos iterations, respectively.

\begin{table}[ht!]
	\centering
	\caption{Number of Lanczos iterations required for computing homo and lumo eigenvectors using the shift-and-square method applied to the matrix $X_0=p_0(F)$. The matrix $F$ is taken from the last  SCF cycle of the \mbox{HF/6-311G*} calculation performed on the alkane chain containing 17 atoms giving 126 basis functions.}
	\label{table:sigma_numiter}
	\begin{tabular}{|l|l|l|l|l|l|l|}
		\hline
		$\sigma$ & -0.419804          & -0.381555 & -0.343306 & -0.305057          & -0.266808 & -0.228558 \\
		Iters    & 5000               & 2606      & 2972      & 4012               & 666       & 3568      \\
		Eigv     & --                 & -0.437706 & -0.437706 & -0.437706          & -0.437706 & -0.437706
		\\
		\hline\hline
		$\sigma$ & \textbf{-0.190309} & -0.15206  & -0.113811 & \textbf{-0.075562} & -0.037313 & 0.000936  \\
		Iters    & \textbf{409}       & 456       & 1074      & \textbf{438}       & 3102      & 5000      \\
		Eigv     & -0.437706          & -0.437706 & 0.159324  & 0.159324           & 0.159324  & --
		\\
		\hline\hline
		$\sigma$ & 0.039185           & 0.077434  & 0.115684  & 0.153933           &           &           \\
		Iters    & 1222               & 1322      & 4750      & 5000               &           &           \\
		Eigv & 0.159324 &	0.159324 &	0.159324 &	--
		& &\\
		\hline
	\end{tabular}
\end{table}

\begin{table}[ht!]
	\centering
	\caption{Number of Lanczos iterations required for computing homo and lumo eigenvectors using the shift-and-square method applied to the matrix $X_0=p_0(F)$. The matrix $F$ was taken from the last  SCF cycle of the \mbox{HF/6-311G*} calculation performed on the alkane chain containing 317 atoms giving 2526 basis functions.}
	\label{table:sigma_numiter2}
	\begin{tabular}{|l|l|l|l|l|l|l|}
		\hline
		$\sigma$ & -0.318091          & -0.291864 & -0.265637          & -0.23941  & -0.213183 & -0.186956 \\
		Iters    & 5000               & 5000      & 4847               & 5000      & 3975      & 5000      \\
		Eigv     & --                 & --        & -0.385602          & --        & -0.385602 & --        \\
		\hline\hline
		$\sigma$ & \textbf{-0.160729} & -0.134502 & \textbf{-0.108275} & -0.082048 & -0.055821 & -0.029594 \\
		Iters    & \textbf{1951}      & 3624      & \textbf{4936}      & 5000      & 5000      & 5000      \\
		Eigv     & -0.385602          & -0.385602 & 0.156964           & --        & --        & --        \\
		\hline\hline
		$\sigma$ & -0.003366          & 0.022861  & 0.049088           & 0.075315  &           &           \\
		Iters    & 5000               & 5000      & 5000               & 5000      &           &           \\
		Eigv     & --                 & --        & --                 & --        &           &           \\
		\hline
	\end{tabular}
\end{table}

\subsubsection{Shift-and-project method}
\label{sec:shift_and_project}

The density matrix $D$ is a projection matrix onto the occupied subspace.
The homo eigenvector is equal to the eigenvector corresponding to the
largest eigenvalue of the matrix $D(F-\lambda_{\textrm{min}} I)$ and the lumo eigenvector is equal to the eigenvector corresponding to the
smallest eigenvalue of the matrix $(I-D)(F-\lambda_{\textrm{max}} I)$~\cite{xiang2007linear}.
In exact arithmetics the matrix $D$ is also a projection matrix onto the
occupied subspace of $X_i$ for every $i$. The occupied subspace of the
matrices $X_i$ is spanned by the eigenvectors corresponding to the largest $n_\textrm{occ}$
eigenvalues, and the homo eigenvector corresponds to the smallest
eigenvalue of the matrix $D(X_i-I)$ and the lumo eigenvector corresponds
to the largest eigenvalue of $(I-D)X_i$. The original shift-and-project method applied to $F$ is essentially equivalent to the shift-and-project method applied to $X_0$.

In general, the projection matrix onto the occupied subspace of $X_i$ is not known in advance.  One may use the density matrix approximation from the previous SCF cycle.
However, it is unclear what impact projection using an approximate density matrix has on eigenvector accuracy.
Another possibility is to use the final matrix $X_n$ in the recursive
expansion as an approximation of the density matrix.  However, this
approach requires saving two matrices $X_i$, $i = i_{\textrm{homo}},
i_{\textrm{lumo}}$. After performing the whole expansion, one may use the density matrix approximation $\widetilde{D}:=X_n$.

\begin{figure}[h!]
	\centering
	\captionsetup[subfigure]{justification=centering}
	\begin{subfigure}[b]{0.45\textwidth}
		\includegraphics[width=\textwidth]{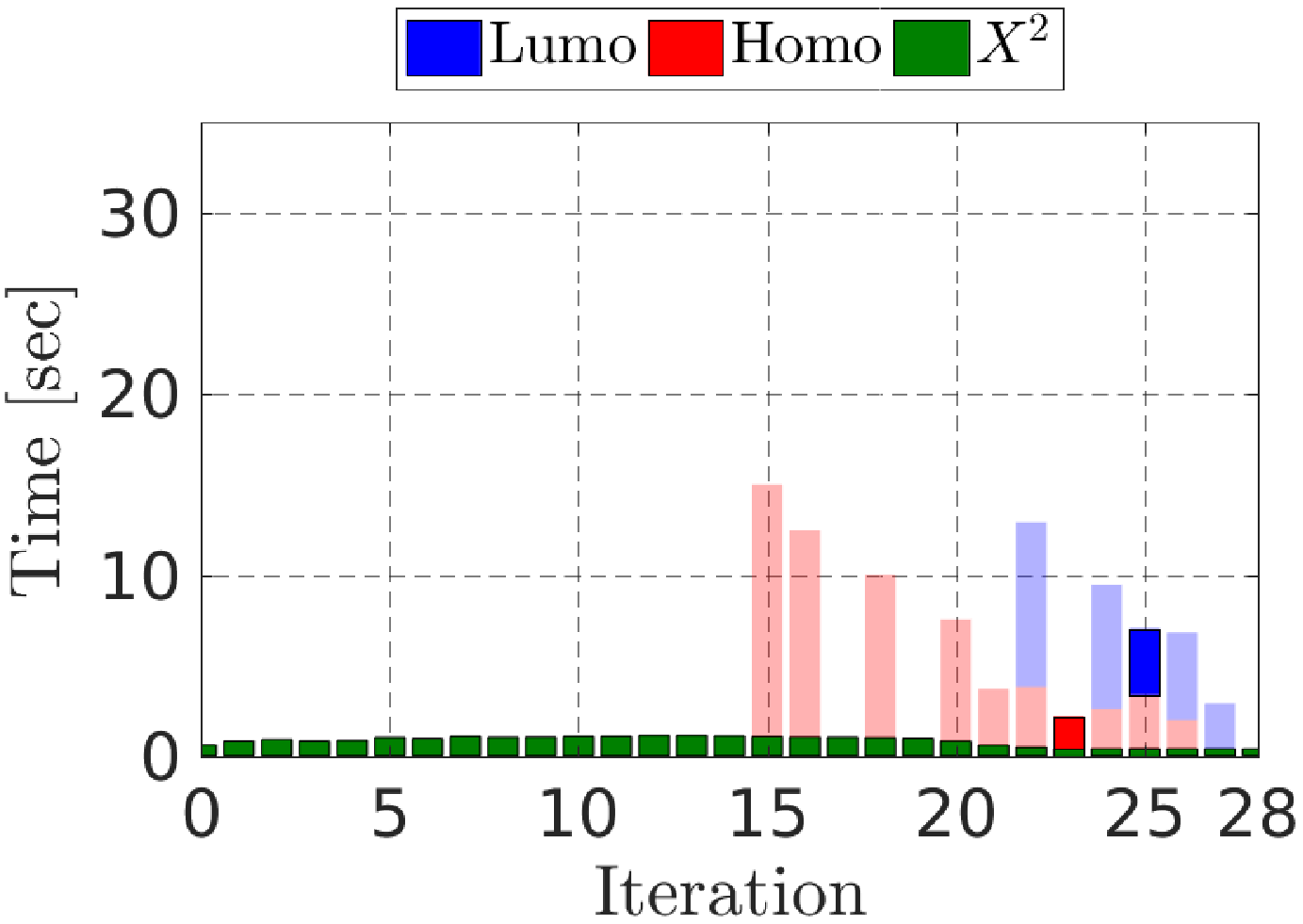}
		\caption{Purify-shift-and-square method}
	\end{subfigure}
	\begin{subfigure}[b]{0.45\textwidth}
		\includegraphics[width=\textwidth]{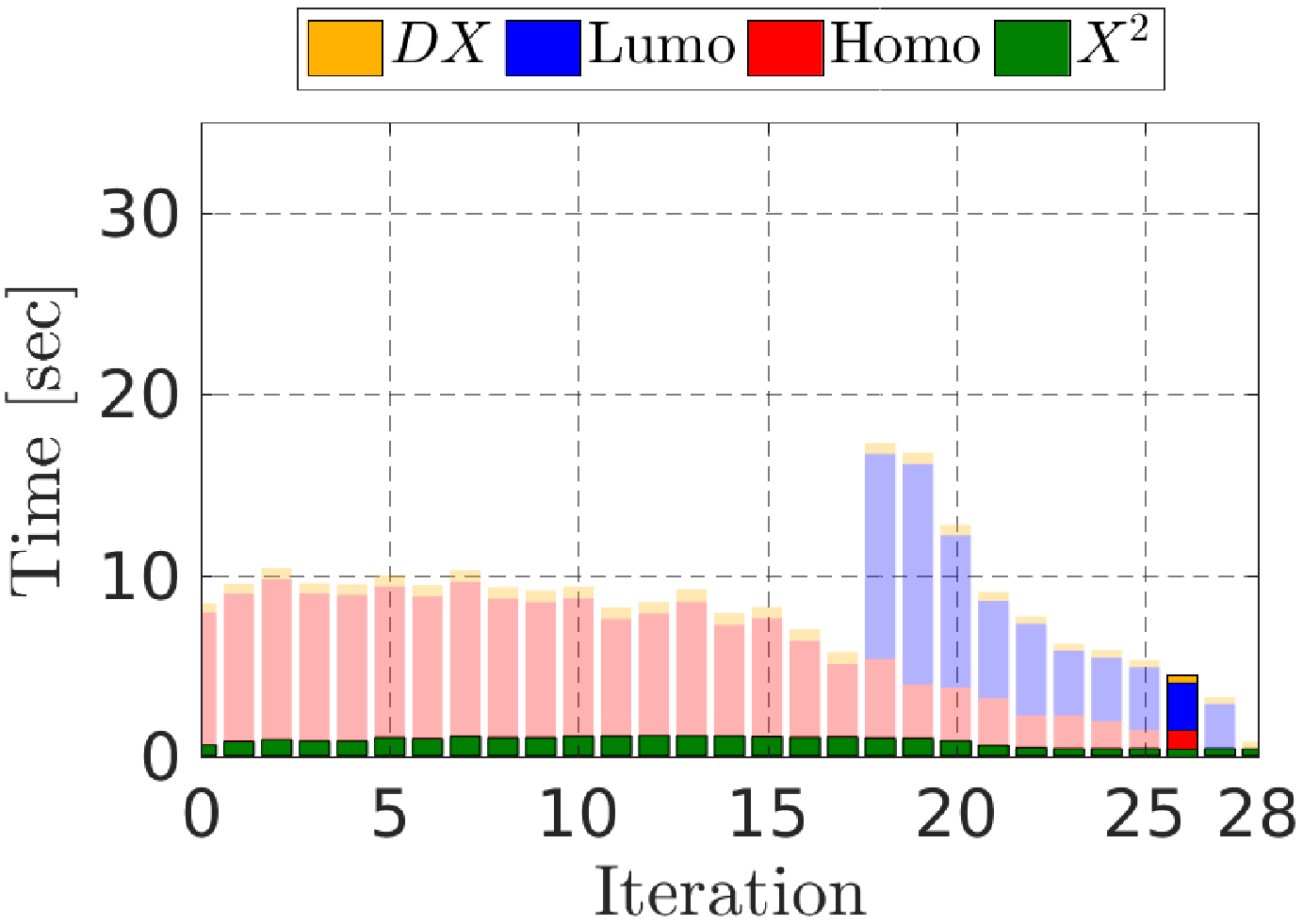}
		\caption{Purify-shift-and-project method}
	\end{subfigure}
	\caption{Recursive expansion in the last SCF cycle of \mbox{HF/6-311G*} calculations performed for the alkane chain containing 917 atoms, see section~\ref{sec:alkanes_and_mol}. The homo and lumo eigenvectors are computed in each iteration. Highlighted iterations for computing eigenvectors in the purify-shift-and-square method are chosen using Algorithm~\ref{alg:determine_iter}. The iteration which is highlighted for the purify-shift-and-project method requires the smallest amount of time for computing both eigenvectors. Note that we skip computation of eigenvectors in the last iteration due to accuracy loss.}
	\label{fid:alkanes_square_vs_proj}
\end{figure}

\begin{figure}[h!]
	\centering
	\captionsetup[subfigure]{justification=centering}
	\begin{subfigure}[b]{0.45\textwidth}
		\includegraphics[width=\textwidth]{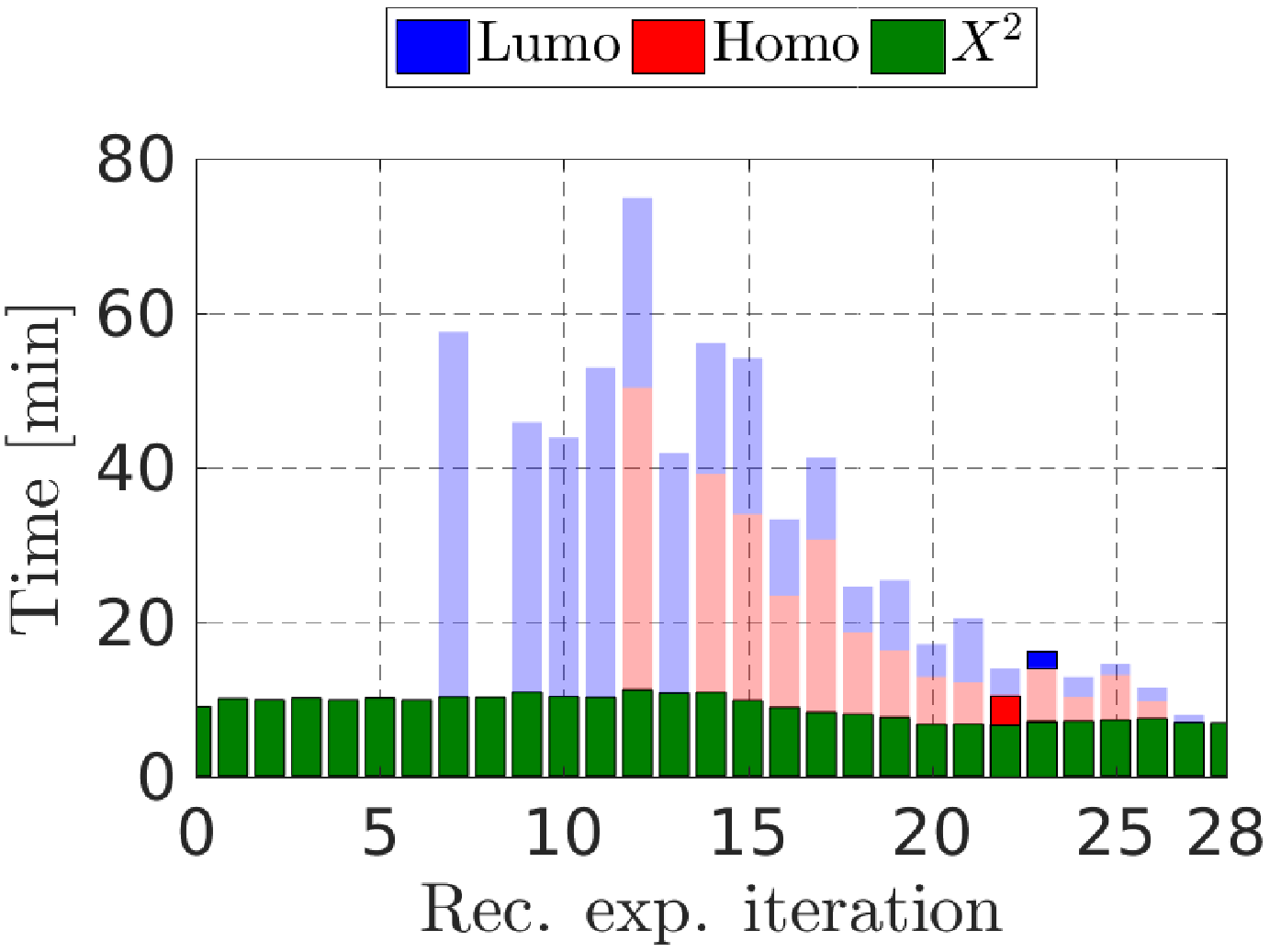}
		\caption{Purify-shift-and-square method}
	\end{subfigure}
	\begin{subfigure}[b]{0.45\textwidth}
		\includegraphics[width=\textwidth]{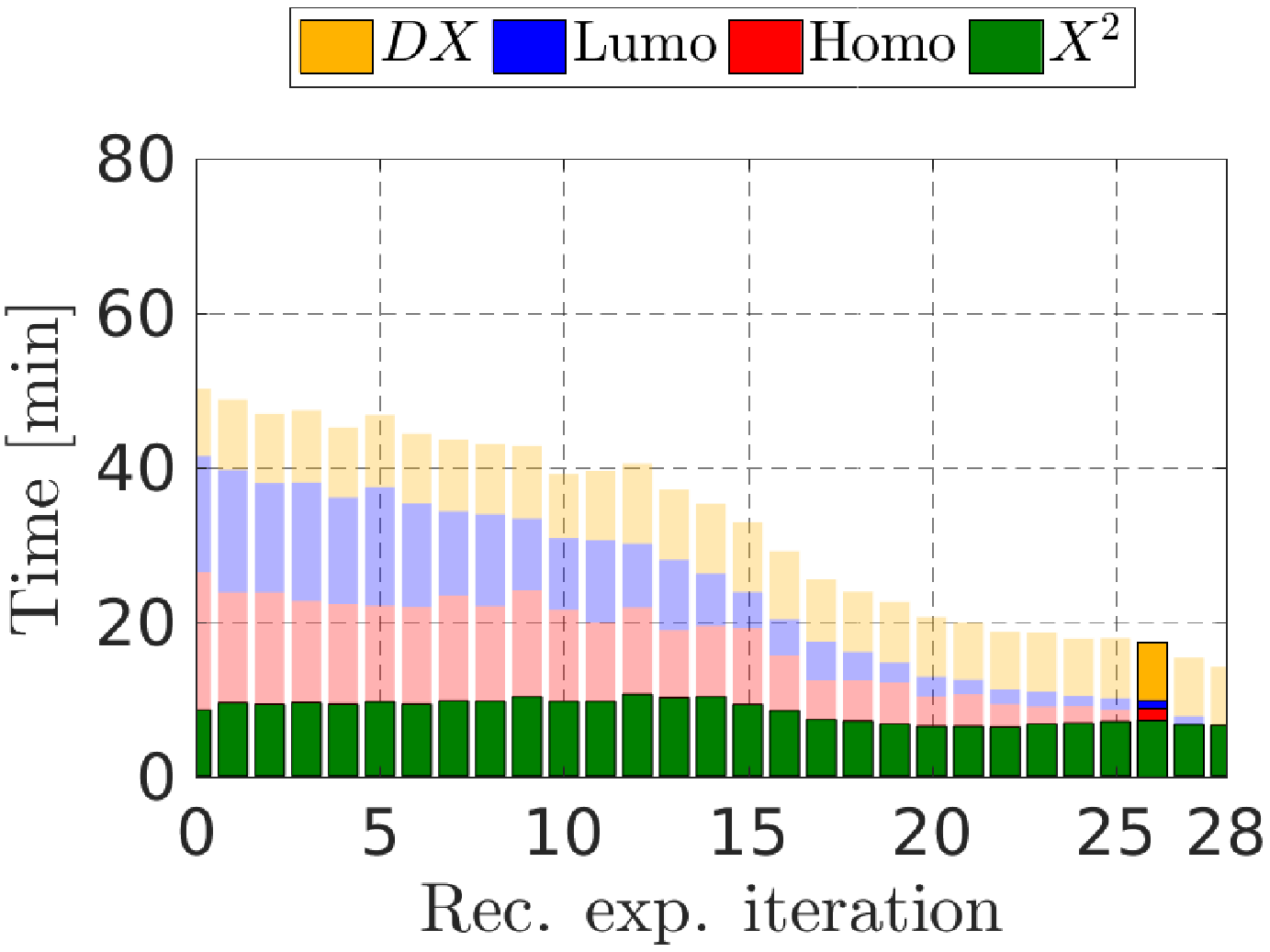}
		\caption{Purify-shift-and-project method}
	\end{subfigure}
	\caption{Recursive expansion in the last SCF cycle of \mbox{HF/3-21G} calculations performed for the water clusters containing 22950 atoms. The homo and lumo eigenvectors are computed in each iteration. Highlighted iterations for computing eigenvectors in the purify-shift-and-square method are chosen using Algorithm~\ref{alg:determine_iter}. The iteration which is highlighted for the purify-shift-and-project method requires the smallest amount of time for computing both eigenvectors. Note that we skip computation of eigenvectors in the last iteration due to accuracy loss.}
	\label{fid:water_square_vs_proj}
\end{figure}

Hartree-Fock calculations are performed for the alkane chain containing 917 atoms with basis set 6-311G* and for the water cluster containing 22950 atoms with basis set 3-21G. In Figures~\ref{fid:alkanes_square_vs_proj} and~\ref{fid:water_square_vs_proj} we present the execution time required for performing Lanczos iterations for computing homo and lumo eigenvectors using purify-shift-and-square and purify-shift-and-project methods in the last SCF cycles. We attempt to compute eigenvectors in each iteration of the recursive expansion. The maximum allowed number of Lanczos iterations is set to 500. Iterations where the Lanczos algorithm did not converge are not shown. In addition, in the purify-shift-and-square method conditions \eqref{eq:sigma_lumo} and \eqref{eq:sigma_homo} should be satisfied. The issue of selecting iteration in the purify-shift-and-project method has not been addressed here, but it could be done by maximizing the slope of the approximation polynomial at the eigenvalue of interest.

% 038

The projection in the purify-shift-and-project method does not affect the eigenvalue separation. Therefore one would in general expect a smaller time spent on Lanczos iterations in the purify-shift-and-project method compared to the purify-shift-and-square method. However, evaluation of a matrix product $\widetilde{D}X_i$ is required in the purify-shift-and-project method.
As shown in the example in Figure~\ref{fid:water_square_vs_proj} the time spent on the evaluation of this matrix product can be significantly larger than the reduction of time required for Lanczos iterations compared to the purify-shift-and-project method. 
Instead of performing one expensive matrix-matrix multiplication $\widetilde{D}X_i$, one may incorporate it into the Lanczos procedure and do two matrix-vector products instead. However, this will increase twice the number of the matrix-vector products in the Lanczos algorithm. This idea is not explored further here. Moreover, as stated above, the purify-shift-and-project method requires saving one or two matrices in memory and computation of eigenvectors is delayed until the end of the recursive expansion. We want to stress that with the purify-shift-and-square approach, eigenvectors are computed on-the-fly in the course of the recursive expansion without any need to store matrices in memory.

\section{Discussion}
\label{sec:discussion}
% computation of many eigenvectors near homo-lumo gap

In practice, the computation of more than one eigenvector near the homo-lumo gap may be needed. Let the homo eigenvector be computed in iteration $j$ using the purify-shift-and-square method. To compute the eigenvector corresponding to the second occupied eigenvalue (homo-1) one can use a deflation technique on the matrix $X_k$, where $k \leq j$.
Instead of working with the matrix $X_k^{\sigma_k} := g(X_k, \sigma_k)$ with chosen shift $\sigma_k$ one works with the deflated matrix
\begin{align}
	\widehat{X}_k^{\sigma_k} = X_k^{\sigma_k}-\delta y_{\homoF}y_{\homoF}^T,
\end{align}
where $y_{\homoF}$ is the computed homo eigenvector and $\delta = \frac{y_{\homoF}X_k^{\sigma_k}y_{\homoF}^T}{y_{\homoF}y_{\homoF}^T} - \lambda_{\max}(X_k^{\sigma_k})$.
Such rank-1 update of the matrix $X_k^{\sigma_k}$ will in general
destroy sparsity if any, and therefore one should not compute the matrix
$\widehat{X}_k^{\sigma_k}$ explicitly. Instead the deflation part
should be incorporated into the Lanczos procedure.

Recall that $\homo^{i}$ and $\lumo^{i}$ are the computed homo and lumo eigenvalues, respectively, in each iteration $i$ of the recursive expansion.
The eigenvalue bounds for homo-1 and lumo+1 can be obtained by applying Algorithm~\ref{alg:homo_lumo_bounds} with modified input values: $v_i \rightarrow u_i$ , $m_i \rightarrow u_i$ , $w_i \rightarrow \omega_i$, where
\begin{align}
	u_i^2    & = v_i^2 - (\homo^i - (\homo^i)^2)^2 - (\lumo^i -
	(\lumo^i)^2)^2, \\
	\omega_i & = w_i - (\homo^i - (\homo^i)^2) - (\lumo^i -(\lumo^i)^2).
\end{align}

The obtained eigenvalue bounds for homo-1 and lumo+1 can be used for selecting shift and recursive expansion iteration for computation of homo-1 and lumo+1 eigenvectors.

In Ref.~\citenum{Rub2011} a method for acceleration of the recursive expansion
(SP2ACC) was proposed. Modified polynomials stretch and fold the eigenspectrum
over itself resulting in an accelerated convergence while making sure that the
occupied and unoccupied parts of the eigenspectrum are not mixed.  Acceleration
may mix states within each of the occupied and unoccupied subspaces, making it
hard to keep track of the location of individual eigenvalues. However, in
practice, acceleration parameters are computed using outer homo and lumo
eigenvalue bounds. Under the assumption~\eqref{eq:assump_eig_bounds} we may
assume that homo and lumo eigenvalues remain the closest eigenvalues to the
homo-lumo gap in each subspace, and they are well separated from the rest of the
spectrum. Therefore, the purify-shift-and-square method can be applied also if
SP2ACC recursive expansion is used, however, this possibility is not discussed
more in this work.

\section{Conclusion}

In this work the purify-shift-and-square method was used for computation of homo and lumo molecular orbitals for large systems in the context of recursive density matrix expansions. The purify-shift-and-square method which was initially discussed in Ref.~\citenum{interior_eigenvalues_2008} is based on two main ideas: 1) the shift-and-square approach, which shifts and folds the spectrum of the matrix for easier access to eigenvalues near the shift, 2) recursive expansion increases the separation of eigenvalues near the homo-lumo gap from the rest of the spectrum. In principle, the shift can be chosen anywhere in the homo-lumo gap. However, the choice of shift may significantly influence the performance of the iterative eigensolver and the accuracy of the computed eigenvector. In this work, shifts and recursive expansion iterations for computing homo and lumo eigenvectors are selected such that fast convergence of the iterative solver and sufficient accuracy of the eigenvector is obtained. Our choices are based on recent homo and lumo eigenvalue bounds~\cite{interior_eigenvalues_2014}.

A key feature of the proposed method is that it makes use of the recursive expansion  as an eigenvalue filter.
We have shown that the presented method allows efficient computation of non-degenerate homo and lumo molecular orbitals for large systems in a small fraction of the total recursive expansion time.

\begin{acknowledgement}

Support from the Swedish research council (grant no.~621-2012-3861) and the
Swedish national strategic e-science research program (eSSENCE) is gratefully
acknowledged. Computational resources were provided by the Swedish National
Infrastructure for Computing (SNIC) at the National Supercomputer Centre (NSC)
in Link{\"o}ping, Sweden.
  
\end{acknowledgement}

\bibliography{biblio}

\end{document}